%% file: Clustering.tex
\documentclass[iop]{emulateapj}
\usepackage{amsmath}
\usepackage{epstopdf}
\usepackage{multirow}

\newcommand{\gil}{GCD03}
\newcommand{\ada}{AMI05}
\newcommand{\sal}{SCP09}
\newcommand{\kms}{km s$^{-1}$}
\shorttitle{Structures in the CDFS up to z=1}
\shortauthors{Dehghan \& Johnston-Hollitt}
\begin{document}

\title{Clusters, Groups, and Filaments in the Chandra Deep Field-South up to Redshift 1}
\author{S. Dehghan and M. Johnston-Hollitt}
\affil{School of Chemical \& Physical Sciences, Victoria University of Wellington, PO Box 600, Wellington, 6140, New Zealand}
\email{siamak.dehghan@vuw.ac.nz}

\begin{abstract}
We present a comprehensive structure detection analysis of the 0.3 square degree area of the MUSYC-ACES field which covers the Chandra Deep Field-South (CDFS). Using a density-based clustering algorithm on the MUSYC and ACES photometric and spectroscopic catalogues we find 62 over-dense regions up to redshifts of 1, including, clusters, groups and filaments. We also present the detection of a relatively small void $\sim$ 10 Mpc$^2$ at $z\sim0.53$. All structures are confirmed using the DBSCAN method, including the detection of nine structures previously reported in the literature. We present a catalogue of all structures present including their central position, mean redshift, velocity dispersions, and classification based on their morphological and spectroscopic distributions. In particular we find 13 galaxy clusters and 6 large groups/small clusters. Comparison of these massive structures with published XMM-Newton imaging (where available) shows that 80\% of these structures are associated with diffuse, soft-band (0.4 - 1 keV) X-ray emission including 90\% of all objects classified as clusters. The presence of soft-band X-ray emission in these massive structures (${\rm M}_{200} \geq 4.9 \times 10^{13} {\rm M}_{\odot}$) provides a strong independent confirmation of our methodology and classification scheme. In the closest two clusters identified (z $<$ 0.13) high quality optical imaging from the Deep2c field of the Garching-Bonn Deep Survey reveals the cD galaxies and demonstrates that they sit at the centre of the detected X-ray emission. Nearly 60\% of the clusters, groups and filaments are detected in the known enhanced density regions of the CDFS at $z\simeq0.13$, $z\simeq0.52$, $0.68$, and $0.73$. Additionally, all of the clusters, bar the most distant, are found in these over-dense redshifts regions. Many of the clusters and groups exhibit signs of on-going formation seen in their velocity distributions, position within the detected cosmic web and in one case through the presence of tidally disrupted central galaxies exhibiting trails of stars. These results all provide strong support for hierarchical structure formation up to redshifts of 1.  

\end{abstract}


\section{Introduction}

Tracing over-dense regions along with the number density of galaxies provides a testbed for verifying hierarchical models and cosmological parameters \citep{mrg08}. Identifying and measuring the abundance of groups and clusters as a function of redshift provide significant constraints on cosmological models \citep{nsp13}. Additionally, studying a large sample of groups and clusters at various redshift ranges plays an important role in understanding cluster and galaxy formation and evolution. However, generation of such samples are rare and subject to availability of multiwavelength data and redshift information. For this reason, many studies have been conducted on legacy fields such as the Chandra Deep Field-South (CDFS) due to the abundance of deep multiwavelength observations: from the optical and near infrared \citep{avb01} up to X-ray \citep{gzw02} observations. There is a growing number of reliable spectroscopic and photometric redshift surveys of the CDFS, e.g. the Great Observatories Origins Deep Survey \citetext{GOODS, \citealp{pdn09,bmp10}}, which allow discovery of large-scale structures.

From observational methods such as X-ray and Sunyaev-Zel'dovich (SZ) detections, to hierarchical and partitioning clustering analysis, there is a wide range of detection methods used to identify the overdensity traces in fields such as the CDFS. To date, several groups and clusters have been detected in the CDFS, by a variety of methods. A cluster at $z \sim 0.15$ \citep{wmk04} and two large-scale structures at $z \sim 0.67$ and $z \sim 0.73$ \citetext{\citealp{gcd03}, here after \gil} have been discovered in intermediate redshift ranges, by analysing photometric and spectroscopic samples respectively. Studying the spatial clustering of extremely red objects (EROs) revealed a structure at $z=1.10$ \citep{dia07}. \citet{ami05}, here after \ada, detected 21 compact structures up to $z=1.4$ by applying a friend of friend-based algorithm to a spectroscopic catalogue, and an adaptive kernel galaxy density and color map to a photometric sample. \citet{scp09}, here after \sal, identified high-density peaks up to $z \sim 2.5$ by applying a (2+1)D algorithm \citep{tcf07}. In addition, \citet{ki09} reported the discovery of possible over-dense region at $z \sim 3.7$ by examining $BV z$ and photometric selected samples.

However, despite the existence of various statistical and data mining algorithms, the best methods merely provide the means of structure detection within the constraints of redshift surveys. As a result, confident photometric detection of smaller groups is a challenging task due to the uncertainty attained by photometric redshifts. On the other hand, detection of high redshift structures is beyond the capability of the analysis methods of the shallower spectroscopic surveys. This situation leads to incomplete structure detection in important multiwavelength fields such as the CDFS. Clearly, this motivates, as a high priority, more complete analysis of structures in such fields.

Recently, a highly-complete survey of spectroscopic redshifts, with depth of $R < 23$, has been released by The Arizona CDFS Environment Survey \citetext{ACES, \citealp{cyd12}} team, containing 5080 secure redshifts across the $\sim30^{\prime} \times 30^{\prime}$ extended CDFS region (ECDFS). This catalogue enabled us to search for the signs of groups and clusters in unprecedented detail. We confirm the detection of 9 over-dense regions and report the discovery of 53 new structures, including filaments, groups, and clusters at $0.1<z<1$. We also include a comparison provided by data analysis based on the Multiwavelength Survey by Yale-Chile \citetext{MUSYC, \citealp{cdm10}} public photometric redshift catalogue.

The paper is laid out as follows; In Section 2, we describe the samples we constructed and used in this study. In Section 3, we explain the clustering method used in our analysis and its basic features. In Section 4, we provide the results of our analysis and the catalogue of detected overdensities. Section 5 provides a comparison with published ultra-deep X-ray observations of the CDFS. Finally, the conclusions and discussion are laid out in Section 6. Throughout this paper, we assumed a standard Lambda-CDM model, with $H_{0}=71$ \kms Mpc$^{-1}$, $\Omega_{m}=0.27$, and $\Omega_{\Lambda}=0.73$.

\section{Redshift Data}

We constructed a spectroscopic and a photometric sample, in order to identify and verify the high-density regions in the ECDFS.

\subsection{Spectroscopic Sample}

The ACES catalogue contains 7277 unique heliocentric redshifts of which 5080 have secure redshifts within the ECDFS region. The spectroscopic sample used herein, is made of 4692 objects with $z < 0.1$, $R < 24$, and secure flags \citetext{i.e. 3 and 4 quality flags corresponding to $\sim 90\%$ and $95\%$ confidence values, \citealp{cyd12}}, extracted from the preliminary ACES dataset. We convert the entire spectroscopic redshift sample to the reference frame of the 3K background \citep{fcg96}. The ACES' completeness level is nearly $80\%$ at $R=23$ and $I=22$, and it drops sharply below those magnitudes. Due to the angular extent $(\sim30^{\prime} \times 30^{\prime})$ of the ECDFS, it was unlikely to detect any structure at $z<0.1$. For this reason the redshift's lower range of the sample is fixed to $z=0.1$. The spectroscopic sample of 4692 galaxies was analysed to detect the over-dense regions to the depth limit of the ACES catalogue, regardless of completeness level of the survey.

\begin{deluxetable}{cccc}
\tabletypesize{\scriptsize}
\tablecolumns{5} \tablewidth{0pt} \tablecaption{Photometric redshift accuracy versus spectroscopic redshift\label{tab:nmad}}
\tablehead{ \colhead{Range} & \colhead{No. of Objects\tablenotemark{a}} & \colhead{$\sigma_{NMAD}$} & \colhead{$\delta z{p}$\tablenotemark{b}}}
\startdata 
\input{TabA.tex}
\enddata
\tablenotetext{a}{Number of photometric objects, in the specified bin, which are cross-identified with spectroscopic redshifts.}
\tablenotetext{b}{Extent of photometric slices.}
\end{deluxetable}

\subsection{Photometric Sample}

The photometric sample was extracted from the MUSYC catalogue, which provides photometric redshifts for $\sim 80,000$ galaxies in the ECDFS down to $R \sim 27$. However, the accuracy and completeness level of the MUSYC catalogue are limited for objects with magnitude of $R>25.5$ and redshift quality flag of $Qz>1$. Our photometric sample contains 5522 galaxies with conservative selection criteria of $Qz<0.2$, $R<24$ within $0.1<z<1.2$. The selection criteria of the photometric sample were chosen to select sources with higher redshift quality. Limiting the catalogue to the sample of objects with $0.1<z<1.2$ yields a dispersion error of $\sigma_{z} \sim 0.008$ (comparing to high quality spectroscopic redshifts), whereas for objects with $1.2<z<3.7$, $\sigma_{z}$ dramatically increases to $0.027$ \citep{cdm10}. For this reason we adopt the redshift limit of $0.1<z<1.2$ in our photometric sample. We limit the magnitude to $R<24$, following the magnitude limit of the spectroscopic sample. Moreover, the magnitude constraint results in more accurate redshifts, since the photometric redshift quality is a strong function of source magnitude \citep{dmd10}. The photometric sample was primarily made to compare the results with the ones attained by spectroscopic data analysis. This enabled us to determine the reliability and efficiency of using the photometric sample in detection of structures in various redshift ranges.

\subsubsection{Impact of photometric redshifts quality on redshift morphology of detected structures}

Our photometric catalogue includes a subset of galaxies with available spectroscopic redshifts. These spectroscopic redshifts were obtained by cross-identification of the MUSYC and ACES catalogues, using a $1.5^{\prime\prime}$ radius. Moreover, some of the sources in the MUSYC catalogue are associated with high quality spectroscopic redshifts available from the literature. These were used to measure the quality of the photometric redshifts in various redshift ranges, assuming there is no error involved in the spectroscopic ones. We expect the accuracy of the photometric sample to deteriorate more rapidly as a function of redshift. We thus need to calculate the quality of the photometric redshifts, $z_{p}$, as a function of spectroscopic redshifts, $z_{s}$. We employ the Normal Median Absolute Deviation (NMAD) to examine the statistical difference in a sample of photometric and spectroscopic redshifts in a series of redshift intervals. NMAD is given by:

\begin{displaymath}
\sigma_{NMAD}(z_{s}) = 1.48 \times median \biggl\lvert \frac{(\Delta z-median (\Delta z))}{1+z_{s}} \biggl\rvert,
\end{displaymath}
with $\Delta z= \lvert z_{s}-z_{p} \rvert$ \citep{iam06}. NMAD is a robust tool to calculate the sample dispersion, with less sensitivity to outliers. We report NMAD values for various redshift slices of the photometric sample in the Table \ref{tab:nmad}. These values were used to estimate the photometric redshift spread of the structures in different redshift bins. We can approximate the dispersion of the photometric values as the convolution of the Gaussian representing the spectroscopic sample, with dispersion of $\sigma_{s}$, and the Gaussian function which corresponds to the error generated by the photometric technique, with dispersion of $\sigma_{Err}$. The result is a broadened normal distribution with photometric dispersion of $\sigma_{p} = \sqrt{{\sigma_{s}}^2+{\sigma_{Err}}^2(z_{s})}$, in which $\sigma_{Err}(z_{s})=(1+z_{s})\sigma_{NMAD}(z_{s})$ has been corrected for our reference frame. Note $\sigma_{Err}$, generated due to use of the photometric technique, is only known with respect to the spectroscopic measurements, which themselves contain error. Therefore, it is not an absolute quantity, but rather a statistical term, which can only be measured over a large sample.

\begin{figure*}
\centering
{\includegraphics[width=3.2in, trim= 0 20 0 0, clip=true]{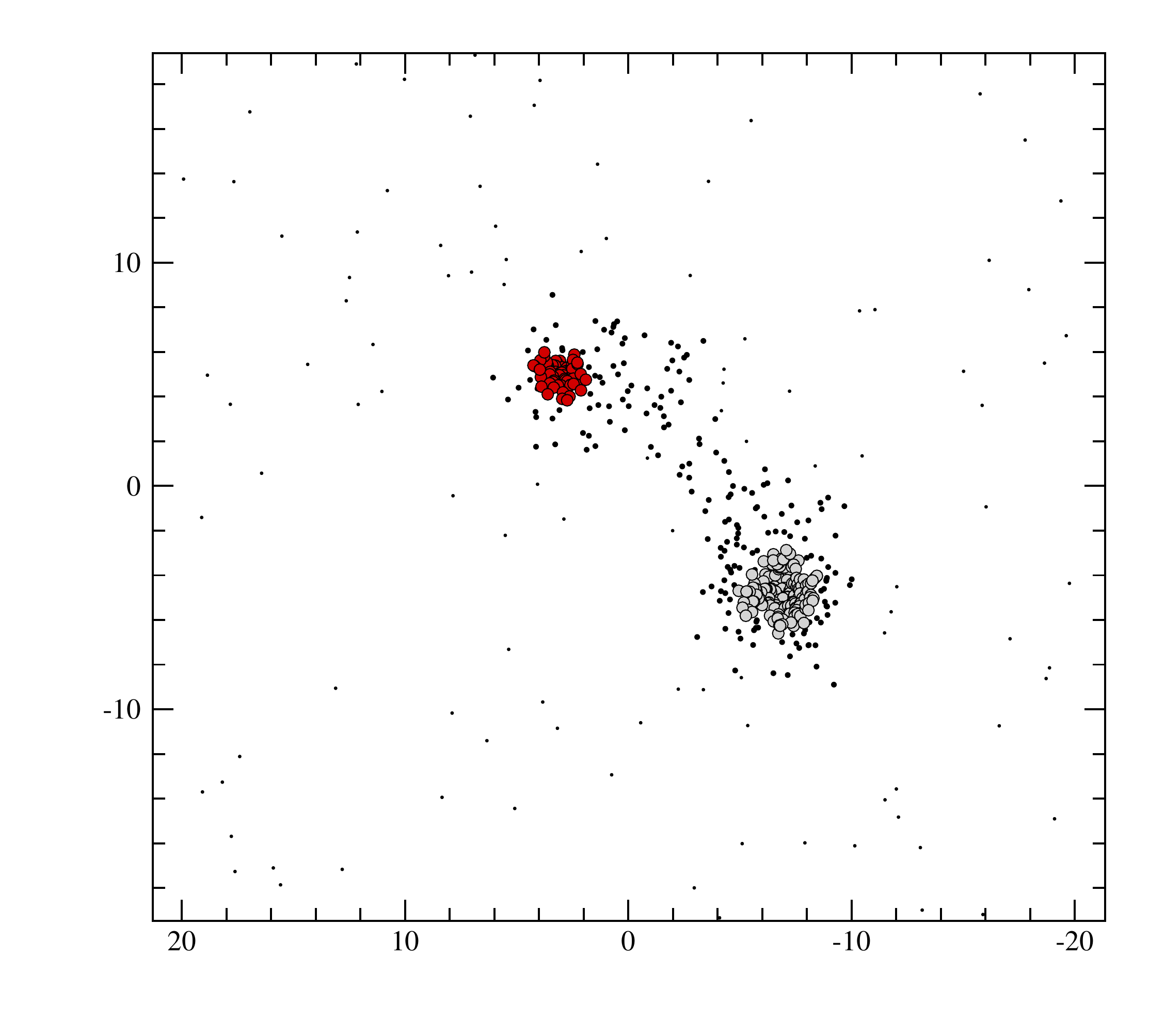}}
{\includegraphics[width=3.2in, trim= 30 20 -30 0, clip=true]{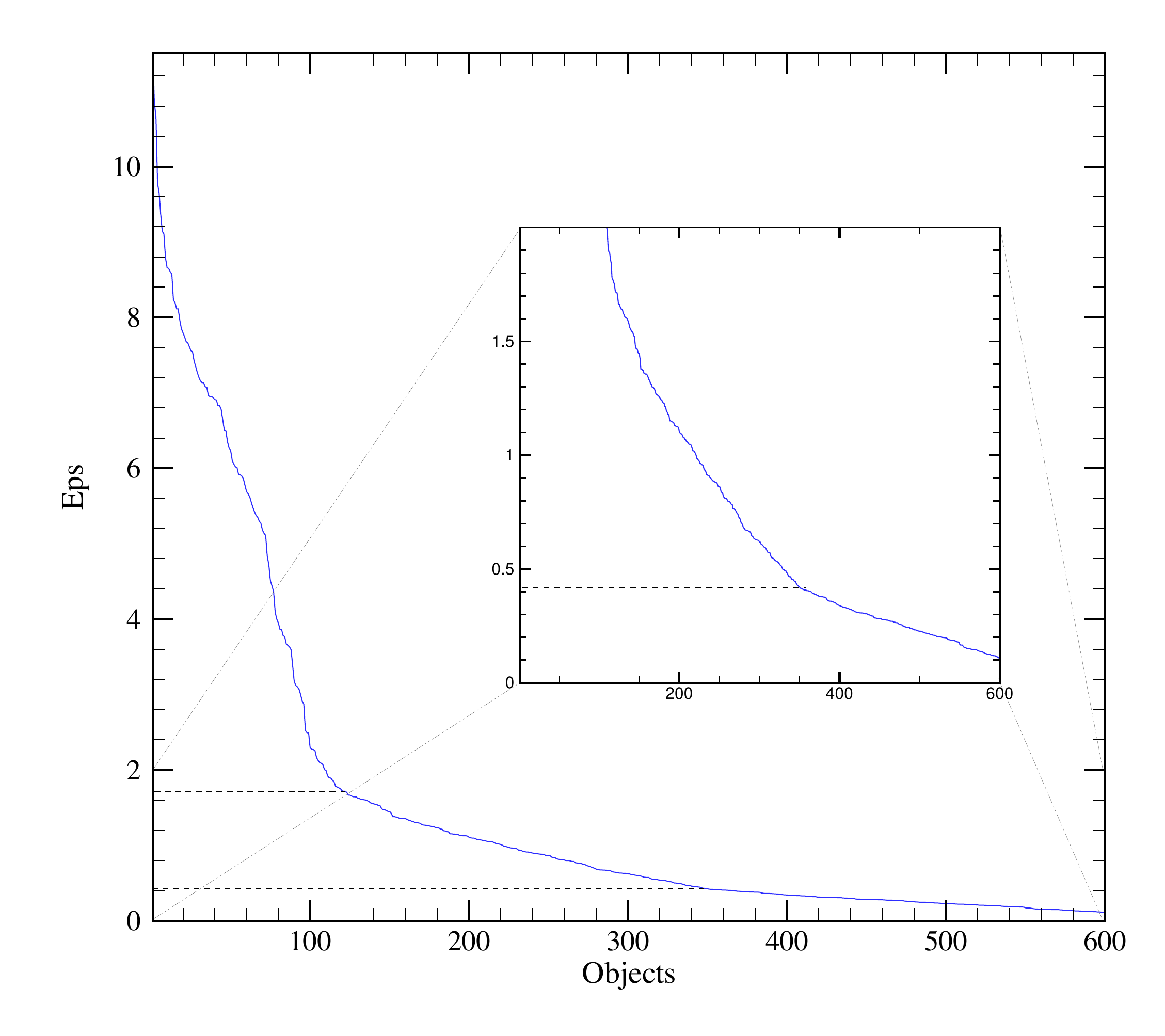}}
\caption{{\bf Left Panel:} Simulated large-scale structures generated to demonstrate suitable parameters for use in the DBSCAN algorithm. We simulated a distribution consisting of two Gaussian clusters, and a connecting filament-shape structure. The dataset consists of 600 data points shown with black points. The clustering result for the distance factor $(Eps) \simeq 0.4$ are shown with large colored dots. Each color (shade) correspond to a separate group or cluster. The large scale structure found with $Eps \simeq 1.7$ is shown with black dots. Note the axis scale is arbitrary. {\bf Right Panel}: Sorted 6-dist graph of the distribution galaxies shown in the left panel. Note the Y-axis scale is arbitrary, however, it has the same units as the scale defined in the left panel. The difference in slope of the graph is significant at $Eps \simeq 0.4$ and $Eps \simeq 1.7$, which corresponds to the structures present.}
\label{fig:eps}
\end{figure*}

\section{Detection Method}

We developed a method to detect the spatial (2-Dimensional) overdensities in the CDFS. This method was applied to both the spectroscopic and photometric sample. The procedure starts with dividing the entire sample volume into redshift bins, in such a way that it prevents, to some extent, the contamination of the member population by neighbouring galaxies, and at the same time, it avoids disposal of valuable data points. This becomes particularly important when dealing with relatively inaccurate photometric redshifts, where the structures expand in redshift space. In the next step we perform our clustering algorithm on each of these slices. We implemented a density-based clustering algorithm known as Density-Based Spatial Clustering of Applications with Noise \citetext{DBSCAN, \citealp{eks96}}. The DBSCAN fundamentals are briefly described in \ref{sec:dbscan}.
 
\subsection{Input data}

The procedure starts with slicing the redshift space in each redshift sample. In the spectroscopic case, we divide the entire sample into a series of redshift bins of width $\delta z_{s} = 0.03$. Assuming a Normal distribution of galaxies located at the center of a bin at $z$, the adopted $\delta z_{s}$ value corresponds to a velocity dispersion of 1500 \kms or $\sigma_{s}=0.005$, at low redshifts, covering all the members within $3 \sigma_{s}$ of the central $z$. This dispersion value covers a range of structures, from compact groups, e.g. Hickson Compact Groups with $\sigma \leq 300$ \kms, to rich clusters like Coma Cluster with $\sigma \simeq 1000$ \kms.

Likewise, for the photometric detection of the same distribution at $z$, we adopt the redshift slice width of $\delta z_{p}$. We note that the extent of photometric slices should be increased in higher redshifts, as the $\delta z_{p}$ value is a function of redshift, i.e. $\delta z_{p} = 6\sqrt{{\sigma_{s}}^2+{(1+z)}^2\sigma_{NMAD}^2(z)}$. In this way, $\delta z_{p}$ is $3\sigma_{p}$ on either side of the central redshift, which is large enough to compensate for the increasingly broadened structures in the photometric sample at higher redshifts. We report $\delta z_{p}$ values in various redshift ranges in the last column of Table \ref{tab:nmad}.

In addition to the preliminary set of slices, another set of slices were also made by shifting the entire initial set by the half-width of each slice. The second set was made to detect structures possibly located on (or near) the slice borders. Following the structure investigation, we made the final set of redshift bins, with their corresponding structures located nearly at the centre of the bin. This enables us to maximize the probability of finding members of structures with symmetrical redshift distribution, i.e. groups and clusters.

\subsection{Clustering algorithm}
\label{sec:dbscan}

DBSCAN is a simple clustering method that has already been employed in astrophysics in several cases: in radio \citep{wtl13} and Gamma Ray \citep{tv12} source detection, along with the detection of the early-type Hipparcos stars \citep{cd08}. Here we outline the fundamentals of the method, including the classification of points and the algorithm. A more detailed description of the algorithm and its features can be found in \citet{eks96}.

DBSCAN utilizes two user-defined factors, a distance factor ($Eps$) and a minimum number of points (MinPts), in order to classify all the data points as either core, border, or noise points with respect to a detected structure. The density of a particular data point is defined as the number of data points, including the point itself, within the radius of $Eps$ of that point, i.e. $Eps$-neighbourhood ($N_{Eps}$). A core point is a point for which its density exceeds the threshold, i.e. MinPts, such that $N_{Eps}\geq MinPts$. A border point is defined as a non-core point that is within the $Eps$-neighbourhood of any core point. Any point not classified as either a core or border point is a noise point.

After considering the previous classification of data points, all the noise points are discarded from the dataset. DBSCAN starts with an unassigned arbitrary core point. Any other core point falling within the neighbourhood of that point belongs to the same structure as the core point. Similarly, the border points of each core point are also assigned to the structure of the first core point. The process continues to the next unassigned core point up to the last unassigned core point in the dataset.

Although DBSCAN is an effective algorithm to detect overdensities of an arbitrary shape, its clustering quality depends on the distance factor, $Eps$. In addition, DBSCAN is unable to detect structures with significant difference in their densities. We adopt the following approach to overcome these issues.

\subsection{Parameter Adjustment}

As described in 3.2, the DBSCAN method uses two user-defined parameters, $Eps$ and $MinPts$. The result dramatically depends on these two factors. The basic approach to determine the parameters, as stated in \citet{eks96}, is to observe the behaviour of the \emph{sorted k-dist graph}. To build the sorted k-dist graph we compute the angular distance of each point to its $k^{th}$ neighbour, the so-called k-dist, and sort the values in descending order. Any change in the slope of the sorted k-dist graph gives an impression of the structure and proper $Eps$ value. Should we adopt this value as $Eps$ and set the $MinPts=k+1$, all the points conforming to k-dist$<Eps$ become core points. This interactive approach reduces the required number of input parameters to one, namely $MinPts$. We adopted a reasonable value of $k=6$ in our DBSCAN analysis. Setting smaller $k$ values results in detection of false or insignificant groups (i.e. groups with 6 or less galaxy members, $N_{s}\leq6$).

The right panel of Figure \ref{fig:eps} shows an example of a sorted k-dist graph for a simulated distribution (shown with black points in the left panel). The distribution consists of 600 field and structure objects of the same redshift. The simulated structure is made of two clusters and a filament in between. By examining the associated k-dist plot, two possible $Eps$ values were found in this dataset, each of which results in detection of a class of objects with different physical extents. The large scale structure (shown with larger black circles) corresponds to $Eps \simeq 1.7$, and the lower value of $Eps \simeq 0.4$ results in detection of two clusters (shown with colored circles). Consequently, the value of $Eps$ adjusts the sensitivity of DBSCAN to detect the structures of different scale.
 
\section{Results}

We applied DBSCAN to various redshift bins of the spectroscopic sample. Using $MinPts=7$, all possible values for $Eps$ were determined and used as the input parameter of DBSCAN. In total, 62 filaments, groups, and clusters were found, some of which are embedded in the four significant large scale structures, located at $z\simeq0.13$, $z\simeq0.52$, $z\simeq0.68$, and $z\simeq0.73$. Redshift location and the rest-frame velocity dispersion of the structures, along with their $68\%$ confidence intervals, were calculated by applying the the biweight, gapper, and jackknife methods \citep{bfg90} to structure members.

The classification was initially made based on the spectroscopic velocity dispersion of the structures; Explicitly, we considered $v_{d} \sim 400$ \kms as a typical lower velocity dispersion limit for clusters \citetext{\citealp{sr91} or \citealp{ogp13}}. In addition, spectroscopic sampling rate and morphology of the spatial and redshift distribution were considered in the classification of structures. The richness estimate is a crucial factor in classification of detected structures. Observed galaxies appear increasingly dimmer at higher redshifts, and finally fall below the magnitude limit of the sample, which in turn result in structures with less members with respect to structures of the same class and size at lower redshifts. We estimate the sampling rate of a cluster at the desired redshift, by performing DBSCAN on 100 random simulated clusters, at each region of interest, according to the method provided by \citet{rab04}. Members of the clusters were assigned a magnitude down to the completeness level of ACES ($I=22$), and conforming to a Schechter luminosity function with $M^{*}=-22.5$ and $\alpha=-1.25$, following \citet{rab04}. We report the average numbers of the simulated cluster members detected by DBSCAN as $N_{c}(z)=39$, 15, 11, 10, and 10 at $z=0.13$, 0.52, 0.68, 0.73, and 0.83 respectively. Therefore, we take $N_{c}(z)$ as the lower number limit required for cluster detection at different redshifts. In addition, the 3D morphological distribution of the detected structures was subdivided into filaments, groups, and clusters.

We define five classes among the detected overdensities:

1- Radial filaments or fake structures: Class 1 objects have very broad velocity distribution $(v_{d}>>400$ \kms) without any significant peak in their velocity distribution. Occasionally, class 1 objects consist of a number of very small groups lined up in the redshift distribution.

2- Filaments on the plane of sky: Class 2 objects have filamentary morphologies in the plane of sky often with low velocity dispersions.

3- Groups: Structures of nearly Gaussian redshift distribution with $v_{d}<400$ \kms are classified as groups.

4- Massive groups or small clusters: Class 4 objects consist of structures with $v_{d} \gtrsim 400$ \kms, though with insufficient spectroscopic sampling rate to be considered as clusters ($N_{s} < N_{c}(z)$).

5- Clusters: Structures conforming to $v_{d} \gtrsim 400$ \kms and $N_{s} \geq N_{c}(z)$ are classified as clusters.

All the structures with their properties including identification number, average coordinates, number of galaxies, redshift, spectroscopic velocity dispersion, $Eps$ value, specified redshift bin, and designated class are summarized in Table \ref{tab:catalogue}.

\begin{deluxetable*}{lcccccccccc}
\centering
\tablewidth{0pt} \tablecaption{Catalogue of overdensities in the CDFS\label{tab:catalogue}}
\tablehead{ \multirow{2}{*}{ID} & \multirow{2}{*}{Type} & \colhead{RA\tablenotemark{a}} & \colhead{Dec\tablenotemark{a}} & \multirow{2}{*}{N} & \multirow{2}{*}{$z$} & \multirow{2}{*}{Error$(z)$\tablenotemark{b}} & \colhead{$\sigma _{v}\tablenotemark{b}$} & \colhead{$Eps$} & \multirow{2}{*}{Bin Range} & \multirow{2}{*}{Class}\\
\colhead{} & \colhead{} & \colhead{(J2000.0)} & \colhead{(J2000.0)} & \colhead{} & \colhead{} & \colhead{} & \colhead{\kms} & \colhead{Mpc}}
\startdata
\input{Tab1.tex}
\enddata
\end{deluxetable*}

\begin{deluxetable*}{@{}l@{}cccccccccc}
\centering
\tablewidth{0pt} \tablecaption{Catalogue of overdensities in the CDFS \em -- Continued} \tablenum{2}
\tablehead{ \multirow{2}{*}{ID} & \multirow{2}{*}{Type} & \colhead{RA\tablenotemark{a}} & \colhead{Dec\tablenotemark{a}} & \multirow{2}{*}{N} & \multirow{2}{*}{$z$} & \multirow{2}{*}{Error$(z)$\tablenotemark{b}} & \colhead{$\sigma _{v}\tablenotemark{c}$} & \colhead{$Eps$} & \multirow{2}{*}{Bin Range} & \multirow{2}{*}{Class}\\
\colhead{} & \colhead{} & \colhead{(J2000.0)} & \colhead{(J2000.0)} & \colhead{} & \colhead{} & \colhead{} & \colhead{\kms} & \colhead{Mpc}}
\startdata
\input{Tab2.tex}
\enddata
\tablenotetext{a}{Coordinates of the midpoint.}
\tablenotetext{b}{This is merely a statistical error due to the sampling error and deviance from the Normal distribution. Note that the intrinsic photometric errors, and possible contaminations from other populations are not included.}
\tablenotetext{c}{Spectroscopic velocity dispersion.}
\tablenotetext{$\star$}{Structure is detected along with a significant $(N\geq 3)$ spectroscopic substructure. The first and second redshift and velocity dispersion values correspond to the structure and its sub-structure, respectively.}
\tablenotetext{$\ddagger$}{We do not present velocity dispersion and classification, since the structure is not spectroscopically detected.}
\tablenotetext{$\dagger$}{Velocity dispersion is unreliable, since the structure is located on the edge of the spectroscopic frame.}
\end{deluxetable*}

\begin{deluxetable*}{ccccc}
\tabletypesize{\scriptsize}
\tablecolumns{5} \tablewidth{0pt} \tablecaption{Comparison with previously detected structures at $z<1$.\label{tab:pre}}
\tablehead{ \multicolumn{2}{l}{\hspace{14ex}\ada\tablenotemark{a}} & \multirow{2}{*}{\gil\tablenotemark{b}} & \colhead{\sal\tablenotemark{c}} & \colhead{This work}\\
\colhead{ID} & \colhead{Class\tablenotemark{d}} & \colhead{} & \colhead{ID} & \colhead{ID}}
\startdata
1 & proto-cluster/group & \nodata & \nodata & 9\\
2 & very early formation or fake structure & \nodata & \nodata & \nodata\\
3 & proto-cluster/group & \nodata & \nodata & \nodata\\
4 & proto-cluster/group & \nodata & \nodata & 22\\
5 & very early formation or fake structure & \nodata & \nodata & \nodata\\
6 & very early formation or fake structure & \nodata & \nodata & \nodata\\
7 & proto-cluster/group & \nodata & \nodata & 35\\
8 & very early formation or fake structure & \nodata & \nodata & 34\\
9 & real group & \nodata & \nodata & 44\\
10 & proto-cluster/group & \nodata & \nodata & \nodata\\
11-1 & low mass structure & \checkmark & 4 & 56\\
11-2 & proto-cluster/group & \nodata & \nodata & 55\\
11-3 & proto-cluster/group & \nodata & \nodata & \nodata\\
11-4 & proto-cluster/group & \nodata & \nodata & 58\\
11-5 & proto-cluster/group & \nodata & \nodata & \nodata\\
12 & very early formation or fake structure & \nodata & \nodata & \nodata\\
\nodata & \nodata & \nodata & 1 & \nodata\\
\nodata & \nodata & \checkmark & 2 & 41\\
\nodata & \nodata & \nodata & 3 & \nodata\\
\nodata & \nodata & \nodata & 5 & \nodata\\
\enddata
\tablenotetext{a}{Structures found by \citet{ami05}}
\tablenotetext{b}{Structures found by \citet{gcd03}}
\tablenotetext{c}{Structures found by \citet{scp09}}
\tablenotetext{d}{Low mass structures, real groups, proto-cluster/groups, and structures at very early formation or fake, respectively correspond to class 1-4 from \citet{ami05}.}
\end{deluxetable*}

We compare our results with structures previously found by AMI05 and \sal, along with two structures detected by \gil. We confirm the detection of half (8/16) of the compact structures found by AMI05 at $z<1$ (see Table \ref{tab:pre}). Four of the remaining structures, which are not detected in the current study, are either labelled, by AMI05, as fake or at very early formation stage structures, or have very small sampling rate (5 or 6) beyond the threshold level ($MinPts$) used in this work. Table \ref{tab:pre} also represents five detections by SCP09 at $z<1$, of which two structures were confirmed in this work, both of which were previously reported in GCD03. The remaining three structures, detected by SCP09, are either only photometrically detected, or unreliable due to poor sampling rates.

\begin{figure*}
\centering
{\includegraphics[width=3.73in, trim= -5 0 5 0, clip=true]{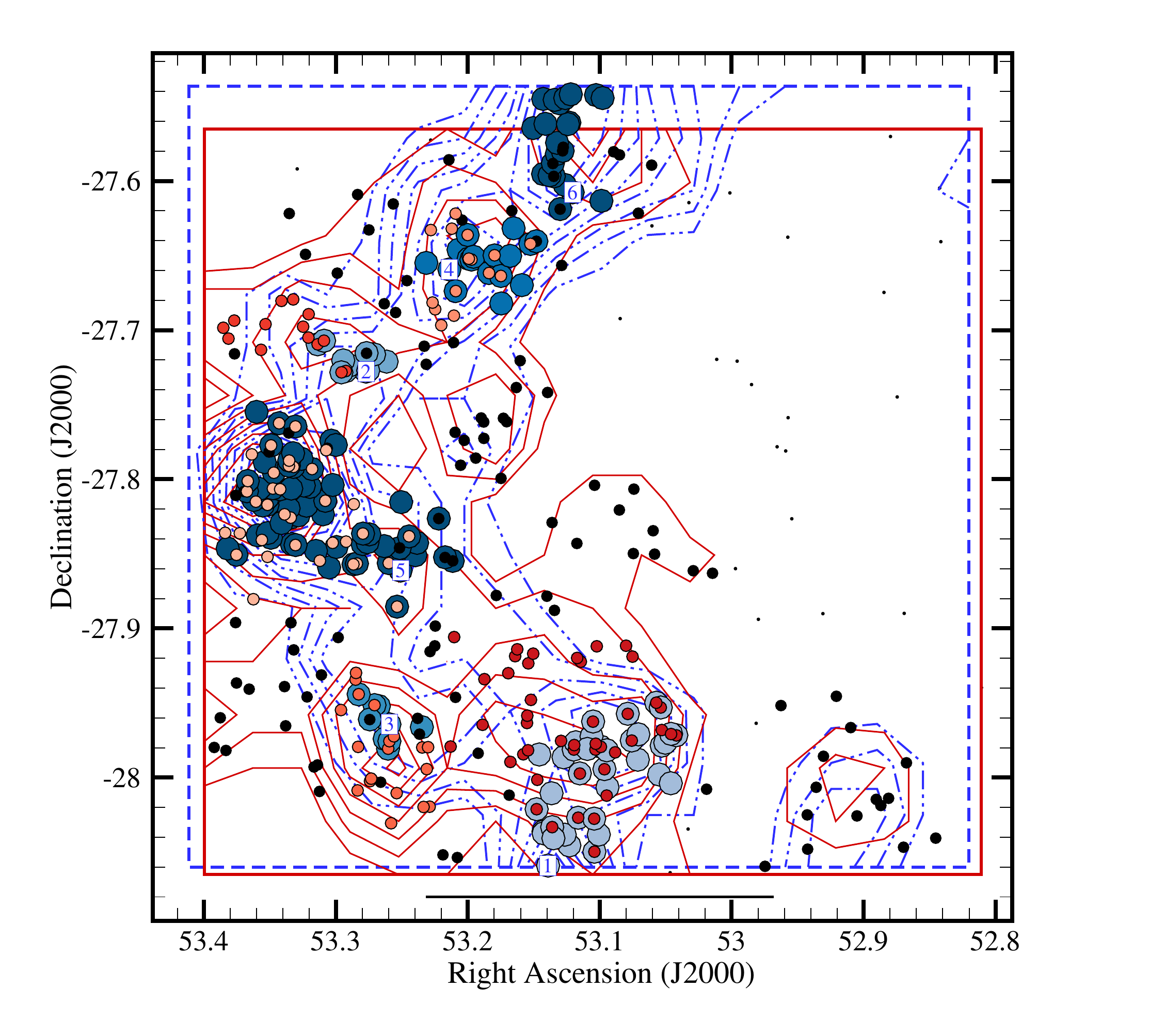}}
{\includegraphics[width=3.333in, trim= 45 1 -45 0, clip=true]{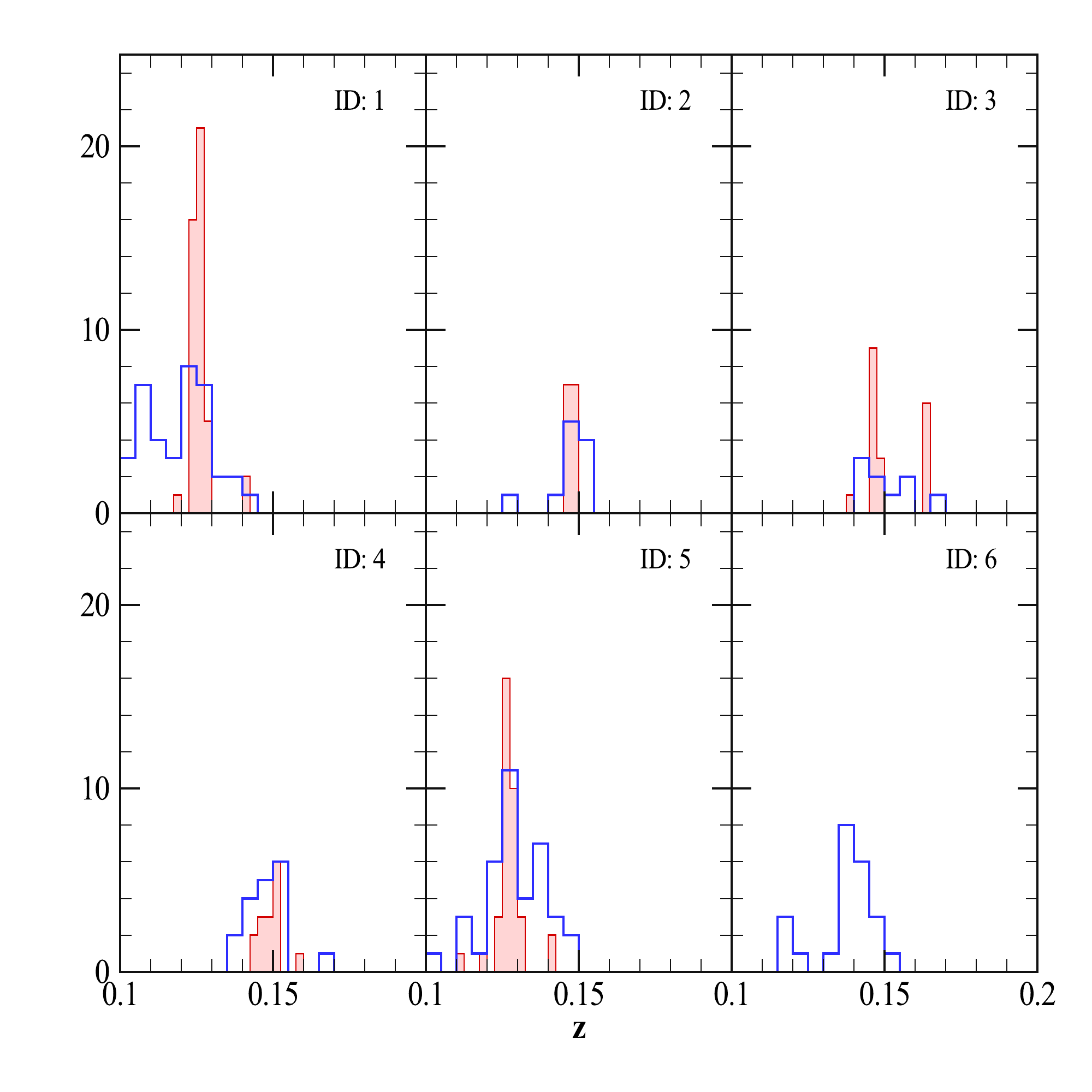}}
%{\includegraphics[width=3.16in, trim= 30 2 -30 0, clip=true]{Histos1-eps-converted-to.pdf}}
\caption{{\bf Left Panel}: Spatial distribution of detected structures and field galaxies within $ 0.11 \leq z_{s} \leq 0.165 $ and $ 0.10 \leq z_{p} \leq 0.175 $. Spectroscopic and photometric groups shown by small red color scheme and large blue color scheme dots, respectively. Black dots show the large scale structure detected in the spectroscopic sample corresponding to the full redshift range with $Eps = 0.42$ Mpc. Field galaxies or `noise' points are shown by black points. Red solid and blue dot-dashed contours represent the spectroscopic and photometric galaxy surface density starting at $\sim$ 15 and then increasing by steps of $\sim$ 5 galaxies per Mpc$^{2}$. The red solid and blue dashed frames represent the outermost of the spectroscopic and photometric datasets, respectively. The black line at the bottom of the plot represents 2 Mpc angular extent at $z\simeq 0.14$. {\bf Right Panel:} The spectroscopic and photometric redshift distributions of the structures shown in the left panel, are respectively represented as red filled histograms (grey filled in the black \& white version) and blue histograms (black unfilled in the black \& white version). IDs in the top left of each plot give the corresponding structure number in the left hand panel.}
\label{fig:first}
\end{figure*}

We present structures detected in rich redshift slices in Figure \ref{fig:first}, \ref{fig:second} \& \ref{fig:histo3}-\ref{fig:sixth}. In each figure, we show the angular extent of the spectroscopic and photometric samples as red solid line and blue hashed line boxes, respectively. These structures are overlaid on both photometric and spectroscopic density (galaxy number) isosurfaces. The contour maps of each redshift bin are made based on the bilinear interpolation of density values to the faces of the cells with a constant extent of $\sim$ 300 kpc at the corresponding mean redshift of the bin. Red solid and blue dot-dashed contours denote the galaxy isodensity values obtained from the spectroscopic and photometric samples, respectively. Small dots in shades of red represent discrete groupings of galaxies obtained by DBSCAN analysis of the spectroscopic data. Whereas, larger dots in shades of blue show groupings detected in the photometric sample. Thus, when a structure is detected in both photometric and spectroscopic analyses, it will appear as a grouping of larger dots in a particular shade of blue overlaid with smaller dots in a particular shade of red or orange. Such structures are labelled with their numerical ID from Table \ref{tab:catalogue}. Large black dots show large-scale structure found in the spectroscopic sample. Small black dots correspond to field galaxies, which are not associated with detected structures. In each figure the black line at the bottom represents 2 Mpc at the average redshift of the slice. In Figures \ref{fig:third}, \ref{fig:fifth} \& \ref{fig:sixth}, the abundant filamentary structures detected are highlighted, with grey shading. Each figure is presented with the corresponding velocity histograms for the numbered structures. The histograms show the detected structures in both the spectroscopic (red filled bins in the color version) and photometric (blue unfilled bins in the color version) data, where available. In the black and white version of the paper the spectroscopic and photometric histograms appear as grey filled bins and black unfilled bins, respectively. We used larger bins in the histogram of photometric overdensities or the groups with less number of galaxies.

\begin{figure*}
\centering
{\includegraphics[width=6.50in, trim= 0 0 -70 -10, clip=true]{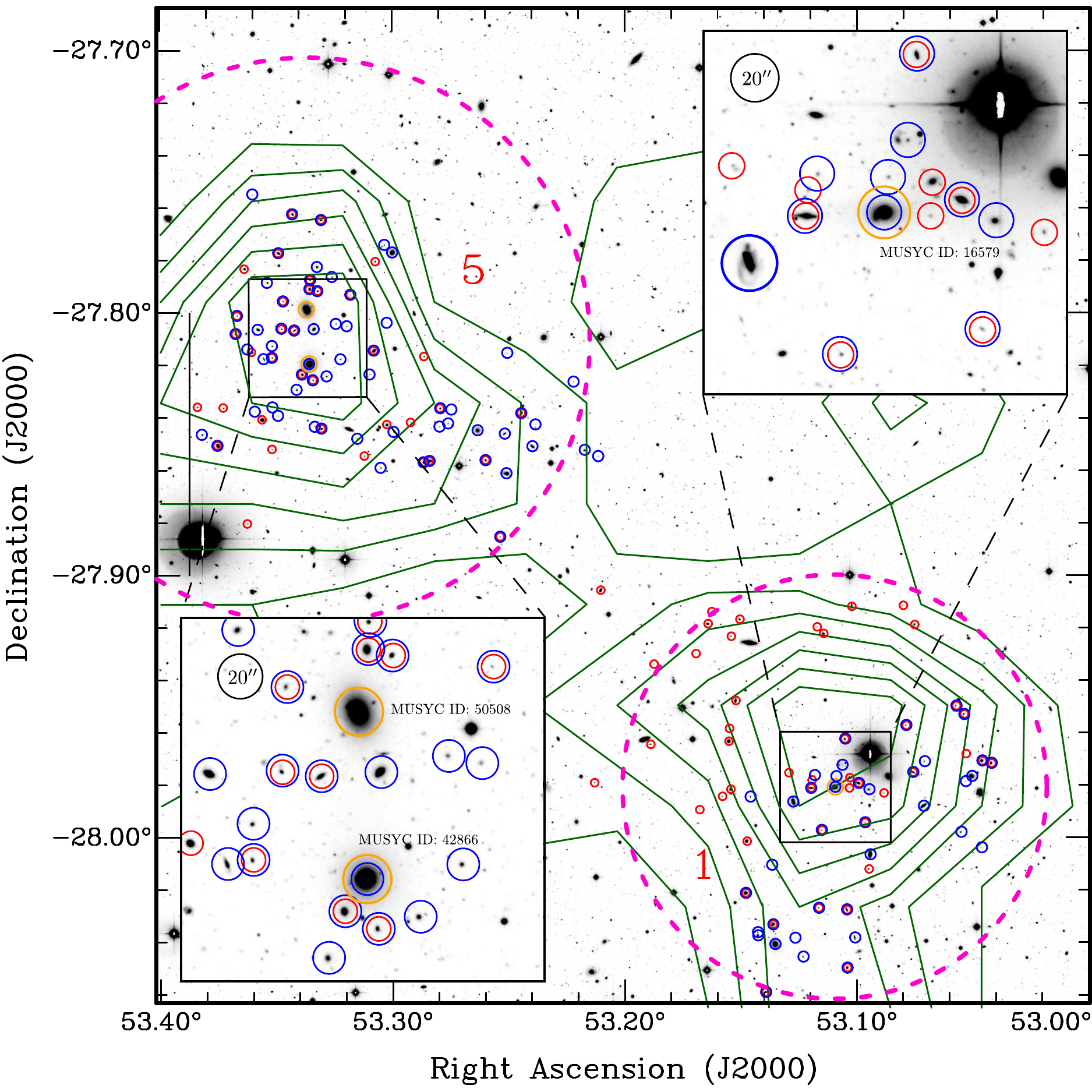}}
\caption{Structure 1 and 5 are overlaid on the optical image of the Deep2c field of the GaBoDS \citep{hed06,tsd05}. Spectroscopic and photometric members shown by small red and large blue circles, respectively. cD galaxies located at proximity of the cluster cores are shown by larger orange circles. Green contours represent the spectroscopic galaxy surface density within $ 0.11 \leq z_{s} \leq 0.14 $ starting at $\sim$ 15 and then increasing by steps of $\sim$ 5 galaxies per Mpc$^{2}$. The approximated extent of the clusters are shown by magenta dashed circles with radius of $R_{200}$, centered at the cD galaxy for Structure 1 and mid-way between the two detected for Structure 5. Insets show an enlarged view of the core of each cluster.}
\label{fig:1-5}
\end{figure*}

\subsection{Structures at $0.11<z<0.17$}

In the left panel of Figure \ref{fig:first} we plot the DBSCAN results applied to the bins within $0.11<z<0.17$, showing the location of six detected structures. We found two major density peaks at $z \simeq 0.126$ and $z \simeq 0.146$ as parts of a larger arc-shaped structure (shown with black circles in the left panel of Figure \ref{fig:first}). The first peak includes two structures with 45 and 36 detected spectroscopic members, located at $z_{s}=0.1253\pm0.0003$ and $z_{s}=0.1267\pm0.0004$, respectively (Structures 1 \& 5). The velocity dispersions of Structures 1 and 5, respectively $428^{+40}_{-40}$ and $575^{+76}_{-76}$ \kms, are within the typical range of velocity dispersion in rich clusters \citep{sr91}. Structure 1's angular distance of $\sim 1-2$ Mpc to the Structure 5, and a disturbed (broadened) morphology along the connecting axis, suggest a possible ongoing merger process.

\begin{figure*}
\centering
{\includegraphics[width=3.73in, trim= -5 0 5 0, clip=true]{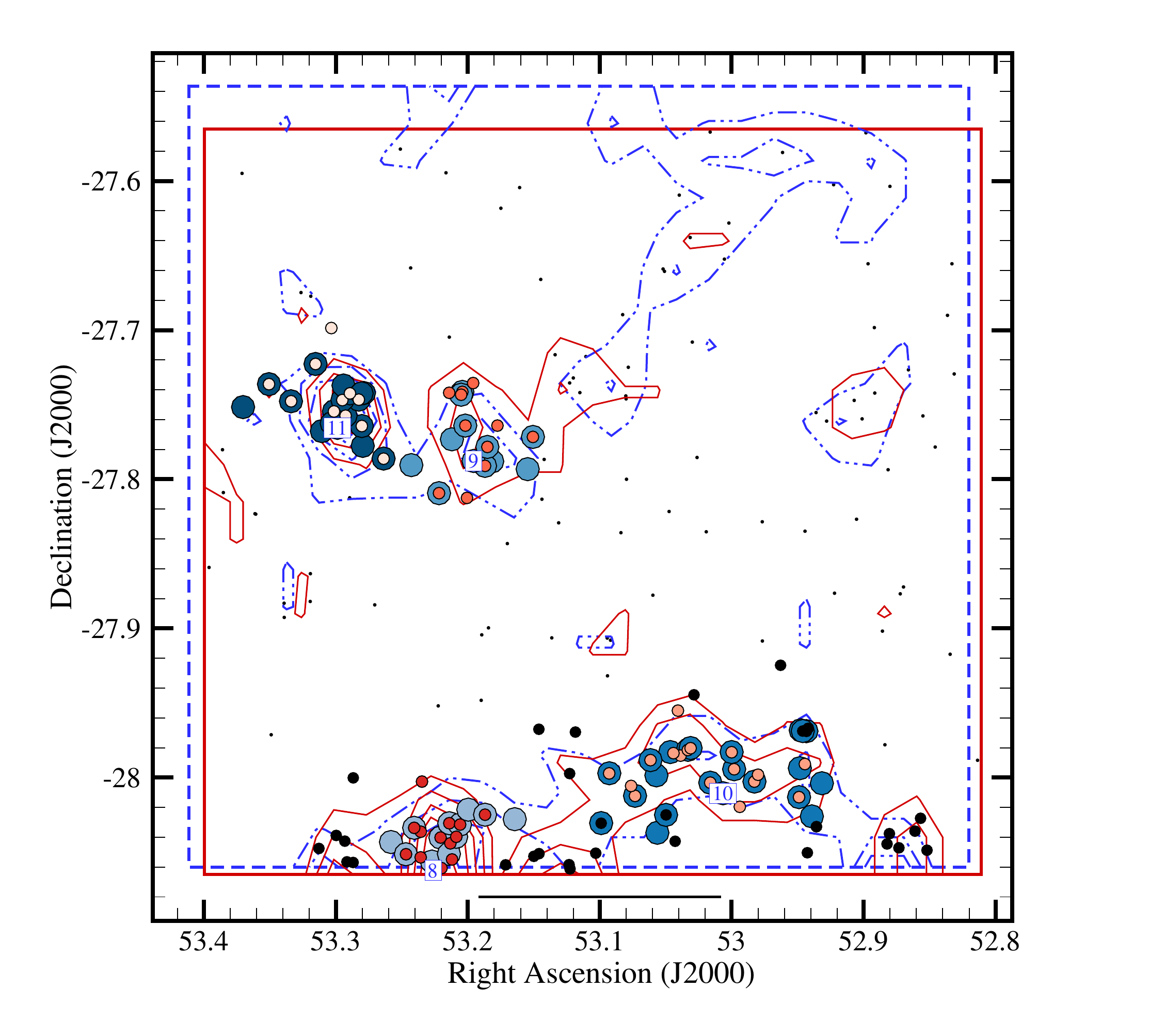}}
{\includegraphics[width=3.333in, trim= 45 1 -45 0, clip=true]{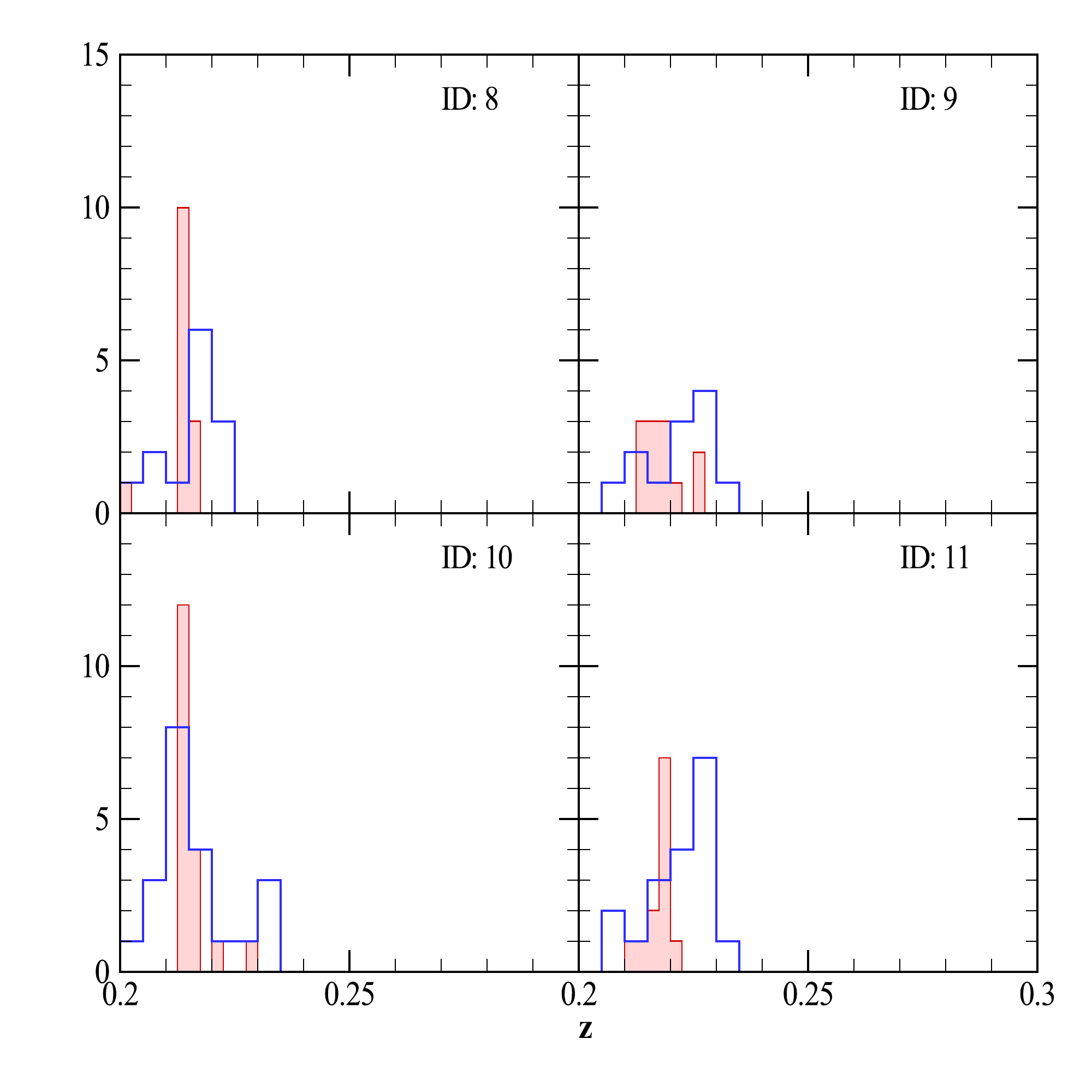}}
%{\includegraphics[width=3.53in, trim= 30 0 -30 0, clip=true]{Histos2-eps-converted-to.pdf}}
\caption{Legends same as Figure \ref{fig:first}. {\bf Left Panel}: Detected structures within $ 0.200 \leq z_{s} \leq 0.230 $ and $ 0.195 \leq z_{p} \leq 0.235 $. The large scale structure is detected by tuning $Eps = 0.62$ Mpc at full redshift range. Spectroscopic and photometric density contours start at $\sim$ 8 and then increasing by steps of $\sim$ 5 galaxies per Mpc$^{2}$. The black line represents 2 Mpc in angular extent at $z\simeq 0.22$. {\bf Right Panel:} The spectroscopic and photometric redshift distributions of the structures shown in the left panel, are respectively represented as red filled histograms (grey filled in the black \& white version) and blue histograms (black unfilled in the black \& white version).}
\label{fig:second}
\end{figure*}

We over-plotted the galaxy density contours and spectroscopic and photometric members for Structure 1 \& 5 on the Deep2c field of the Garching-Bonn Deep Survey \citetext{GaBoDS, WFI Data Release: Version 1.0, \citealp{hed06,tsd05}} \footnote{Based on data obtained from the ESO Science Archive Facility under request number SDEHGHAN173380.} in Figure \ref{fig:1-5}. Figure \ref{fig:1-5} clearly shows two massive elliptical galaxies along the north-south galaxy density axis. The southern most has a MUSYC ID of 42866 and a measured spectroscopic redshift of 0.1264 \citetext{GOODS, \citealp{bmp10}}. The northern galaxy is identified as MUSYC 50508 and has a spectroscopic redshift of 0.1270, which is extracted from the Southern Abell Redshift Survey \citetext{SARS, \citealp{wqi05}}. Both galaxies, shown at the left hand inset of Figure \ref{fig:1-5}, exhibit morphological characteristics of cD galaxies and have a number of smaller members around them. The redshifts are also nearly at the average cluster redshift ($z_{s}=0.1267$), suggesting they sit at the bottom of the gravitational well. In the case of Structure 1, we find another massive elliptical galaxy (MUSYC ID 16579) at the center of the galaxy density. This system also sits close to the cluster average redshift of 0.1253 with a photometric redshift of 0.122 (extracted from MUSYC). Interestingly, the morphology of both the putative cD galaxy and several of the galaxies surrounding it, are distorted with either large tidal tails or fragmentary structures. In particular, the cD galaxy has a tidal tail extending east and then south, and nearby galaxy (MUSYC ID 16882) shown to the left of the top inset in Figure \ref{fig:1-5}, has a pronounced spiral tail of stars. These features are strong evidence of on-going hierarchical structure formation.

The second velocity peak includes four smaller groups (Structures 2,3,4, and 6), of which the Structure 6 was only detected in the photometric sample, due to its location being close to the edge of the spectroscopic frame. The redshift histograms of the Structures 1-6, shown in the right panel of the Figure \ref{fig:first}, indicate that, despite the expected broadened photometric distribution, the photometric and spectroscopic peaks do not differ considerably. In addition, Structure 7 is an isolated group located at $z_{s}=0.1796\pm0.0003$, which is not shown in Figure \ref{fig:first}, as it lies slightly beyond this redshift range.

\begin{figure}
\centering
{\includegraphics[width=3.73in, trim= -5 0 5 0, clip=true]{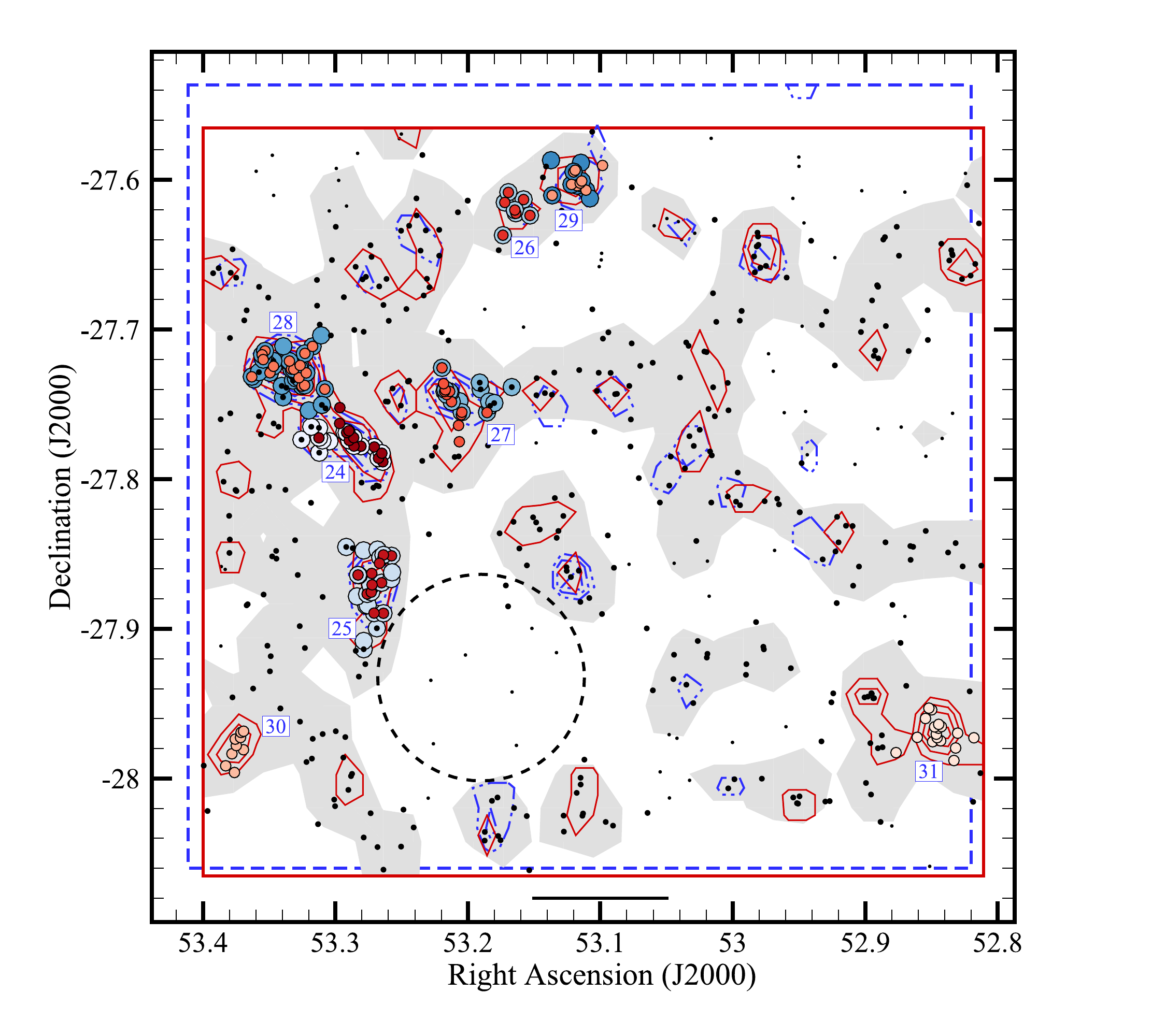}}
\caption{Detected structures within $ 0.505 \leq z_{s} \leq 0.550 $ and $ 0.495 \leq z_{p} \leq 0.560 $. The large scale structure is detected by tuning $Eps = 0.69$ Mpc at full redshift range. Spectroscopic and photometric density contours starting at $\sim$ 8 and then increasing by steps of $\sim$ 5 galaxies per Mpc$^{2}$. The highlighted areas corresponds to the regions with minimum spectroscopic density of $\sim$ 3 galaxies per Mpc$^{2}$. The dashed circle represents the approximate location and extent of the detected void. The scale line represents 2 Mpc angular extent at $z\simeq 0.53$. Legends same as Figure \ref{fig:first}.}
\label{fig:third}
\end{figure}

\begin{figure*}
\centering
{\includegraphics[width=5in, trim= -35 20 135 0, clip=true]{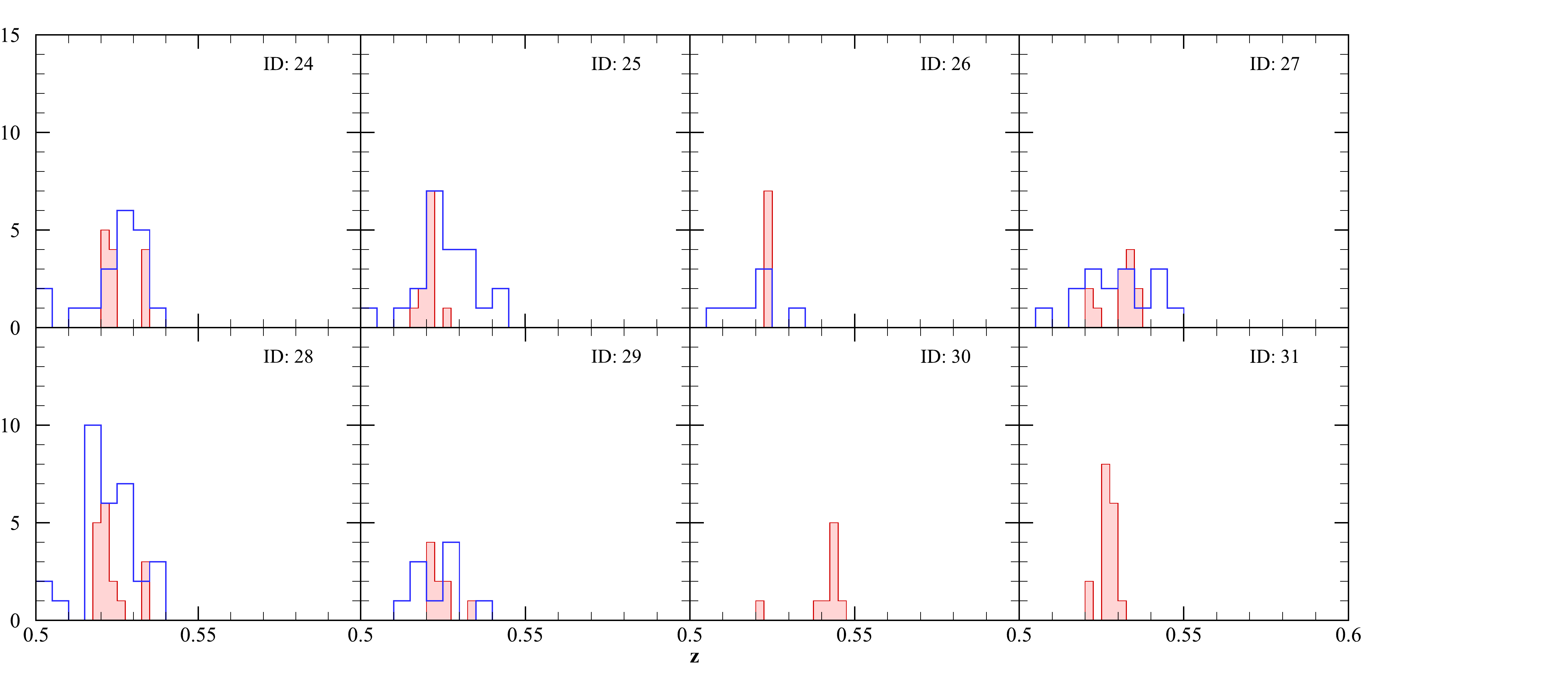}}
\caption{The spectroscopic and photometric redshift distributions of the structures shown in the the Figure \ref{fig:third}, are respectively represented as red filled histograms (grey filled in the black \& white version) and blue histograms (black unfilled in the black \& white version).}
\label{fig:histo3}
\end{figure*}

\subsection{Structures at $0.20<z<0.28$}

At $0.20<z<0.28$ we identified ten high density regions of which four are located at $z \simeq 0.21$ (see the left panel of Figure \ref{fig:second}). Structure 10 has a very narrow velocity dispersion of $100^{+68}_{-68}$ \kms, located at $z_{s}=0.2142\pm0.0001$, and appears to be a filament-like structure in the plane of sky falling to the nearby Structure 8 located at $z_{s}=0.2136\pm0.0002$ with $v_{d}=183^{+47}_{-47}$ \kms; a typical velocity dispersion of a normal group. The overall group-filament structure extends to about 6 Mpc, and was detected by adopting the higher $Eps$ value of 0.62 Mpc. Structures 9 and 11, located at $z_{s}=0.2167\pm0.0013$ and $0.2176\pm0.004$, appear to be an interacting system, resulting in a distorted shape for both spatial and redshift distributions of Structure 9 (see the upper right histogram of the Figure \ref{fig:second}). Structure 12 (not shown in Figure \ref{fig:second}) located at $z_{s}=0.2465\pm0.0003$ was detected with a tiny companion group located at $z_{s}=0.2600\pm0.0002$. The remaining overdensities in the region include two groups possibly in a merging process located at $z_{s}=0.2498\pm0.0002$ and $0.2487\pm0.0002$ (Structures 13 \& 14), and three isolated overdensities at $z_{s}=0.2778\pm0.0009$, $0.2771\pm0.0003$, and $0.2772\pm0.0003$ (Structures 15-17, not shown in Figure \ref{fig:second}).

\begin{figure*}
\centering
{\includegraphics[width=3.73in, trim= -5 0 5 0, clip=true]{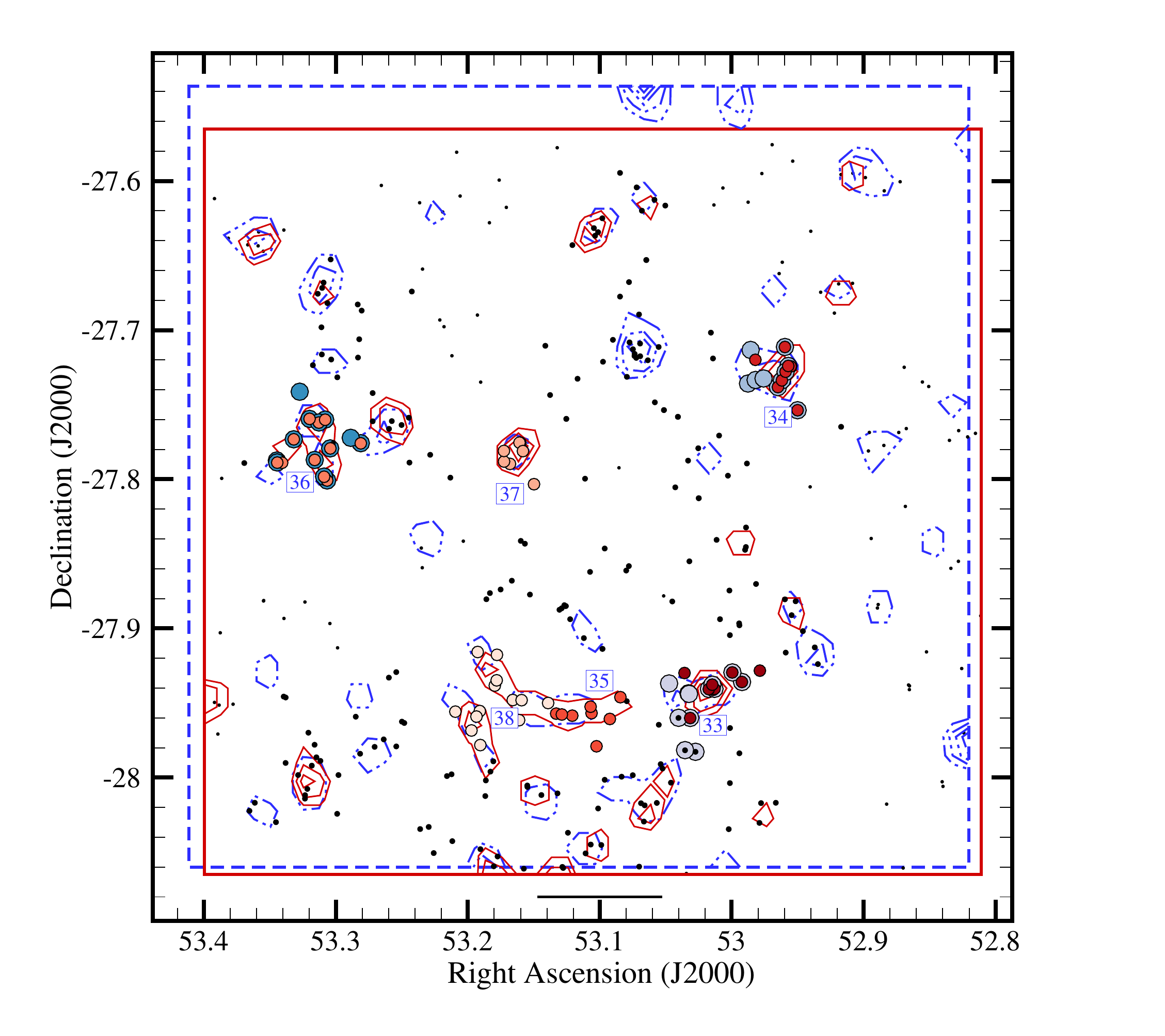}}
{\includegraphics[width=3.333in, trim= 45 1 -45 0, clip=true]{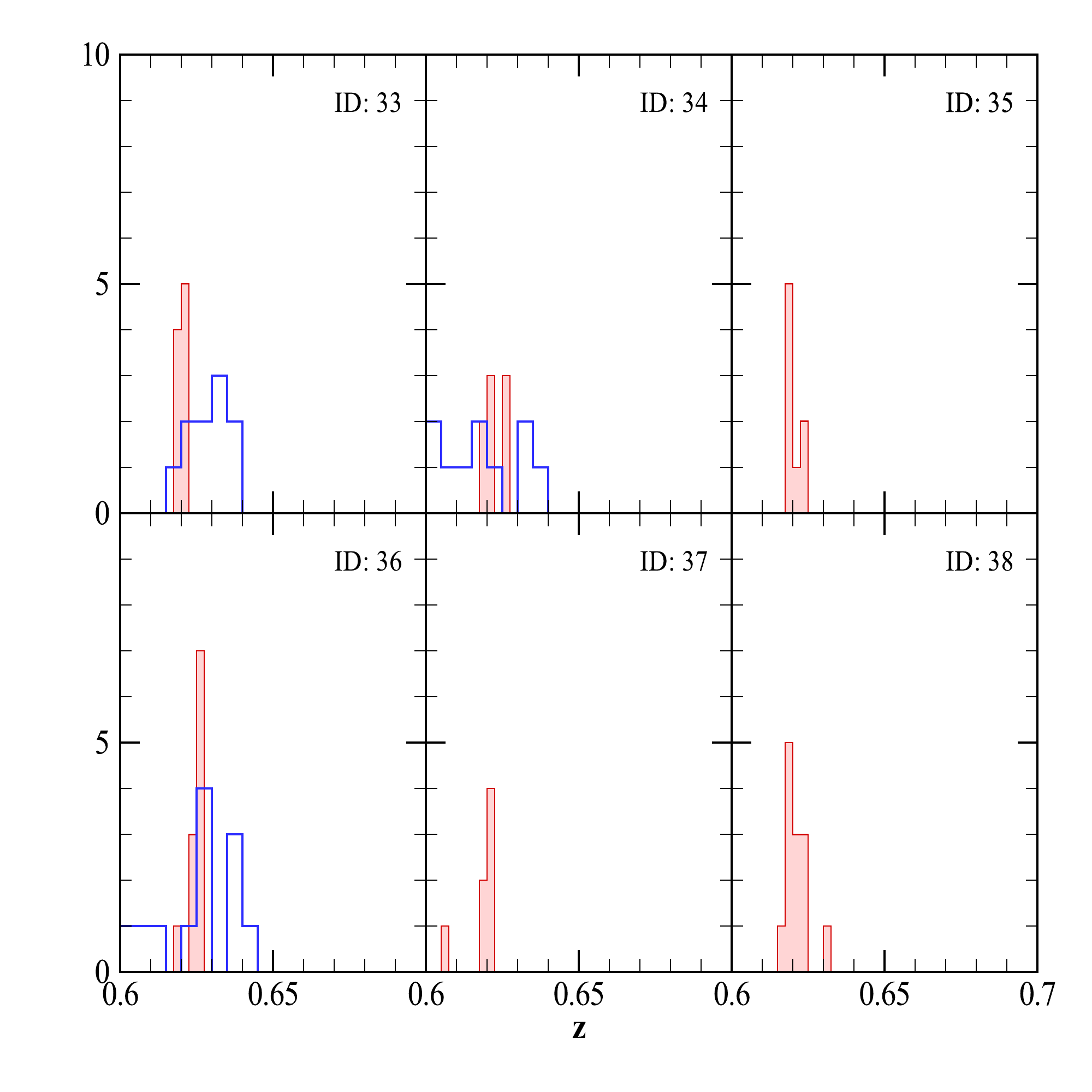}}
%{\includegraphics[width=3.53in]{Forth-eps-converted-to.pdf}}
%{\includegraphics[width=3.53in, trim= 30 0 -30 0, clip=true]{Histos4-eps-converted-to.pdf}}
\caption{Legends same as Figure \ref{fig:first}. {\bf Left Panel}: Detected structures within $ 0.605 \leq z_{s} \leq 0.640 $ and $ 0.590 \leq z_{p} \leq 0.655 $. The large scale structure is detected by tuning $Eps = 1.17$ Mpc at full redshift range. Spectroscopic and photometric density contours starting at $\sim$ 6 and then increasing by steps of $\sim$ 3 \& 5 galaxies per Mpc$^{2}$, respectively. The scale line represents 2 Mpc angular extent at $z\simeq 0.62$. {\bf Right Panel:} The spectroscopic and photometric redshift distributions of the structures shown in the left panel, are respectively represented as red filled histograms (grey filled in the black \& white version) and blue histograms (black unfilled in the black \& white version).}
\label{fig:forth}
\end{figure*}

\begin{figure}
\centering
{\includegraphics[width=3.74in, trim= 30 0 -30 0, clip=true]{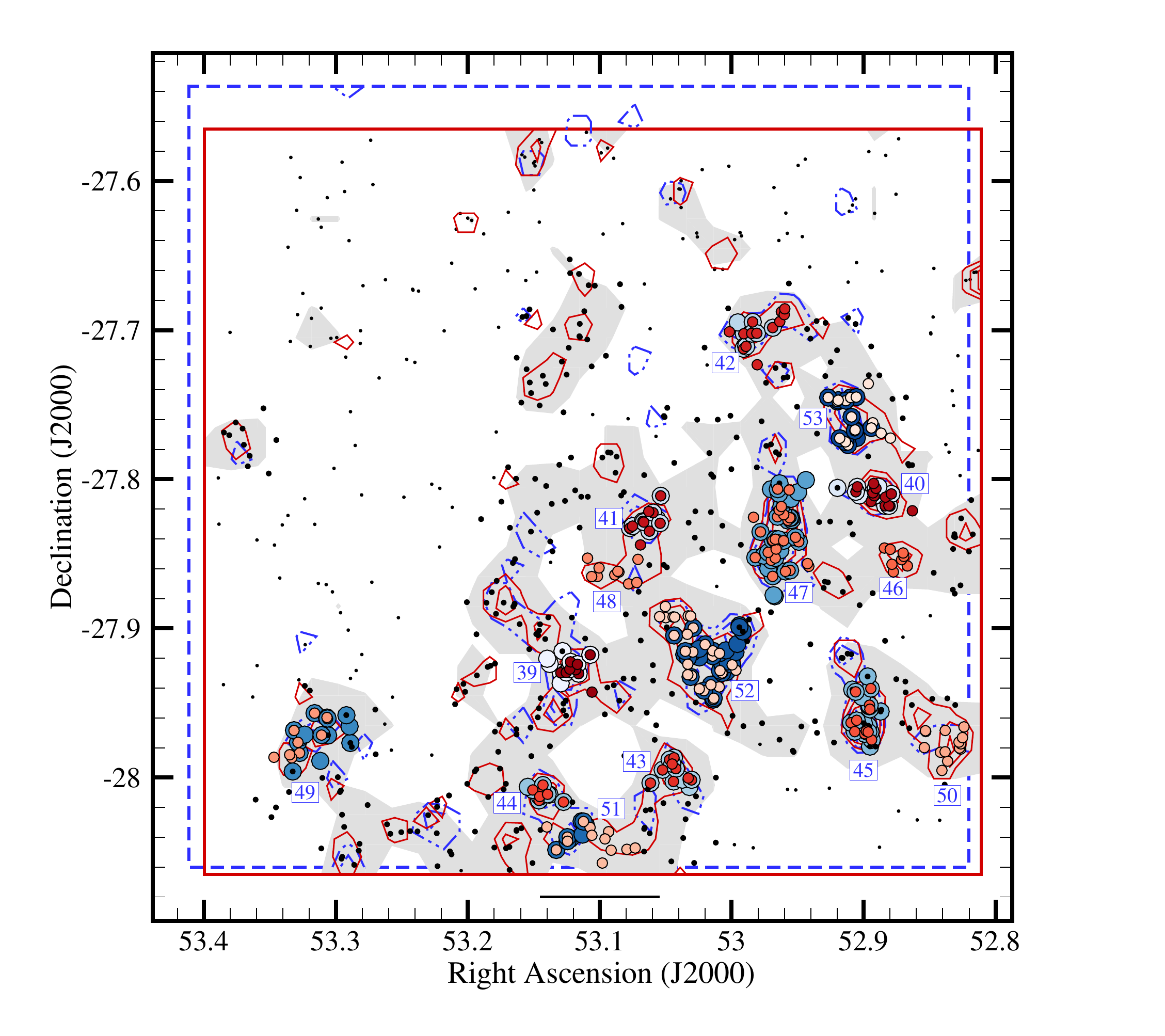}}
\caption{Detected structures within $ 0.655 \leq z_{s} \leq 0.700 $ and $ 0.640 \leq z_{p} \leq 0.715 $. The large scale structure is detected by tuning $Eps = 0.62$ Mpc at full redshift range. Spectroscopic and photometric density contours starting at $\sim$ 9 \& 11, respectively, and then increasing by steps of $\sim$ 5 galaxies per Mpc$^{2}$. The highlighted areas corresponds to the regions with minimum spectroscopic density of $\sim$ 5 galaxies per Mpc$^{2}$. The scale line represents 2 Mpc angular extent at $z\simeq 0.68$. Legends same as Figure \ref{fig:first}.}
\label{fig:fifth}
\end{figure}

\begin{figure*}
\centering
{\includegraphics[width=5.1in, trim= -30 20 0 50, clip=true]{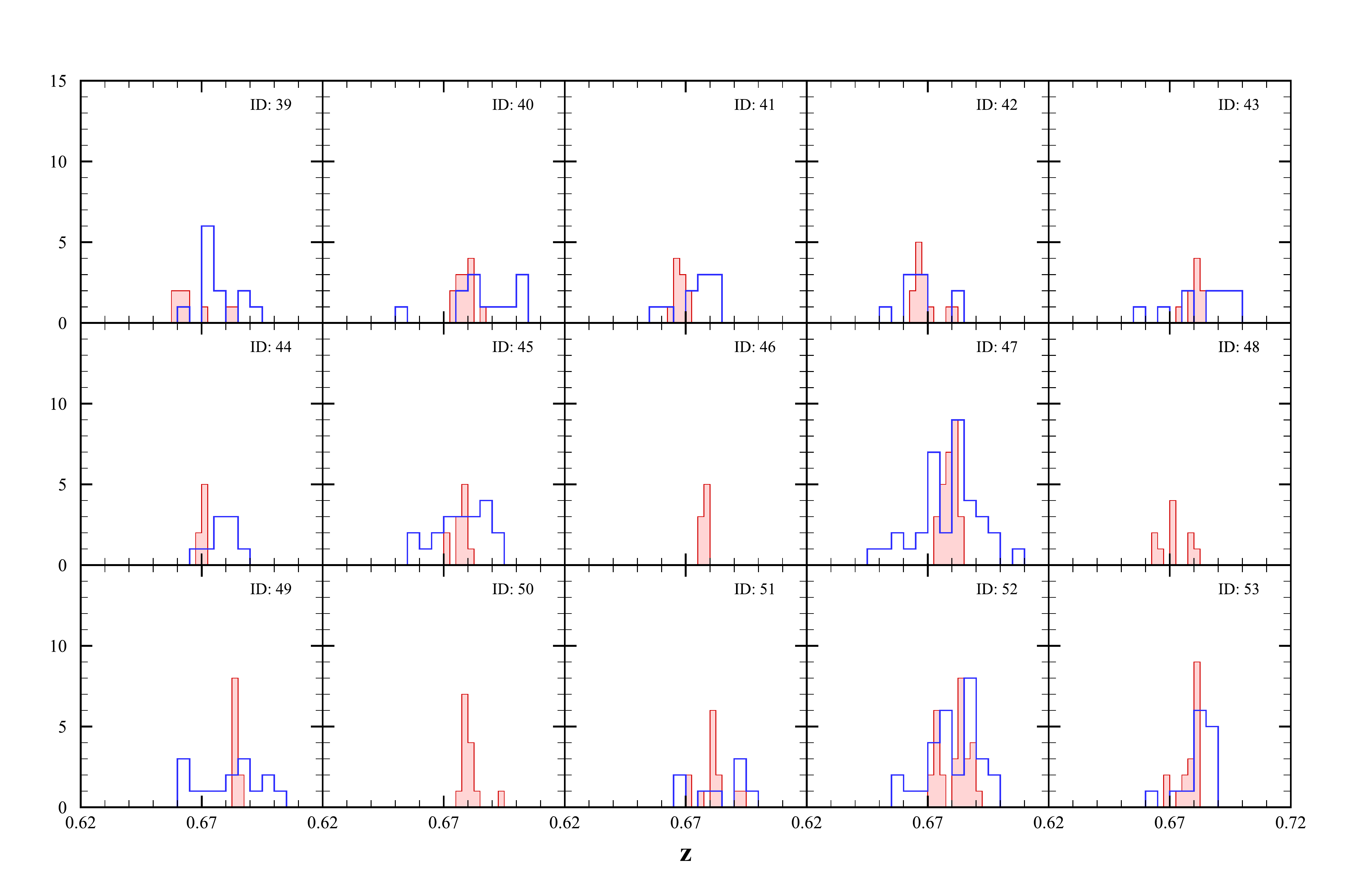}}
\caption{The spectroscopic and photometric redshift distributions of the structures shown in the the Figure \ref{fig:fifth}, are respectively represented as red filled histograms (grey filled in the black \& white version) and blue histograms (black unfilled in the black \& white version).}
\label{fig:histo5}
\end{figure*}

\subsection{Structures at $0.30<z<0.42$}

There are six structures (18-23) detected in this comparatively large isolated region of the CDFS. Four structures (18-21) are classified as groups, and Structures 23 \& 24 with broad distributions of velocities are classified as radial filaments or fake structures. All structures in this redshift range are unremarkable and therefore not shown.

\begin{figure*}
\centering
{\includegraphics[width=3.73in, trim= -5 0 5 0, clip=true]{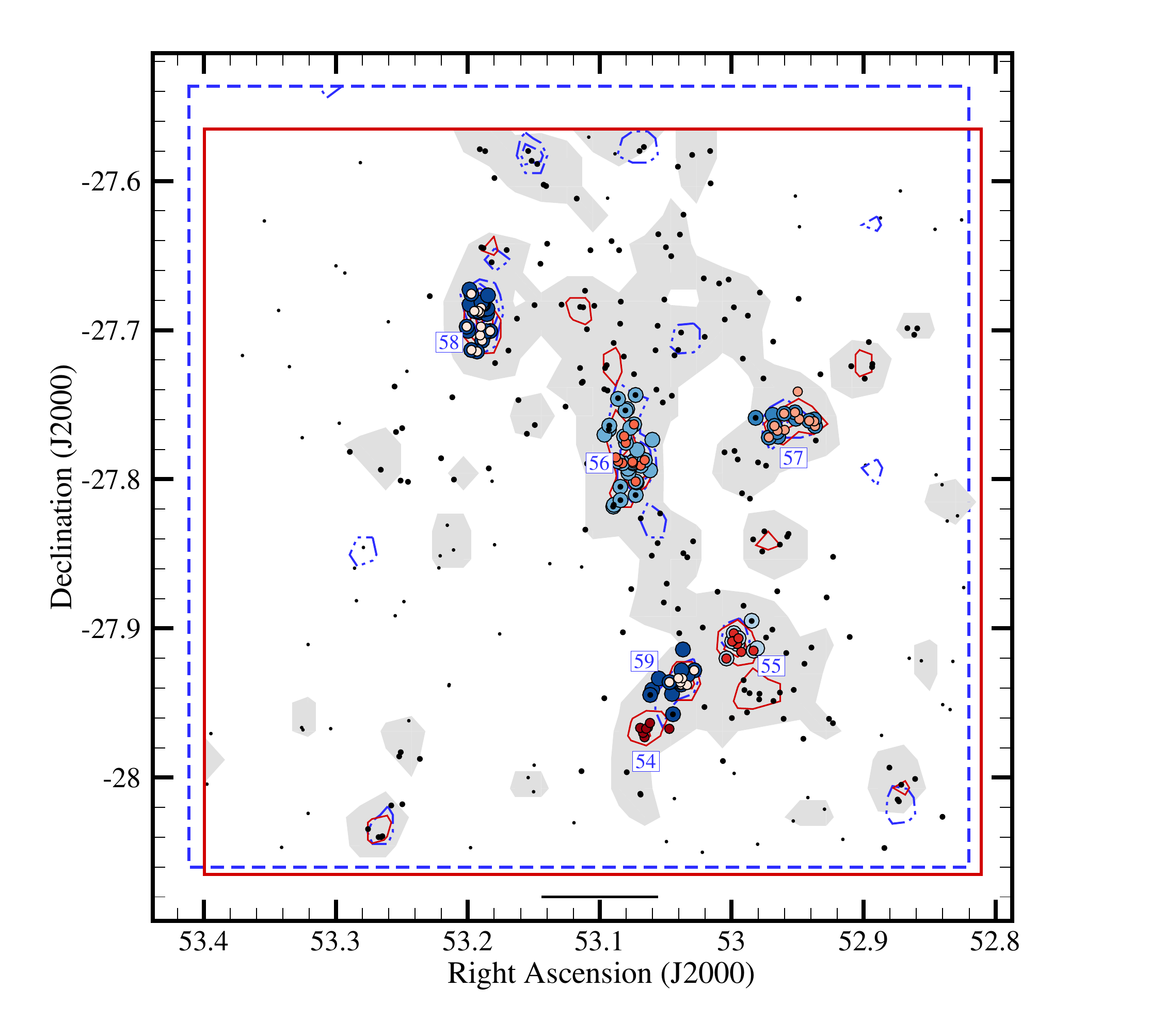}}
{\includegraphics[width=3.333in, trim= 45 1 -45 0, clip=true]{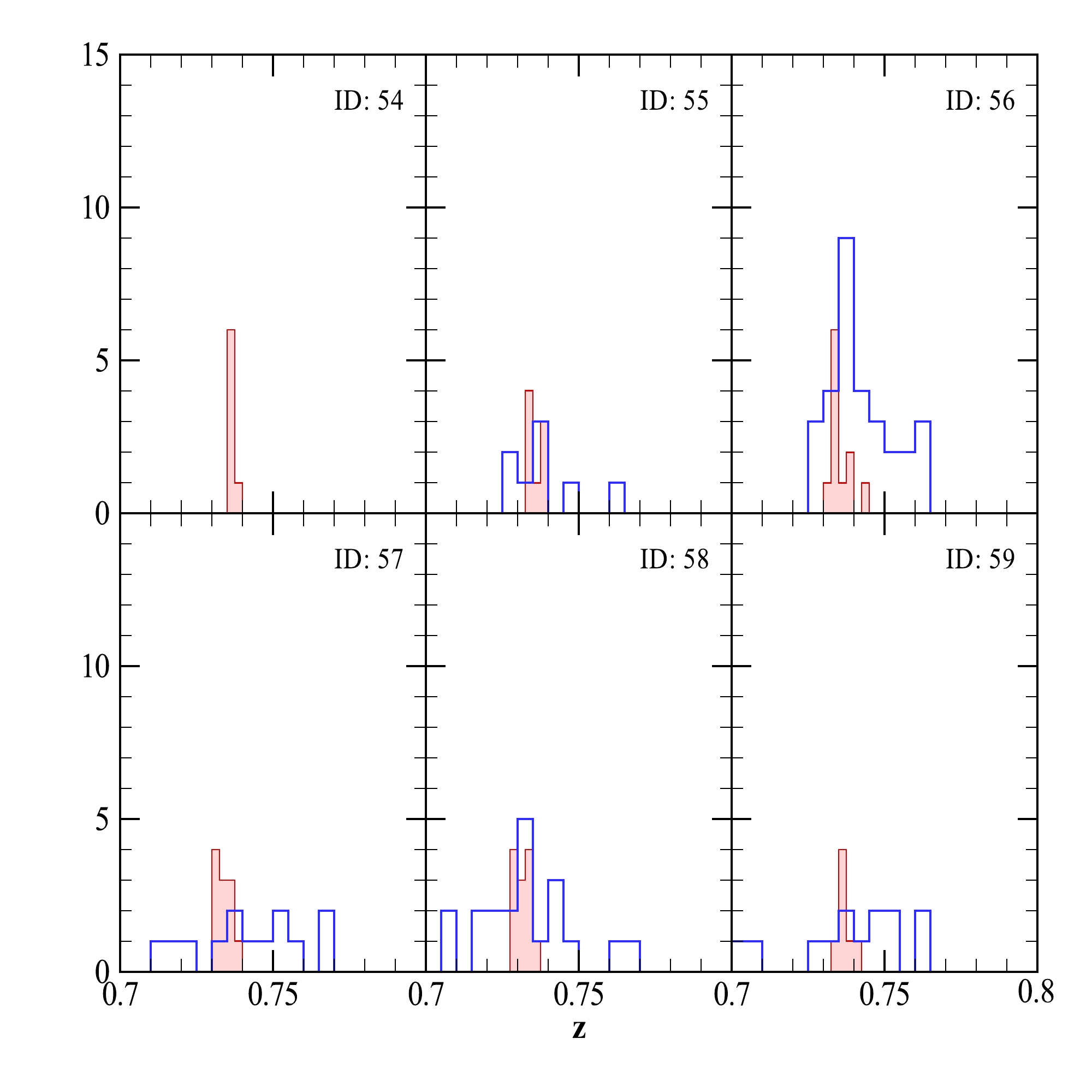}}
%{\includegraphics[width=3.53in]{Sixth-eps-converted-to.pdf}}
%{\includegraphics[width=3.53in, trim= 30 0 -30 0, clip=true]{Histos6-eps-converted-to.pdf}}
\caption{Legends same as Figure \ref{fig:first}. {\bf Left Panel}: Detected structures within $ 0.720 \leq z_{s} \leq 0.750 $ and $ 0.700 \leq z_{p} \leq 0.770 $. The large scale structure is detected by tuning $Eps = 0.92$ Mpc at full redshift range. Spectroscopic and photometric density contours starting at $\sim$ 9 \& 5, respectively, and then increasing by steps of $\sim$ 5 galaxies per Mpc$^{2}$. The highlighted areas corresponds to the regions with minimum spectroscopic density of $\sim$ 2.5 galaxies per Mpc$^{2}$. The scale line represents 2 Mpc angular extent at $z\simeq 0.735$. {\bf Right Panel:} The spectroscopic and photometric redshift distributions of the structures shown in the left panel, are respectively represented as red filled histograms (grey filled in the black \& white version) and blue histograms (black unfilled in the black \& white version).}
\label{fig:sixth}
\end{figure*}

\subsection{Structures at $0.51<z<0.55$}

At $z \sim 0.5$ the ACES richness and field of view, extending over 140 Mpc$^{2}$, is sufficiently large to reveal a clear picture of hierarchical structure formation with a wealth of groups and clusters embedded in a field of galaxies. We tuned the scale parameter, the larger $Eps$ value in k-dist graph, to detect the large scale structure as a whole in the DBSCAN. In addition, we increased the density cell sizes to 0.5 Mpc to reveal the large scale structure in the density map (shown with the grey highlighted area in Figure \ref{fig:third}).  We note that all the structures, except Structure 33, are located within the filamentary structure.

In this region of the ECDFS we found structures of all scales, from small group, and filaments to a $\sim$ 10 Mpc$^{2}$ void located in the South-East section of the ACES frame. This void with a circular outline in the plane of sky, shown by dashed line in Figure \ref{fig:third}, is located at the West side of the Structure 25 and roughly centered at ($03^\text{\textrm{h}}32^\text{\textrm{m}}45.6^\text{\textrm{s}}$, $-27^\text{\textrm{d}}55^\text{\textrm{m}}57^\text{\textrm{s}}$), with a diameter of $\sim 4$ Mpc. Three structures (24, 27, and 28), embedded in the intersection of two major filaments (North-South and East-West), present with double peaks at $z\simeq0.522$ and $z\simeq0.534$ in their redshift distribution (see Figure \ref{fig:histo3}). Structure 24 appears to be a part of the filament connecting two nearby overdensities (Structures 25 and 28). Structure 28 with its 17 associated spectroscopic members and $v_{d}=411^{+82}_{-82}$ \kms is a cluster located at $z_{s}=0.5205\pm0.0006$. The interacting Structures 26 and 29, are both located at $z\simeq0.523$ and have an angular separation of about 1 Mpc. Structure 30 with 9 members and $v_{d}=410^{+169}_{-74}$ \kms is a massive group located at the edge of the ACES field. Structure 31 with a sampling rate of 17 galaxies and velocity dispersion of $454^{+105}_{-105}$ \kms at $z_{s}=0.5264\pm0.0006$ is a typical cluster, which appears to be in an interaction with neighbouring galaxy concentrations to the East (as can be seen in the spectroscopic density map). Both Structures 30 and 31, located at the left and right edges of the field, were only spectroscopically detected, since photometric redshifts are less reliable on the edges and have already been flagged out from the photometric sample by constraining the $Qz$ factor. In addition, Structure 32 is an isolated group at $z_{s}=0.5615\pm0.0007$, which is not shown in the Figure \ref{fig:third}.

\subsection{Structures at $z \sim 0.62$}

In Figure \ref{fig:forth} we plot six detected structures at $z \sim 0.62$ and their corresponding redshift distribution. A large, although weak, web-like structure was detected by tuning the $Eps$ parameter to 1.17 Mpc. This structure is considerably less populated than the filamentary structure at $ z \sim 0.52$, partially due to the higher redshift. Structure 35, already detected by \ada, along with structures 33 and 38 are particularly interesting, as they appear to be segments of a triangular-shape filament structure located in the South section of the field. Structure 33, with $v_{d}=171^{+63}_{-29}$ \kms, is a compact group located at the North-West vertex of the filament structure. Structure 36, likewise, is a group that is located on a large scale structure of inverted Y-shape. The density map of Structure 36 shows a complex system of three smaller groups, of which two are labelled as Structure 36 in DBSCAN. The undetected part of the system, shown by contours, did not meet the MinPts criteria of the DBSCAN analyses. Structures 34, already detected and classified as a structure in an early formation stage by \ada, and 37 are isolated groups.

\subsection{Structures at $0.66<z<0.70$}

This region of the field, with 704 spectroscopic redshifts and 15 overdensity regions, is the most significant density peak of the ECDFS. This dense sheet of galaxies has already been detected by \gil, \citet{wmk04}, \ada, and \sal. The redshift distribution of the region is dominated by two solid density peaks located at $z_{s} = 0.670 \pm 0.001 $ and $z_{s} = 0.680 \pm 0.001$. Shown by the highlighted grey shading in Figure \ref{fig:fifth}, the large scale structure was detected as a whole (shown by black circles), by adjusting the scale parameter to $Eps=0.62$ Mpc.

The South section of the large scale structure, including Structures 43, 44 (already detected by \ada), along with a filament-like Structure 51, appears to be a continuation of the triangular shape structure detected at $z \sim 0.62$ (see the left panel of Figure \ref{fig:forth}). Structure 39 and 48, especially the former with an apparent compact structure and $v_{d}=1674^{+654}_{-303}$ \kms, are filament structures with a radial (line of sight) elongation. Already detected by GCD03 and SCP09 Structure 41 located at $z_{s}=0.6666\pm0.0008$ is a large group or small cluster possibly interacting with the filamentary Structure 48. Structures 47 and 52 with 27 and 29 spectroscopic members are classified as clusters, of which the latter has a disturbed spatial distribution and presents two major peaks in the histogram. From the velocity histogram we find Structure 52 to consist of a cluster and a group (in a merger process) with $v_{d}=500^{+66}_{-66}$ and $v_{d}=246^{+49}_{-49}$ \kms at $z_{s}=0.6846\pm0.0007$ and $z_{s}=0.6732\pm0.0005$, respectively. Structure 40, 42, and 45 with sufficient sampling rate ($N \geq 11$) and velocity dispersions of $459^{+67}_{-67}$, $398^{+154}_{-154}$, and $495^{+165}_{-165}$ \kms are clusters located at redshifts of $0.6781\pm0.0008$, $0.6661\pm0.0007$, and $0.6773\pm0.0009$, respectively. The remaining structures in the slice (46, 49, 50, \& 53) were all classified as groups, of these Structures 46 and 50 were only spectroscopically detected due to their position being at the edge of the MUSYC field.

\subsection{Structures at $0.72<z<0.75$}

In Figure \ref{fig:sixth} we plot the redshift and spatial distribution of six identified structures at $0.72<z<0.75$. All of the detected structures are located at $z_{s}=0.735\pm0.002$ within the large scale structure shown by large black dots (correspond to $Eps=0.92$ Mpc), and highlighted grey in the left panel of Figure \ref{fig:sixth}. Structure 56 is a well studied cluster (\gil, \ada, \citealp{tcf07}, and \sal), located at the heart of the large scale chain at $z_{s}=0.7341\pm0.0009$. Structures 55 and 58, also already detected by \ada, are isolated overdensities with velocity dispersions of $438^{+178}_{-78}$ and $399^{+191}_{-78}$ \kms, of which the latter with 12 members is sufficiently rich to be considered as a cluster. Structure 54 and 59 are two compact groups located at $z_{s}=0.7360\pm0.0004$ and $z_{s}=0.7362\pm0.0008$ with $\sim$ 1 Mpc angular separation. These are conceivably an interacting system in an ongoing merger process. Structure 57 with sampling rate of 11 and $v_{d}=451^{+78}_{-78}$ \kms is classified as a cluster.

\subsection{Structures at $0.8<z<1$}

We detected three structures beyond redshift 0.8. Two structures, 60 \& 61, are a group and a cluster located at close spectroscopic redshifts of $0.8350\pm0.0011$ and $0.8374\pm0.0006$, receptively. The presence of Structures 60 \& 61 with close spectroscopic redshifts is possibly evidence of another thin and populous sheet at $z\simeq 0.83$. Structure 62 with $142^{+76}_{-28}$ \kms and $z_{s}=0.9666\pm0.0004$, is classified as a group, which is the farthest structure we detected in the ACES field.

\begin{deluxetable}{lcccccc}
\centering
\tabletypesize{\scriptsize}
\tablewidth{0pt} \tablecaption{Properties of massive structures\label{tab:massive}}
\tablehead{\multirow{2}{*}{ID} & \multirow{2}{*}{Class} & \multirow{2}{*}{$z_{s}$} & \multicolumn{2}{c}{$R_{200}$} & \colhead{$M_{200}$} & \colhead{X-ray\tablenotemark{a}}\\
\colhead{} & \colhead{} & \colhead{} & \colhead{Mpc} & \colhead{arcmin} & \colhead{$\times 10^{13} M_{\odot}$} & \colhead{emission}}
\startdata 
\input{TabM.tex}
\enddata
\tablenotetext{a}{Soft band X-ray emission detection for the structure. $\star$ corresponds to the structures obscured by foreground emission. $\dagger$ represents structures located out of the field.}
\end{deluxetable}

\begin{figure*}
\centering
{\includegraphics[width=7.55in, trim= -7 0 -58 0, clip=true]{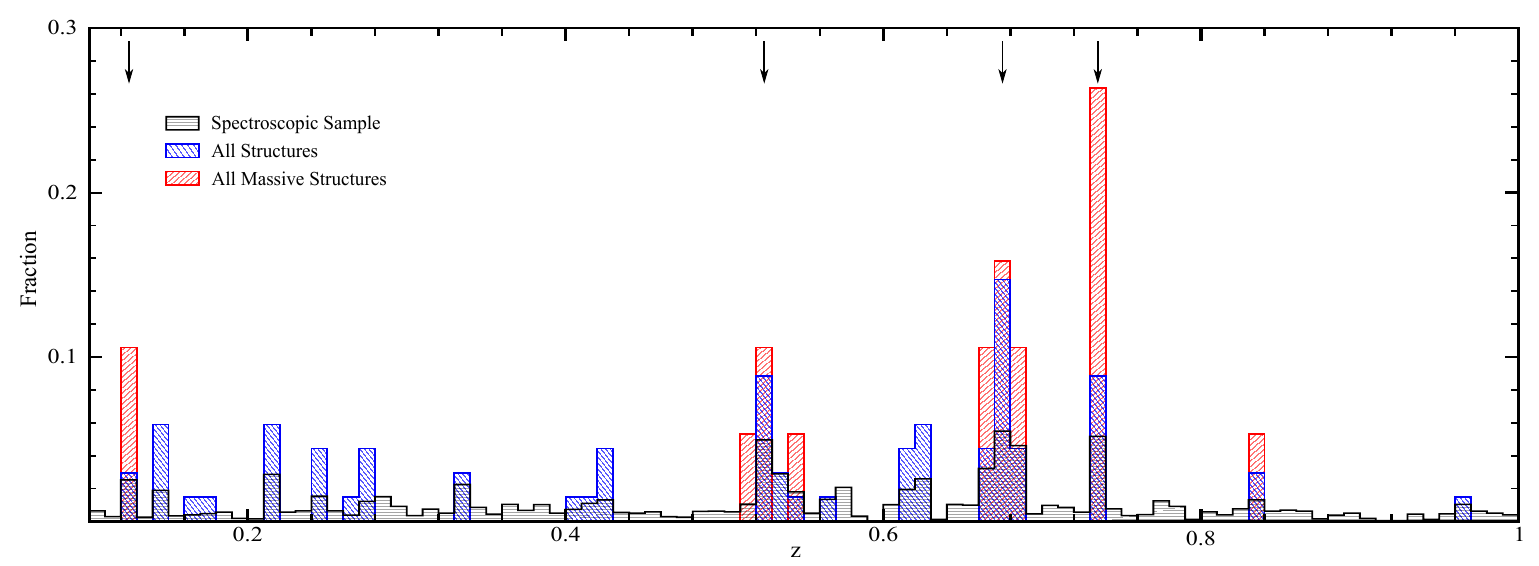}}
\caption{The fractional distribution of galaxies and detected structures in the ACES field. Black, blue, and red histograms respectively show the fractions of all galaxies, the 62 detected structures, and the 19 massive groups and clusters in the corresponding bins with $\delta z =0.01$. The arrows represent the location of detected large scale structures.}
\label{fig:overall}
\end{figure*}

\subsection{Massive Structures}

We now concentrate on the massive structures detected across the whole spectroscopic redshift range. Thirteen structures meet our classification conditions, $v_{d} \gtrsim 400$ \kms and $N \geq N_{c}(z)$, for clusters (class 5). In addition, 6 non-filamentary structures with $v_{d} \gtrsim 400$ \kms do not conform to the richness condition, and are classified as big groups or small clusters (class 4). Assuming that all the class 4 \& 5 objects are virialized structures, we estimate the virial mass based on the empirical $M_{200}-\sigma_{v}$ relation \citetext{\citealp{e04} \& \citealp{v05}}:

\begin{displaymath}
M_{200}=\frac{10^{15}h^{-1}M_{\odot}}{\sqrt{\Omega_{m}(1+z)^3+\Omega_{\Lambda}}} {\left(\frac{\sigma_{v}}{\text{1080 km s}^{-1}}\right)}^{3}
\end{displaymath}

In the case of clusters which are not virialized, as is evident in several cases here, this estimate will be less accurate and should be taken as an indicative mass only \citep{tnm10}. Table \ref{tab:massive} provides the virial radius mass estimation of the 19 massive structures. In addition, Figure \ref{fig:overall} represents the redshift distribution of massive groups and clusters overlaid on the redshift distribution of all 63 the detected structures, along with all the galaxies in the spectroscopic sample. Clearly, the majority of the detected clusters and massive groups (18/19) are located in highly-populated and thin sheets of galaxies at $z\simeq0.13$, $z\simeq0.52$, $z\simeq0.68$, and $z\simeq0.73$. Note that all of the detected clusters and massive groups, within the mentioned ranges, are embedded in larger structures (see the arrows in Figure \ref{fig:overall}), whereas nearly $60\%$ of all the detected structures (36/62) are located in large scale structures.

\begin{figure*}
\centering
{\includegraphics[width=2.3in, trim= 0 20 0 0, clip=true]{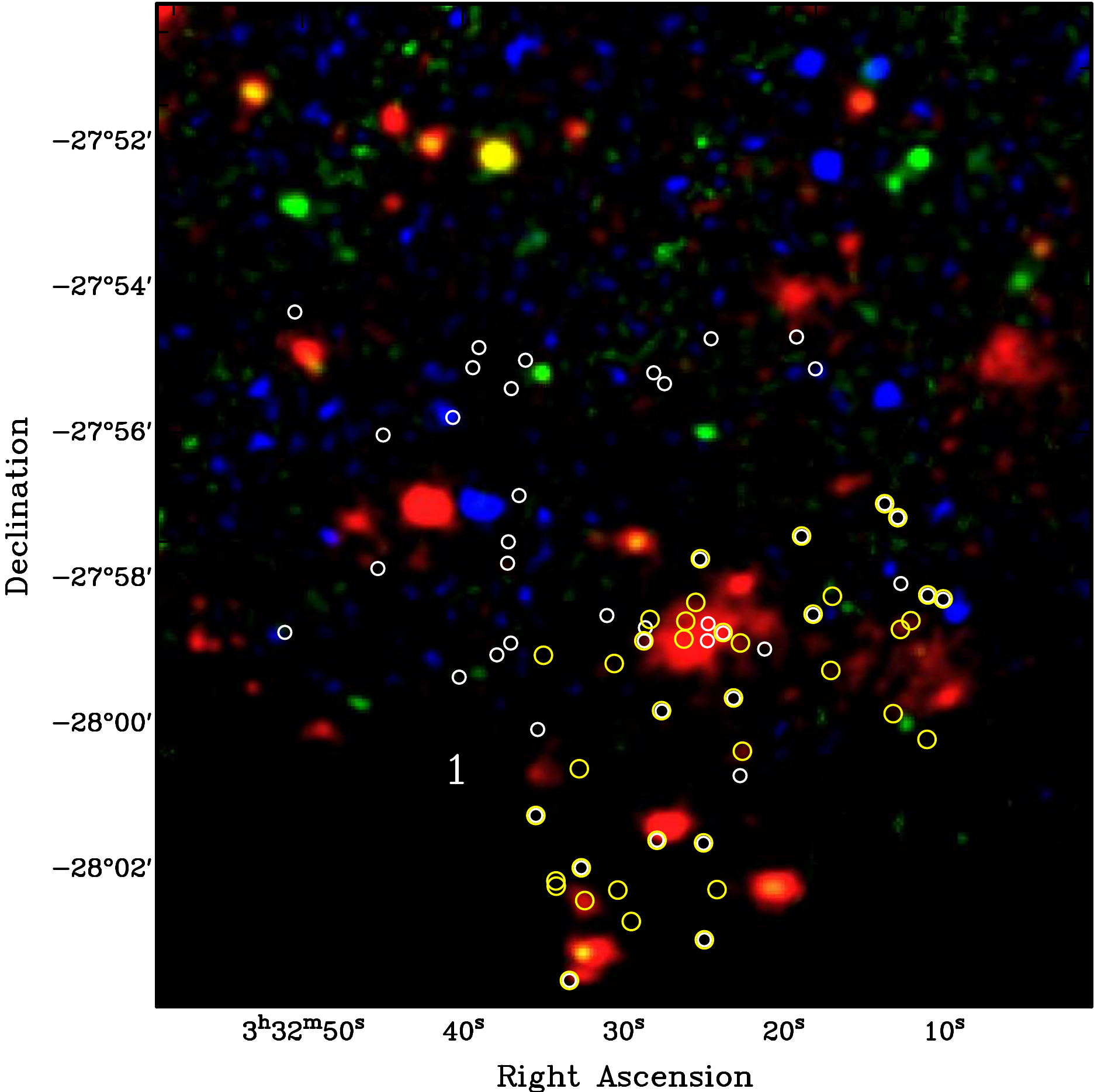}}
{\includegraphics[width=2.174in, trim= 30 24 0 0, clip=true]{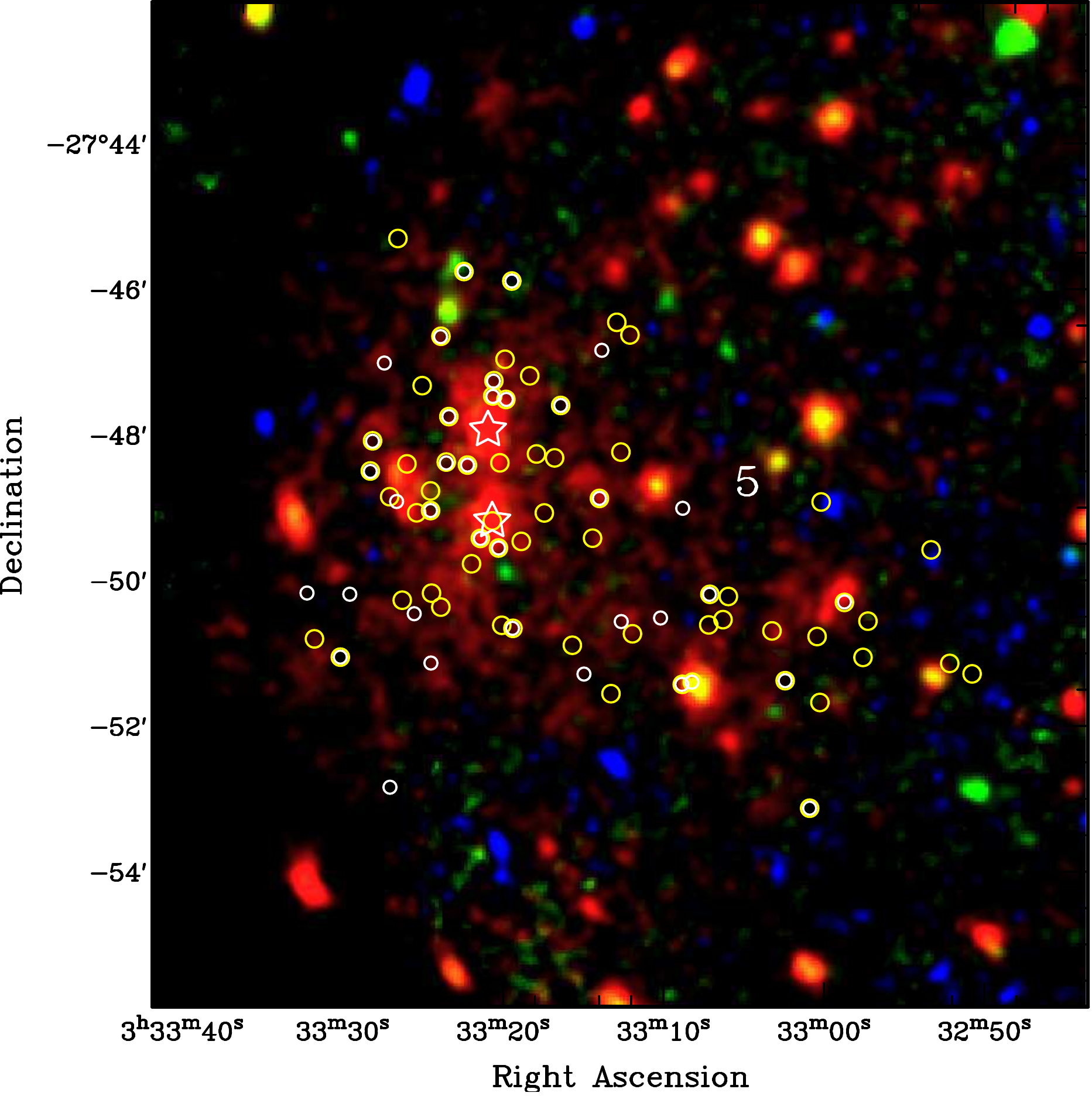}}
{\includegraphics[width=2.174in, trim= 30 20 0 0, clip=true]{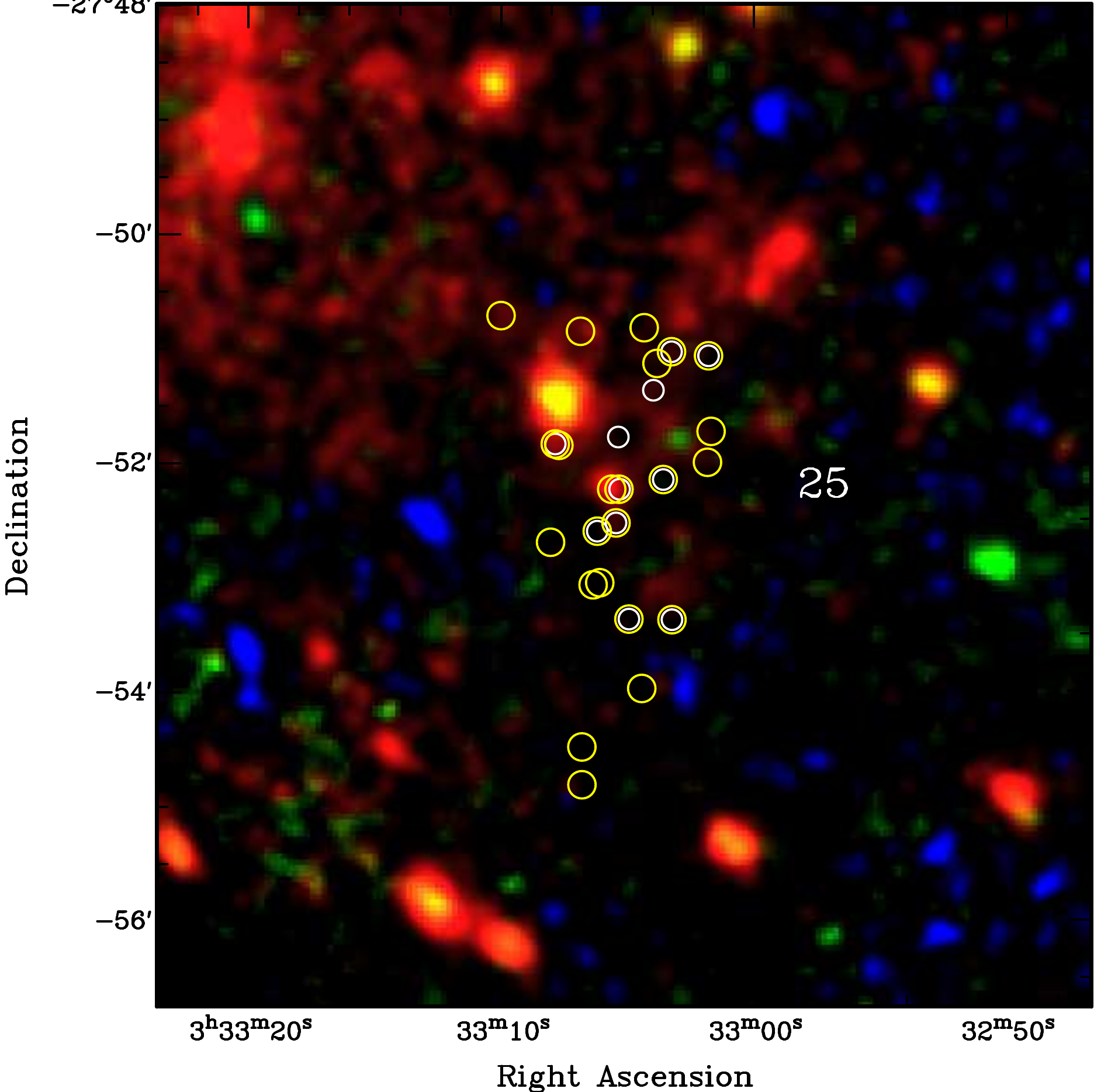}}
{\includegraphics[width=2.3in, trim= 0 20 0 0, clip=true]{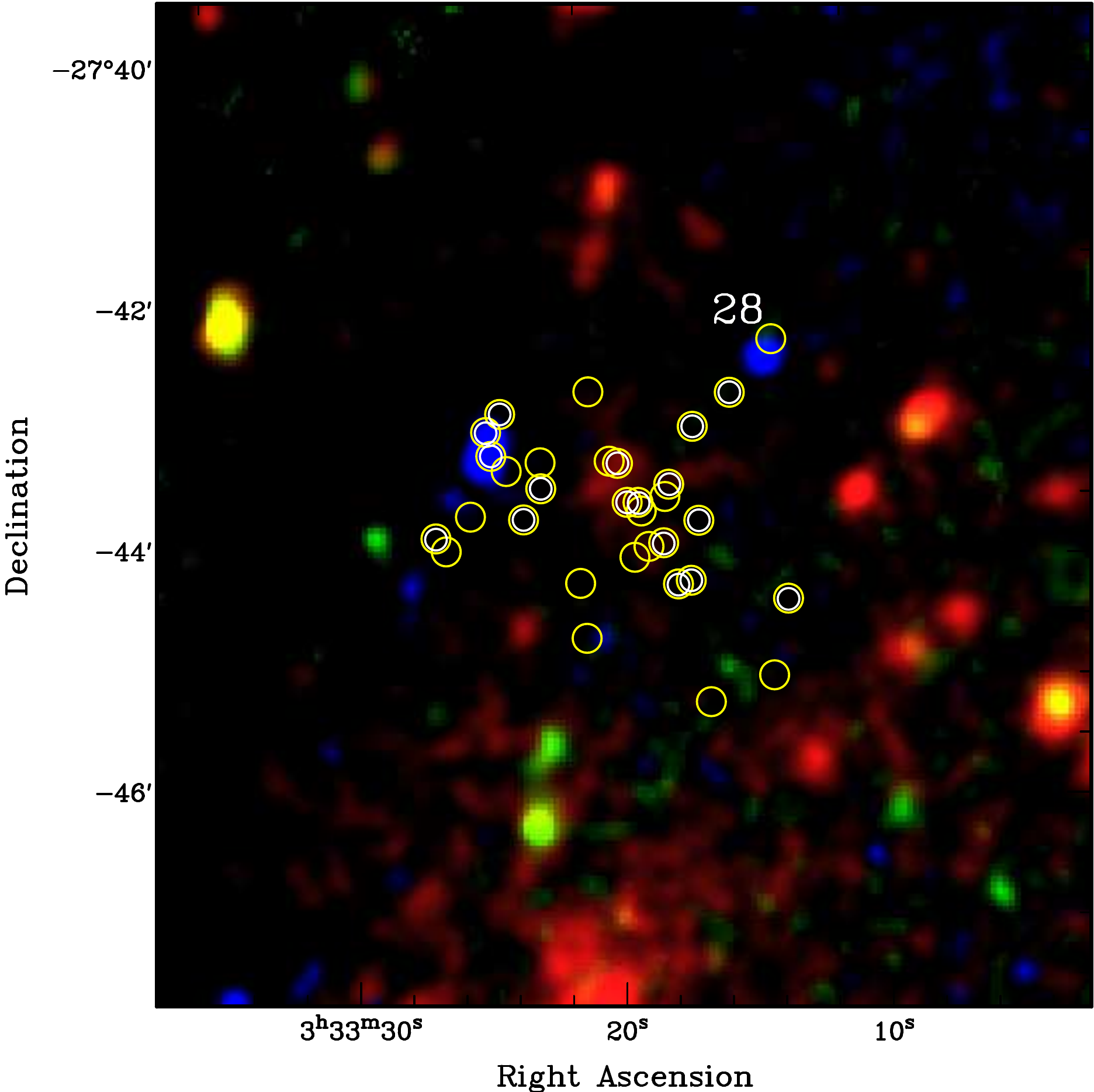}}
{\includegraphics[width=2.174in, trim= 30 20 0 0, clip=true]{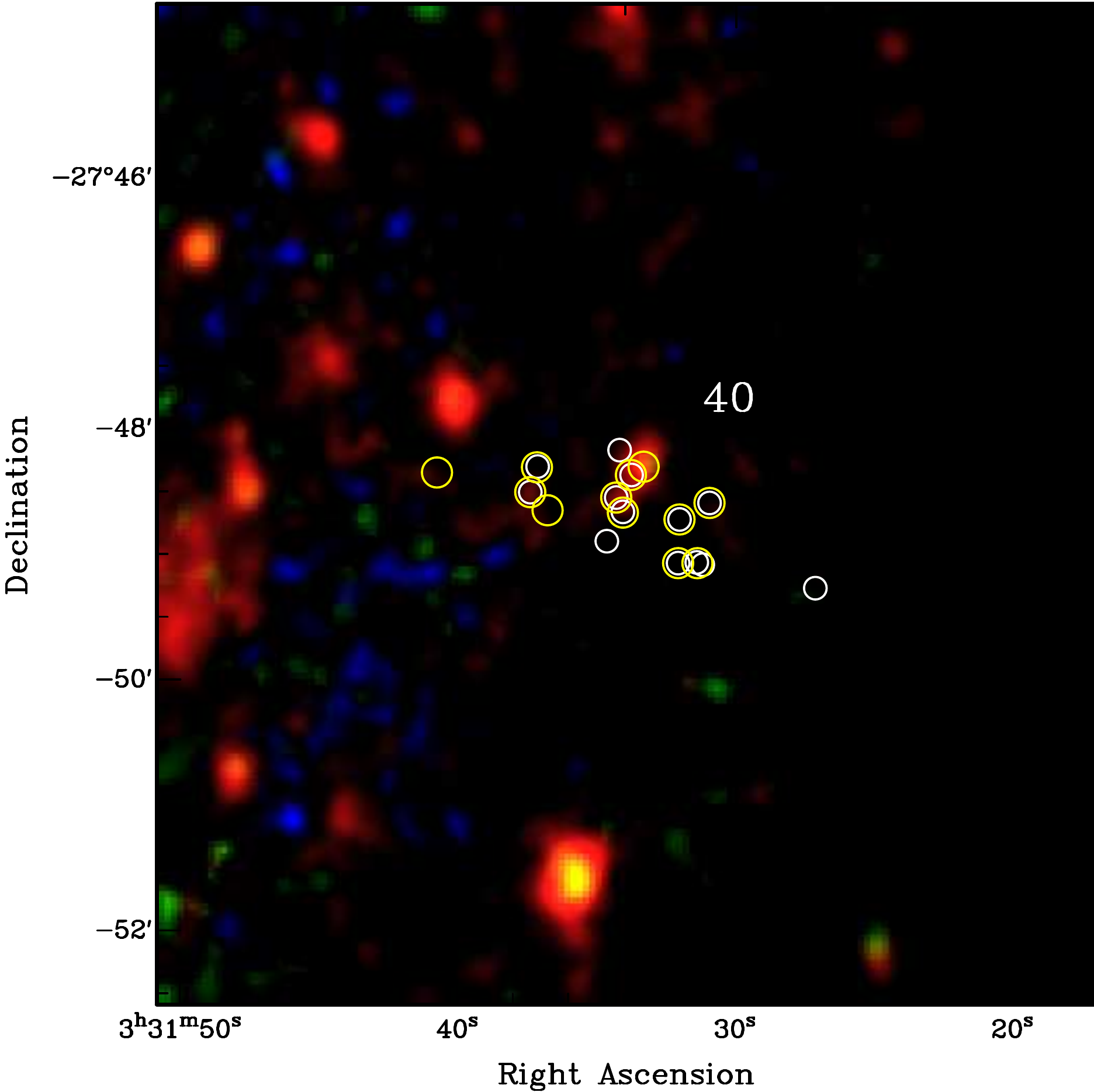}}
{\includegraphics[width=2.174in, trim= 30 20 0 0, clip=true]{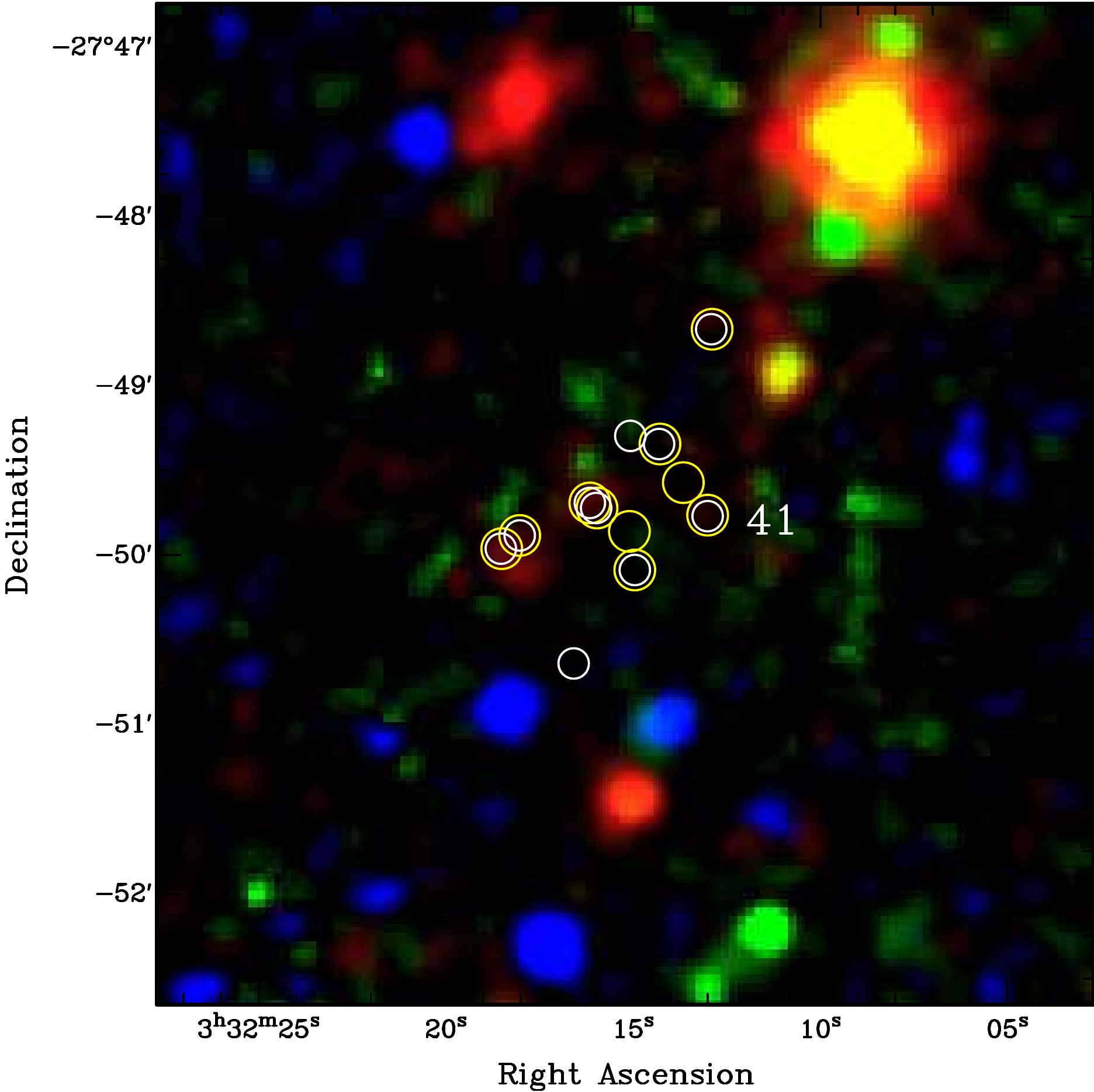}}
{\includegraphics[width=2.3in, trim= 0 20 0 0, clip=true]{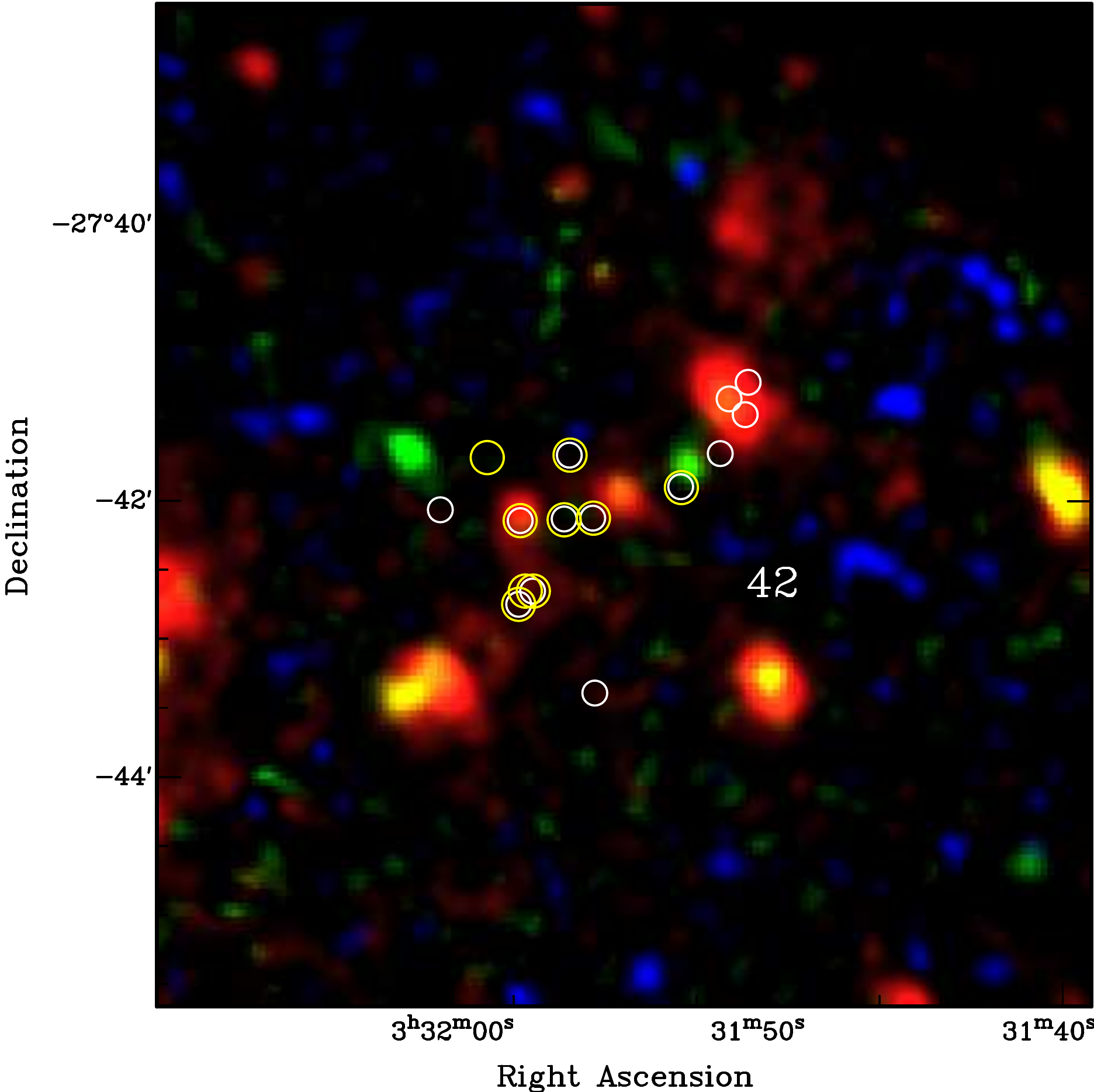}}
{\includegraphics[width=2.174in, trim= 30 20 0 0, clip=true]{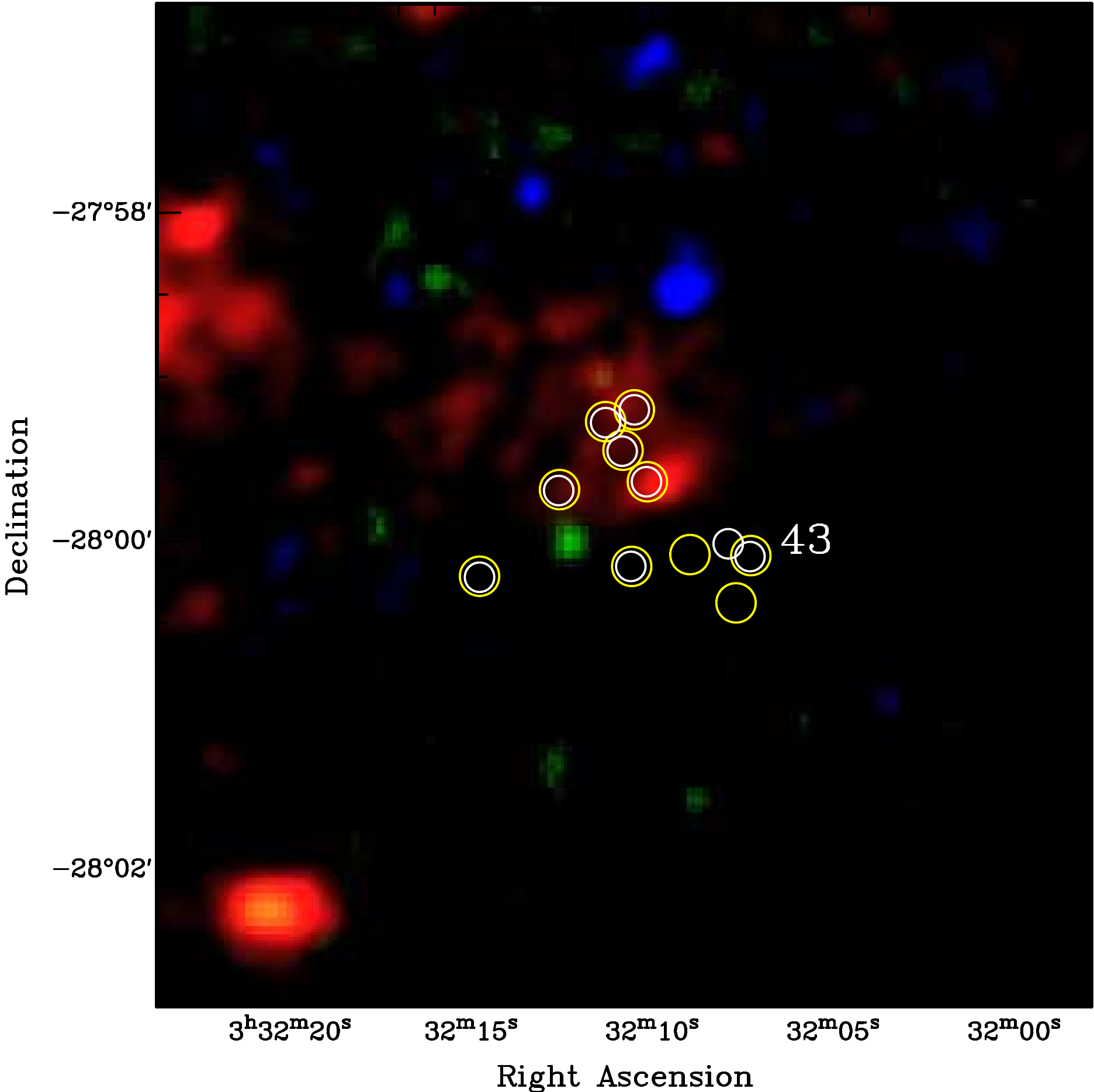}}
{\includegraphics[width=2.174in, trim= 30 20 0 0, clip=true]{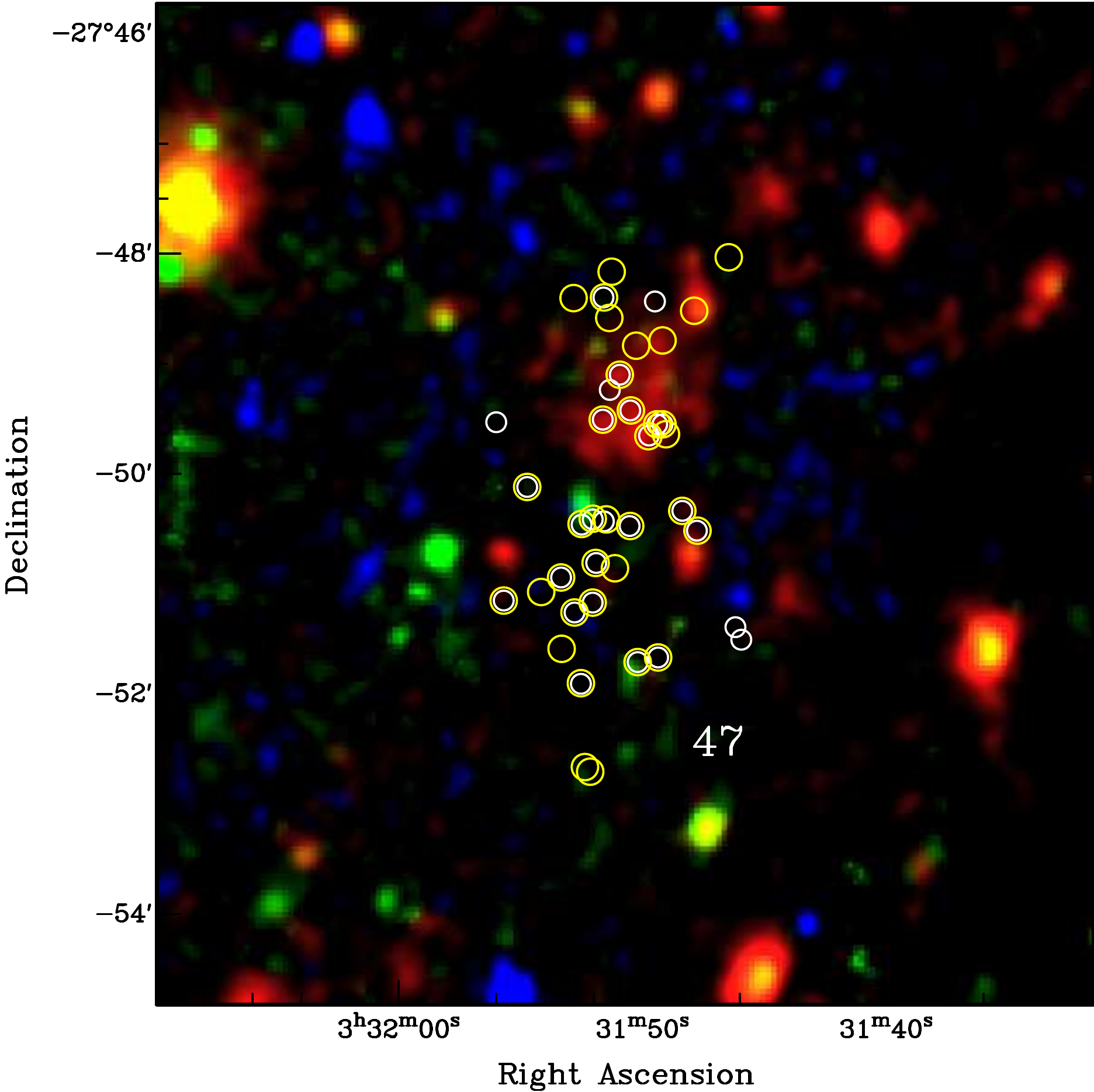}}
{\includegraphics[width=2.3in, trim= 0 20 0 0, clip=true]{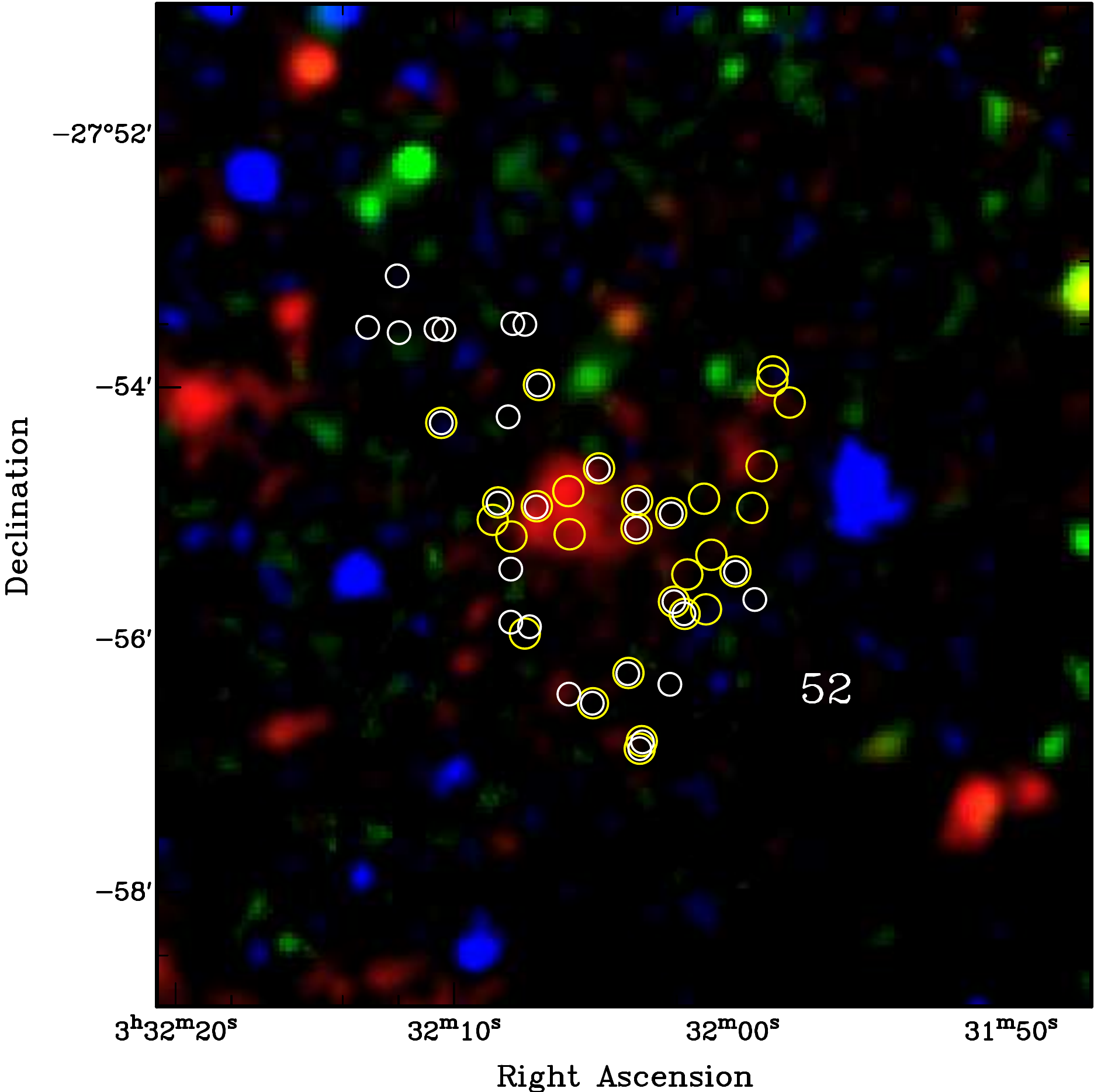}}
{\includegraphics[width=2.174in, trim= 30 20 0 0, clip=true]{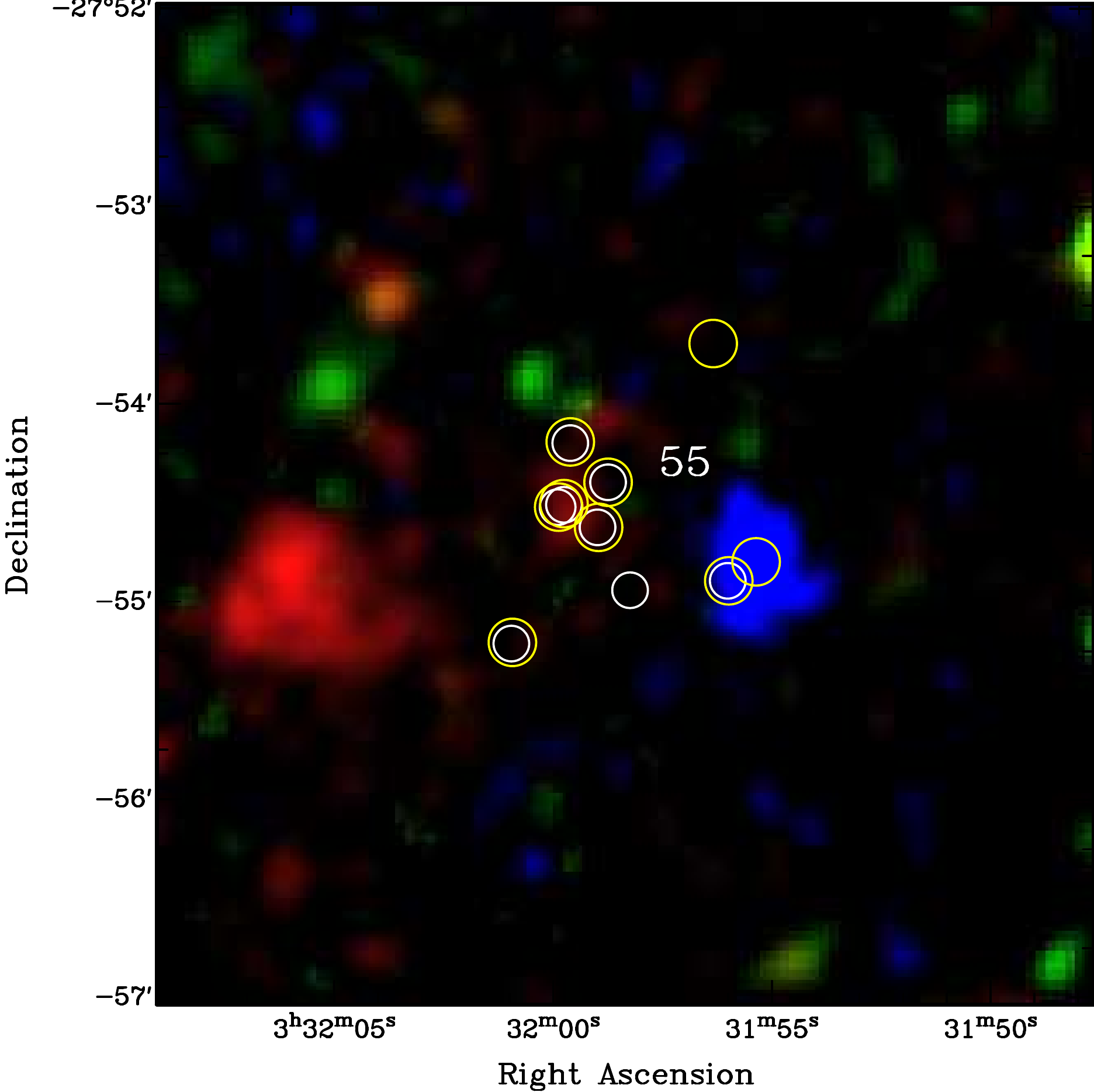}}
{\includegraphics[width=2.174in, trim= 30 20 0 0, clip=true]{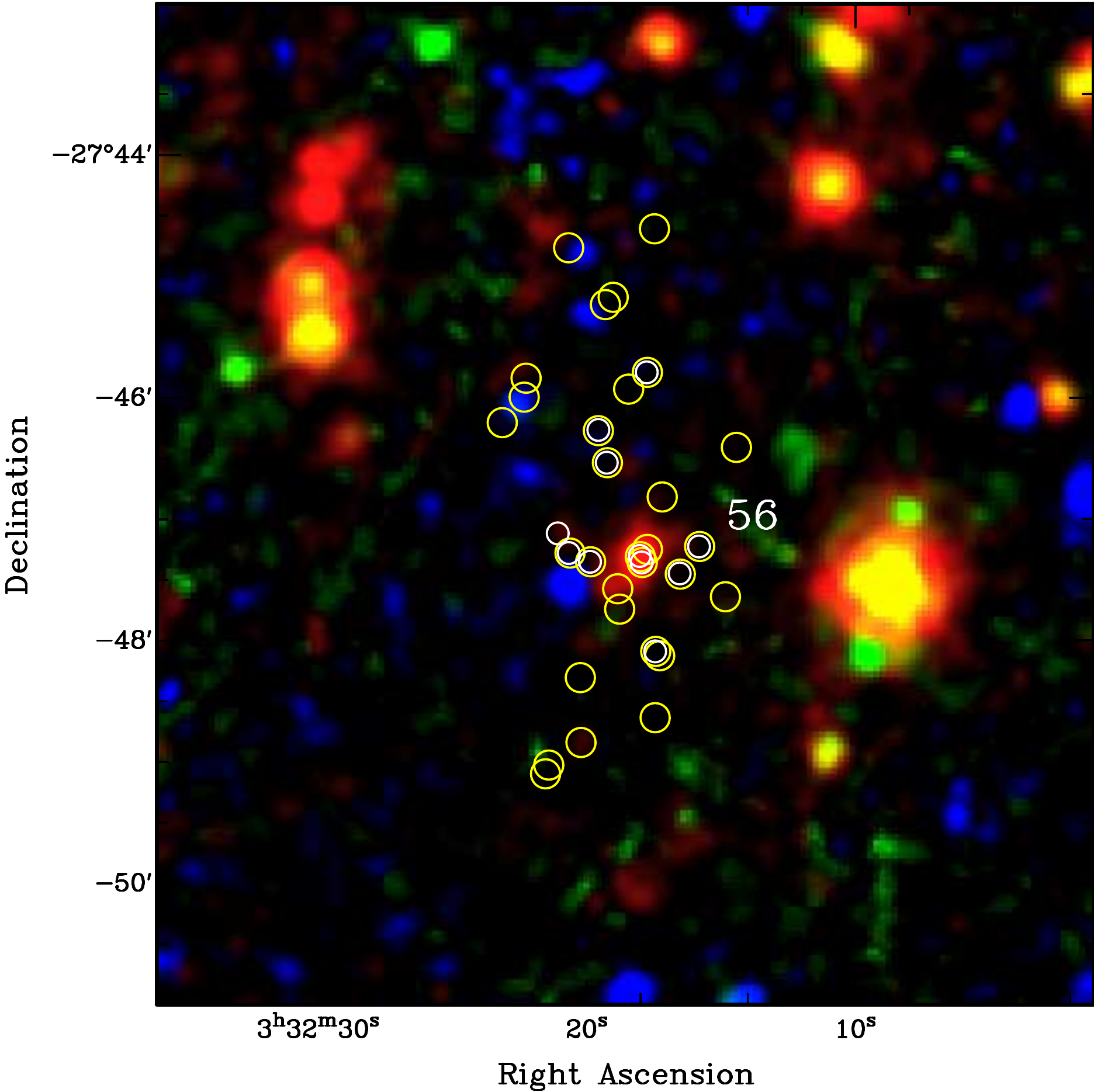}}
\caption{Massive structures, presented in the Table \ref{tab:massive}, overlaid on the composite X-ray image of the CDFS \citep{cri11}. The soft (0.4 - 1 keV), medium (1 - 2 keV), and hard (2 - 8 keV) bands are represented by red, green, and blue colors respectively. Spectroscopic and photometric members of the structures are shown by large (yellow) and small (white) circles, respectively. The IDs correspond to those given in Table \ref{tab:catalogue}. The two stars shown on ID 5 represent the position of the cD galaxies.}
\label{fig:xray}
\end{figure*}

\section{X-ray Comparisons}

X-ray observations of the CDFS are extensive and include both XMM-Newton and Chandra data which is both published and publicly available in the respective instrument archives. In particular a total of $\sim$ 3 Ms of XMM-Newton observations have been collected over a period of 9 years and include both the CDFS and flanking fields. These data, which comprise the deepest XMM exposure of a single field ever undertaken, have been combined into a single impressive multiband image in \citet{cri11} and \citet{rcv13}. The multiband image shows the combined {\it pn} and MOS data over a $\sim$ 32' $\times \sim$ 32' field smoothed with a 4 arcsecond Gaussian kernel in three color coded bands being; soft (0.4 - 1 keV), medium (1 - 2 keV) and hard (2 - 8 keV) in red, green and blue, respectively. These data represent an extensive observational campaign and are a worthy comparison to the overdense structures presented here, particularly for the diffuse, soft band emission evident in several areas.

In order to make such a comparison we restored the coordinate system to the published 3-color image given in Figure 1 of \cite{cri11} by using the original XMM-Newton archival data which still contains the positional information. This was done by using the Koords task in the Karma software package \citep{g96}, which applies a coordinate transformation based on matched pairs of points between two images. In this case we used the positions of 13 point-like active galactic nuclei (AGN) evident across the field in the original, raw archival data and the published image of \citet{cri11} et al. As the published image has been smoothed by a $4^{\prime\prime}$ Gaussian kernel, the astrometric accuracy of the coordinates applied to this image is only precise to of order $0.5^{\prime\prime}$ (half a pixel).\long\def\/*#1*/{}\/* As a cross-check the published coordinates of 20 AGN \cite{rcv13} not used as targets for the coordinate transformation were compared with their positions in the transformed image. The mean difference in coordinates is X and Y.*/ Therefore, the astrometric offset generated by applying the coordinate system after the fact gives astrometric errors which are well within the uncertainties of the spectroscopically and photometrically detected galaxies used in this work.

Of the 19 such structures listed in Table \ref{tab:massive} classified as either a large group (4) or cluster (5), three lie outside of the XMM observations (objects with ID 30, 31 and 45) and therefore cannot be compared to the X-ray images. Two structure (25 and 60) are obscured by foreground objects, in the case of object 60 which is extremely distant (z=0.835), it is impossible to distinguish any features and we therefore do no present it here.

The remaining 15 large groups or clusters are shown in Figure \ref{fig:xray} as sections of the 3-color XMM image from \citet{cri11} and \citet{rcv13} overlaid with the location of the photometric and spectroscopically identified structure members. We note that the color-scale is unaltered and therefore follows that presented in the literature which has been noted to be non-linear and has suppressed the background, see \citet{rcv13} for details.

As seen in Figure \ref{fig:xray} soft band (0.4 - 1 keV), diffuse emission is seen coincident with the centre of the galaxy distributions for many of the clusters and large groups detected here. While it is beyond the scope of this paper to present quantitative results of the X-ray data, the detection of soft band, diffuse emission is a strong independent confirmation of the reality of these overdensities, and furthermore suggests they are clusters hosting thermal, Bremsstrahlung emitting plasma. Details of the comparison for the X-ray emission for each object are discussed below.

{\bf Structure 1} - This is classified as a cluster here at a mean spectroscopic redshift of z$_{s} = 0.1253 \pm 0.0003 $ with a combined total of 66 unique members identified in the photometric and spectroscopic samples. The distribution of all members is roughly circularly symmetric and concentrated in density at the core (see Figure \ref{fig:first}). At the location of the peak galaxy number density, a diffuse soft band X-ray source is clearly evident. There is also some additional, low level diffuse, soft band emission to the west of the core. Additionally, there appears to be 3 AGN associated with members galaxies.

{\bf Structure 5} - This is a large structure in terms of angular size with $\sim 2$Mpc diameter, z$_{s} = 0.1267 \pm 0.0004$, 75 distinct member galaxies, and 2 cD galaxies (see Figure \ref{fig:1-5}). The member galaxy distribution shows a core with north-south elongation and clear signs of substructure to the west, where a large group is detected. The diffuse X-ray emission replicates precisely this distribution showing emission elongated into two strong peaks with north-south orientation, which are aligned with two cD galaxies  at the center of the cluster (shown with the star symbols in the upper middle panel of Figure \ref{fig:xray}). In addition, there is a lower surface brightness emission following the galaxy distribution to the west. This is therefore very likely to be a merging system.

{\bf Structure 25} - We classify Structure 25 as a large group lying at z$_{s} = 0.5200 \pm 0.0007$ and having 25 unique members. Unfortunately, this object lies behind the western sub-structure in Structure 5 and hence we cannot asses any likely diffuse X-ray emission for this structure. We do find one AGN aligns with a member of this group.

{\bf Structure 28} - We determine this to be a cluster with 32 galaxy members and a central core with z$_{s} = 0.5205 \pm 0.0006$ into which an associated subgroup at z$_{s} = 0.5332 \pm 0.0008$ is likely to be in-falling. Despite being just north of Structure 5, we find the peak galaxy number density does correspond to a very faint, small, diffuse, soft band X-ray source, which is suggestive of cluster emission. However, we note that post-processing of the original X-ray data such as wavelet analysis on a image with point sources removed is likely to be required to confirm this detection. Additionally, we find one hard band, X-ray source to be associated with a cluster member.

{\bf Structure 40} -  There are 16 galaxies associated with this cluster which is located at z$_{s} = 0.6781 \pm 0.0008$. At the centre of the object is a slightly elongated soft band source, which seems to trace the distribution of the four central galaxies. Note the diffuse emission on the eastern edge of the sub-image corresponds to the core of Structure 47, discussed below.

{\bf Structure 41} - This is classified here as a large group with 12 members detected at z$_{s} = 0.6666 \pm 0.0008$. It has previously been found in both SCP09 and GCD03. Despite this, we find no evidence of any significant diffuse or point source X-ray emission in this group. 

\begin{figure*}
\figurenum{12}
\centering
{\includegraphics[width=2.3in]{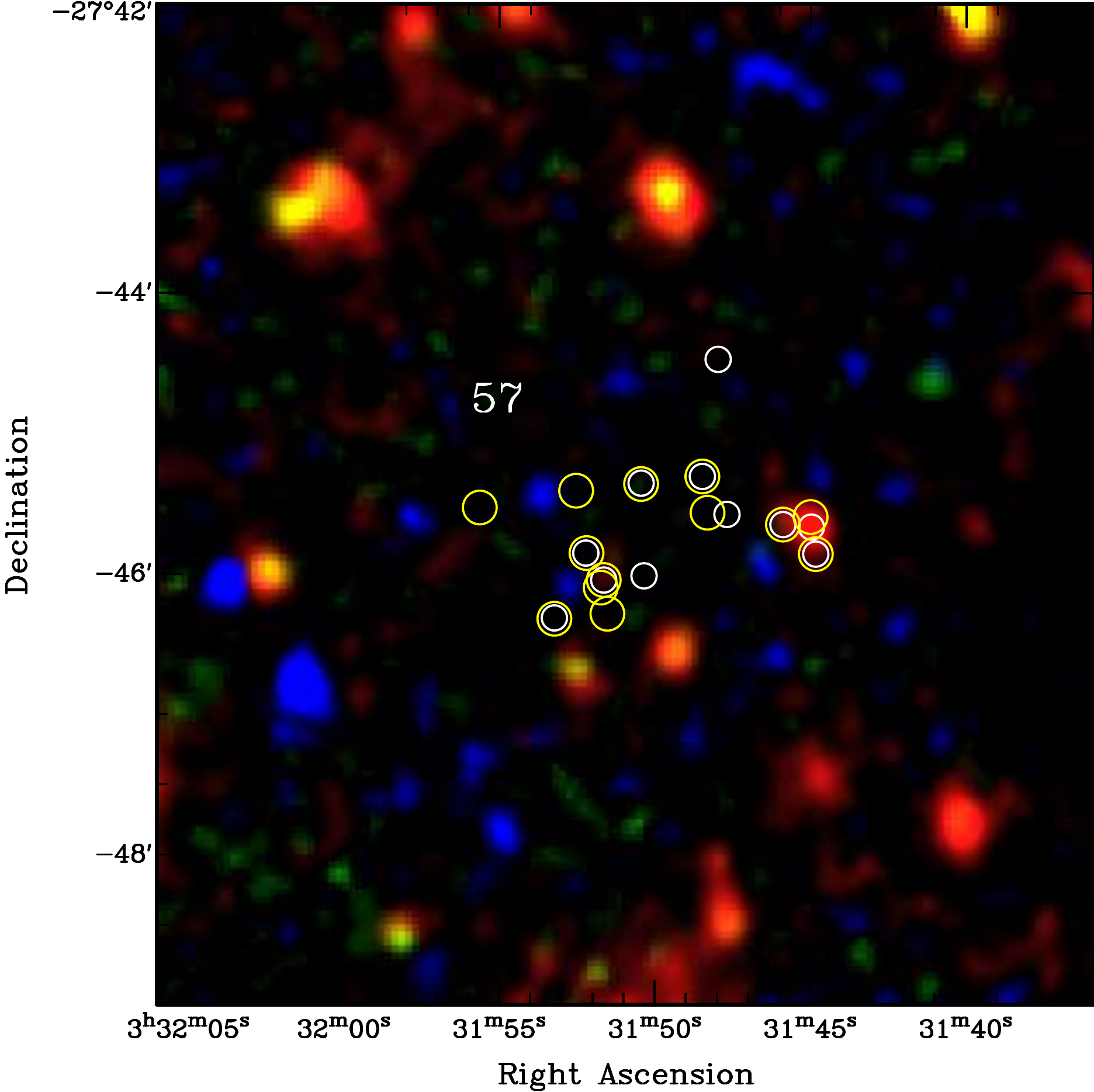}}
{\includegraphics[width=2.174in, trim= 30 0 0 0, clip=true]{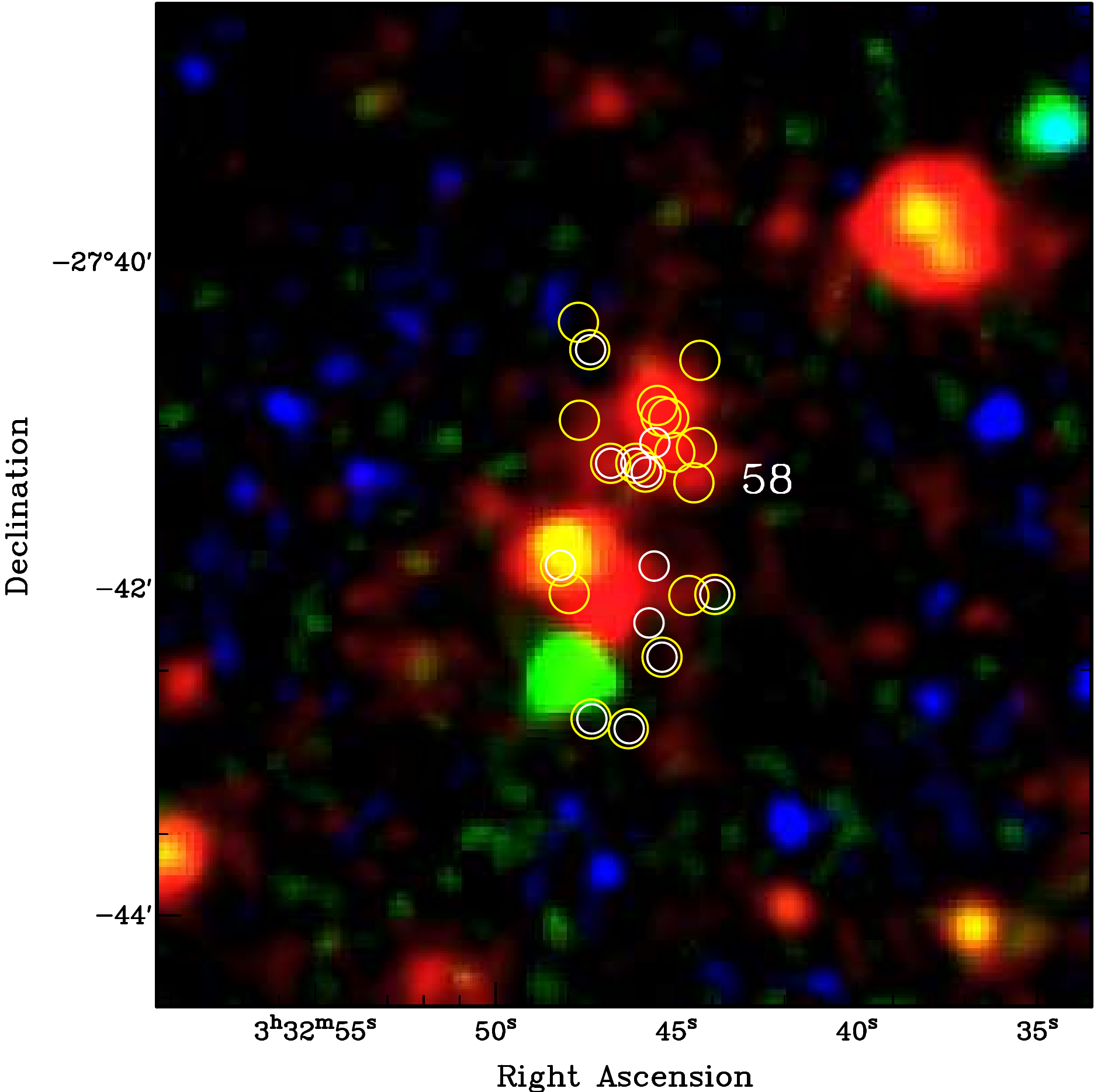}}
{\includegraphics[width=2.174in, trim= 30 0 0 0, clip=true]{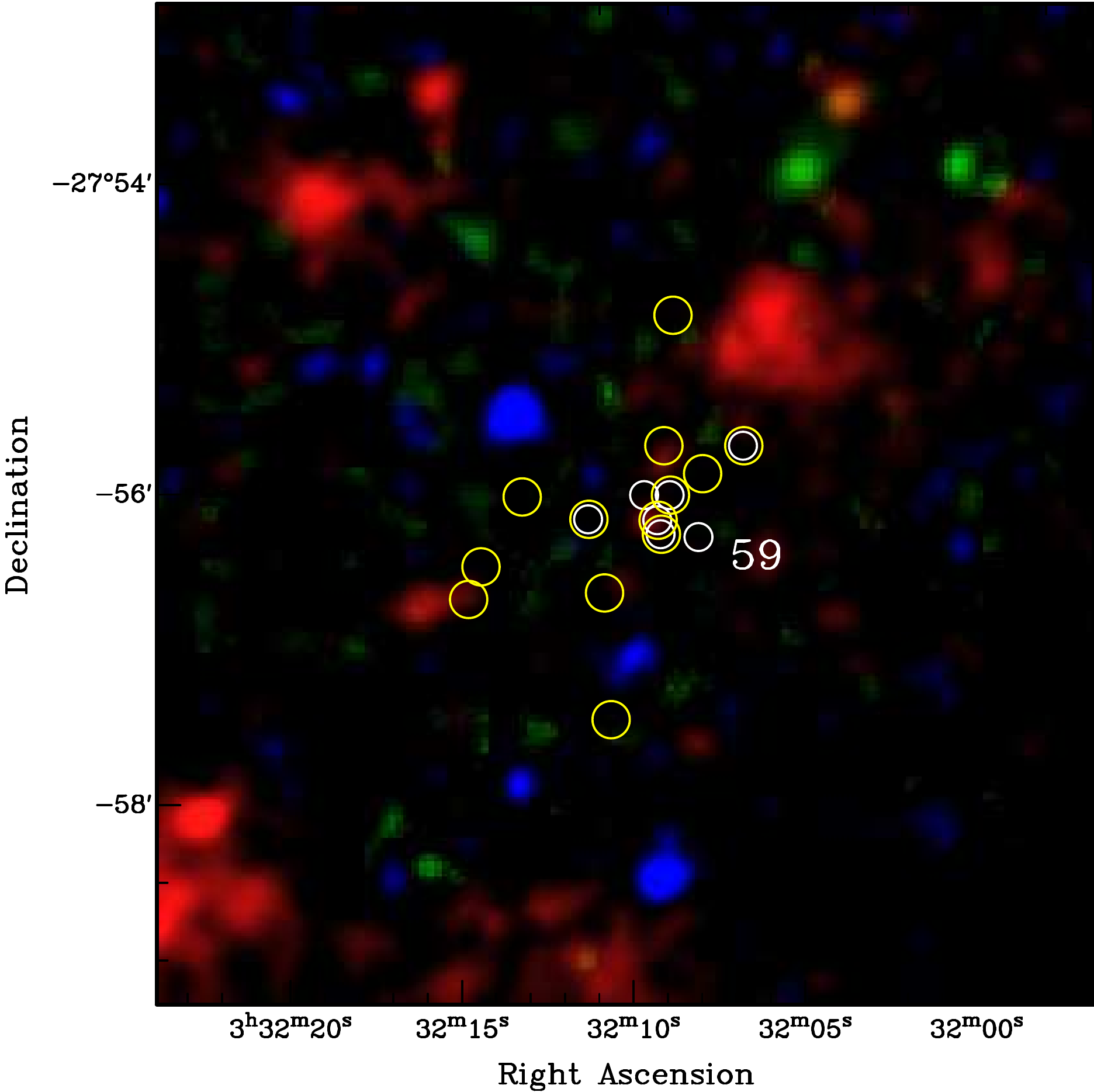}}
\caption{Massive structures, presented in the Table \ref{tab:massive} \em -- Continued}
\end{figure*}

{\bf Structure 42} - This cluster system is seen with 15 unique members at a redshift of z$_{s} = 0.6661 \pm 0.0007$, which seem to follow a curved chain at the core, somewhat reminiscent of a left-facing crescent. The peak galaxy density sits at the top of this crescent and there is clearly a diffuse, soft band X-ray source associated with this region. Furthermore, there is an evidence for diffuse emission along the chain of galaxies suggestive of a very early stage of cluster evolution. 

{\bf Structure 43} - We classify this as a large group at  z$_{s} = 0.6808 \pm 0.0008$ with 11 unique members identified. Again, the peak galaxy density is coincident with an extended soft band source and there is an evidence of further soft band excess to the east of the core. 

{\bf Structure 47} - This is the most massive structure presented here and contains 40 individual member galaxies divide into two north-south aligned sub-structures, which appear to be connected to a filament at the northern extent of the cluster. The filament is seen in Figure \ref{fig:fifth} running east-west and connects to structures 53 (a small group) and 40 a large group at the western end. The core of the cluster is at z$_{s} = 0.6787 \pm 0.0006$ and is located at the northern extend of the structure. An extensive diffuse, soft X-ray source is seen coincident with the core with evidence of an extension to
the west, which corresponds to the location of Structure 40. There is likely to be extensive on-going merging in this region as evidenced in both the velocity and X-ray data.

{\bf Structure 52} - There are 43 individual galaxies associated with this structure in a disturbed arrangement, which has two clear velocity peaks. We classify this as a cluster with the main component at z$_{s} = 0.6846 \pm 0.0007$ and an associated sub-group at z$_{s} = 0.6732 \pm 0.0005$. There is a clearly extended diffuse soft-band source associated with this system. 

{\bf Structure 55} - We find 10 individual objects associated with this group at z$_{s} = 0.6732 \pm 0.0005$, arranged roughly symmetrically. There is a faint extended diffuse soft X-ray source at the centre of these galaxies. Note the larger diffuse source to the eastern edge is coincident with the core of Structure 52. 

{\bf Structure 56} - This cluster has been detected previously in all previous structure examinations of the CDFS (GCD03, AMI05, and SCP09). Here we find 31 member galaxies and confirm the previously reported north-south elongation with the peak galaxy density located slightly to the south. We obtain a spectroscopic redshift of z$_{s} = 0.7341 \pm 0.0009$ and detect extended soft-band emission in the core of the cluster, also elongated in the north-south direction and following the galaxy distribution in the core.  

{\bf Structure 57} - We classify this as a cluster with 17 members at z$_{s} = 0.7334 \pm 0.0009$. The galaxy distribution is unremarkable and beyond one AGN there is no X-ray emission associated with this system.

{\bf Structure 58} - This cluster exhibits a north-south elongation with the 23 unique galaxy members falling into two distinct subgroups both spatially and spectroscopically with an overall redshift of z$_{s} = 0.7311 \pm 0.0007$. There is considerable, extended diffuse emission associated with this system, also seemingly in two subgroups which is likely to denote on-going merging.  

{\bf Structure 59} - We find 15 galaxies associated with this cluster system at z$_{s} = 0.7362 \pm 0.0008$. The galaxy distribution is unremarkable and there is no X-ray emission associated with this object.

\subsection{Incidence of soft-band X-ray emission}

Table \ref{tab:massive} reports the $M_{200}$ mass of the clusters and large groups in addition to if diffuse soft-band X-ray is associated with the cores of these systems. There are 14 systems classified as either a cluster or large group within the XMM field which are unobscured by foreground clusters. We find diffuse soft-band X-ray associated with 80\% of these structures, which have virialized mass of $M_{200}\geq4.9\times10^{13}M_{\odot}$. Of the clusters, 9 out 10 are seen coincident with diffuse soft-band X-ray. This fraction drops to only 2/4 for the large groups. Given that soft-band X-ray is not expected to be detected for large groups, this supports our detection and classification methodology.

\section{Summary}

We applied the DBSCAN method to a spectroscopic and a high quality selected photometric sample of the 0.3 square degree area of the MUSYC-ACES field, in order to detect galaxy concentrations down to $z\sim 1$. We have detected 62 structures, of which 61 and 50 structures were independently verified by spectroscopic and photometric analysis, respectively. These structures were classified as groups, clusters and filaments and their properties were determined. We also discovered a relatively small, $\sim$ 10 Mpc$^{2}$, void at $z\sim0.53$. Additionally, we were able to detect large scale structures, where a secondary $Eps$ parameter was distinguishable in the sorted k-distribution. We detected four large galaxy concentration that correspond to the narrow redshift spikes located at $z\simeq0.13$, $z\simeq0.52$, $0.68$, and $0.73$ (see Figure \ref{fig:overall}), along with a weak and less concentrated web-like structure located at $z\simeq0.62$. Nearly $60\%$ of the detected structures (36/62) are embedded in large scale structures. Based on the structures' velocity distribution and sampling rate, we classified 13 objects as clusters and 6 as massive groups or small clusters. We note that the majority of massive compact structures, 18 out of 19, are located within large scale structures, as expected for current structure formation models \citep{p80}. 

A qualitative examination of the X-ray image of the CDFS was performed, and it revealed that $80\%$ of the clusters and massive groups (9/10 clusters and 2/4 massive groups), that are within the X-ray image coverage, are associated with soft-band (0.4 to 1 keV) diffuse X-ray emission. Additionally, many of the clusters and groups identified here exhibit some indications from on-going merging. This includes, several cases of double systems or clusters which have an in-falling group. In the case of cluster 1, the high quality optical imaging of the Deep2c field of the Garching-Bonn Deep Survey shows member galaxies in the central part of the cluster to be highly disturbed with tidal tails. There can be little doubt that all of these pieces fit together to support hierarchical structure formation. 

Legacy fields such as the CDFS play an important role in furthering our understanding of astrophysical processes. Having presented the most comprehensive analysis of the structure in the CDFS to date, we hope that this will serve as a useful tool for further work on characterising properties of clusters and groups as a function of both wavelength and cosmic redshift. Further X-ray analysis is beyond the scope of this paper, however more structures may be better revealed through post-processing and this should also more fully demonstrate the extent of the X-ray emission \citep{fgh07,fwt10}. Similarly this would provide more accurate mass estimates of these systems are well as details of the temperatures and dynamical histories.  

\begin{acknowledgements}

This work was supported by strategic funding from the Vice-Chancellor \& the Faculty of Science at Victoria University of Wellington.

This research has made use of the NASA/IPAC Extragalactic Database (NED) which is operated by the Jet Propulsion Laboratory, California Institute of Technology, under contract with the National Aeronautics and Space Administration.

\end{acknowledgements}

\end{document}

%% file: TabA.tex
$0.1 \leq z<0.2$ & 128 & 0.005 & 0.05\\
$0.2 \leq z<0.3$ & 125 & 0.004 & 0.04\\
$0.3 \leq z<0.5$ & 138 & 0.005 & 0.05\\
$0.5 \leq z<0.6$ & 139 & 0.004 & 0.05\\
$0.6 \leq z<0.7$ & 180 & 0.005 & 0.06\\
$0.7 \leq z<0.9$ & 182 & 0.006 & 0.07\\

%% file: Tab1.tex
\multirow{2}{*}{1} & $s$ & $03^\text{\textrm{h}}32^\text{\textrm{m}}29.4^\text{\textrm{s}}$ & $-27^\text{\textrm{d}}58^\text{\textrm{m}}07^\text{\textrm{s}}$ & $45$ & $0.1253$ & $0.0003$ & \multirow{2}{*}{$428^{+40}_{-40}$} & $0.30$ & $[0.110 , 0.140]$ & \multirow{2}{*}{5}\\
& $p$ & $03^\text{\textrm{h}}32^\text{\textrm{m}}24.0^\text{\textrm{s}}$ & $-27^\text{\textrm{d}}59^\text{\textrm{m}}54^\text{\textrm{s}}$ & $38$ & $0.119$ & $0.002$ & & $0.21$ & $[0.100 , 0.150]$\\
\multirow{2}{*}{2} & $s$ &$03^\text{\textrm{h}}33^\text{\textrm{m}}20.6^\text{\textrm{s}}$&$-27^\text{\textrm{d}}42^\text{\textrm{m}}08^\text{\textrm{s}}$& $14$ & $0.1471$ & $0.0003$ & \multirow{2}{*}{$267^{+76}_{-76}$} & $0.27$ & $[0.135 , 0.165]$ & \multirow{2}{*}{3}\\
& $p$ &$03^\text{\textrm{h}}33^\text{\textrm{m}}08.9^\text{\textrm{s}}$&$-27^\text{\textrm{d}}43^\text{\textrm{m}}13^\text{\textrm{s}}$& $11$ & $0.149$ & $0.001$ & & $0.20$ & $[0.125 , 0.175]$\\
\multirow{2}{*}{3\tablenotemark{$\star$}} & $s$ &$03^\text{\textrm{h}}33^\text{\textrm{m}}02.9^\text{\textrm{s}}$&$-27^\text{\textrm{d}}58^\text{\textrm{m}}57^\text{\textrm{s}}$& $19$ & $0.1463$,$0.1638$ & $0.0003$,$0.0003$ & \multirow{2}{*}{$213^{+64}_{-64}$,$152^{+79}_{-29}$} & $0.27$ & $[0.135 , 0.165]$ & \multirow{2}{*}{3,3}\\
& $p$ &$03^\text{\textrm{h}}33^\text{\textrm{m}}03.3^\text{\textrm{s}}$&$-27^\text{\textrm{d}}57^\text{\textrm{m}}48^\text{\textrm{s}}$& $9$ & $0.148$ & $0.001$ & & $0.20$ & $[0.125 , 0.175]$\\
\multirow{2}{*}{4} & $s$ &$03^\text{\textrm{h}}32^\text{\textrm{m}}48.5^\text{\textrm{s}}$&$-27^\text{\textrm{d}}39^\text{\textrm{m}}30^\text{\textrm{s}}$& $15$ & $0.1487$ & $0.0010$ & \multirow{2}{*}{$933^{+213}_{-213}$} & $0.27$ & $[0.135 , 0.165]$ & \multirow{2}{*}{1}\\
& $p$ &$03^\text{\textrm{h}}32^\text{\textrm{m}}44.5^\text{\textrm{s}}$&$-27^\text{\textrm{d}}39^\text{\textrm{m}}14^\text{\textrm{s}}$& $18$ & $0.148$ & $0.002$ & & $0.20$ & $[0.125 , 0.175]$\\
\multirow{2}{*}{5} & $s$ &$03^\text{\textrm{h}}33^\text{\textrm{m}}18.8^\text{\textrm{s}}$&$-27^\text{\textrm{d}}49^\text{\textrm{m}}17^\text{\textrm{s}}$& $36$ & $0.1267$ & $0.0004$ & \multirow{2}{*}{$575^{+70}_{-70}$} & $0.30$ & $[0.110 , 0.140]$ & \multirow{2}{*}{5}\\
& $p$ &$03^\text{\textrm{h}}33^\text{\textrm{m}}15.2^\text{\textrm{s}}$&$-27^\text{\textrm{d}}49^\text{\textrm{m}}15^\text{\textrm{s}}$& $63$ & $0.130$ & $0.001$ & & $0.21$ & $[0.100 , 0.150]$\\
\multirow{2}{*}{6\tablenotemark{$\ddagger$}} &$s$& \nodata & \nodata & \nodata & \nodata & \nodata & \nodata & \nodata & \nodata& \nodata\\
& $p$ &$03^\text{\textrm{h}}32^\text{\textrm{m}}30.7^\text{\textrm{s}}$&$-27^\text{\textrm{d}}34^\text{\textrm{m}}20^\text{\textrm{s}}$& $23$ & $0.141$ & $0.002$ & & $0.20$ & $[0.115 , 0.165]$\\
\multirow{2}{*}{7} & $s$ &$03^\text{\textrm{h}}31^\text{\textrm{m}}38.6^\text{\textrm{s}}$&$-27^\text{\textrm{d}}44^\text{\textrm{m}}59^\text{\textrm{s}}$& $10$ & $0.1796$ & $0.0003$ & \multirow{2}{*}{$230^{+52}_{-52}$} & $0.45$ & $[0.165 , 0.195]$ & \multirow{2}{*}{3}\\
& $p$ &$03^\text{\textrm{h}}31^\text{\textrm{m}}41.5^\text{\textrm{s}}$&$-27^\text{\textrm{d}}45^\text{\textrm{m}}50^\text{\textrm{s}}$& $26$ & $0.183$ & $0.002$ & & $0.40$ & $[0.155 , 0.205]$\\
\multirow{2}{*}{8} & $s$ &$03^\text{\textrm{h}}32^\text{\textrm{m}}53.2^\text{\textrm{s}}$&$-28^\text{\textrm{d}}02^\text{\textrm{m}}26^\text{\textrm{s}}$& $14$ & $0.2136$ & $0.0002$ & \multirow{2}{*}{$183^{+47}_{-47}$}\tablenotemark{$\dagger$} & $0.43$ & $[0.200 , 0.230]$ & \multirow{2}{*}{3}\\
& $p$ &$03^\text{\textrm{h}}32^\text{\textrm{m}}51.7^\text{\textrm{s}}$&$-28^\text{\textrm{d}}02^\text{\textrm{m}}17^\text{\textrm{s}}$& $13$ & $0.217$ & $0.002$ & & $0.42$ & $[0.195 , 0.235]$\\
\multirow{2}{*}{9} & $s$ &$03^\text{\textrm{h}}32^\text{\textrm{m}}46.8^\text{\textrm{s}}$&$-27^\text{\textrm{d}}46^\text{\textrm{m}}12^\text{\textrm{s}}$& $12$ & $0.2167$ & $0.0013$ & \multirow{2}{*}{$1010^{+291}_{-291}$} & $0.43$ & $[0.200 , 0.230]$ & \multirow{2}{*}{1}\\
& $p$ &$03^\text{\textrm{h}}32^\text{\textrm{m}}46.8^\text{\textrm{s}}$&$-27^\text{\textrm{d}}46^\text{\textrm{m}}39^\text{\textrm{s}}$& $12$ & $0.224$ & $0.002$ & & $0.42$ & $[0.195 , 0.235]$\\
\multirow{2}{*}{10} & $s$ &$03^\text{\textrm{h}}32^\text{\textrm{m}}04.8^\text{\textrm{s}}$&$-27^\text{\textrm{d}}59^\text{\textrm{m}}41^\text{\textrm{s}}$& $18$ & $0.2142$ & $0.0001$ & \multirow{2}{*}{$100^{+68}_{-68}$} & $0.43$ & $[0.200 , 0.230]$ & \multirow{2}{*}{2}\\
& $p$ &$03^\text{\textrm{h}}32^\text{\textrm{m}}02.3^\text{\textrm{s}}$&$-27^\text{\textrm{d}}59^\text{\textrm{m}}55^\text{\textrm{s}}$& $22$ & $0.215$ & $0.002$ & & $0.42$ & $[0.195 , 0.235]$\\
\multirow{2}{*}{11} & $s$ &$03^\text{\textrm{h}}33^\text{\textrm{m}}12.1^\text{\textrm{s}}$&$-27^\text{\textrm{d}}44^\text{\textrm{m}}51^\text{\textrm{s}}$& $12$ & $0.2176$ & $0.0004$ & \multirow{2}{*}{$341^{+192}_{-192}$} & $0.43$ & $[0.200 , 0.230]$ & \multirow{2}{*}{3}\\
& $p$ &$03^\text{\textrm{h}}33^\text{\textrm{m}}12.3^\text{\textrm{s}}$&$-27^\text{\textrm{d}}45^\text{\textrm{m}}10^\text{\textrm{s}}$& $18$ & $0.224$ & $0.002$ & & $0.42$ & $[0.195 , 0.235]$\\
\multirow{2}{*}{12\tablenotemark{$\star$}} & $s$ &$03^\text{\textrm{h}}31^\text{\textrm{m}}47.7^\text{\textrm{s}}$&$-28^\text{\textrm{d}}01^\text{\textrm{m}}03^\text{\textrm{s}}$& $12$ & $0.2465$,$0.2600$ & $0.0003$,$0.0002$ & \multirow{2}{*}{$221^{+88}_{-38}$,$87^{+69}_{-18}$} & $0.54$ & $[0.235 , 0.265]$ & \multirow{2}{*}{3,3}\\
& $p$ &$03^\text{\textrm{h}}31^\text{\textrm{m}}47.2^\text{\textrm{s}}$&$-28^\text{\textrm{d}}00^\text{\textrm{m}}43^\text{\textrm{s}}$& $12$ & $0.246$ & $0.003$ & & $0.47$ & $[0.230 , 0.270]$\\
\multirow{2}{*}{13} & $s$ &$03^\text{\textrm{h}}32^\text{\textrm{m}}41.8^\text{\textrm{s}}$&$-27^\text{\textrm{d}}37^\text{\textrm{m}}03^\text{\textrm{s}}$& $11$ & $0.2498$ & $0.0002$ & \multirow{2}{*}{$140^{+55}_{-25}$} & $0.54$ & $[0.235 , 0.265]$ & \multirow{2}{*}{3}\\
& $p$ &$03^\text{\textrm{h}}32^\text{\textrm{m}}41.1^\text{\textrm{s}}$&$-27^\text{\textrm{d}}37^\text{\textrm{m}}16^\text{\textrm{s}}$& $10$ & $0.247$ & $0.002$ & & $0.47$ & $[0.230 , 0.270]$\\
\multirow{2}{*}{14} & $s$ &$03^\text{\textrm{h}}32^\text{\textrm{m}}17.0^\text{\textrm{s}}$&$-27^\text{\textrm{d}}35^\text{\textrm{m}}50^\text{\textrm{s}}$& $13$ & $0.2487$ & $0.0002$ & \multirow{2}{*}{$108^{+69}_{-69}$} & $0.54$ & $[0.235 , 0.265]$ & \multirow{2}{*}{3}\\
& $p$ &$03^\text{\textrm{h}}32^\text{\textrm{m}}17.1^\text{\textrm{s}}$&$-27^\text{\textrm{d}}35^\text{\textrm{m}}10^\text{\textrm{s}}$& $19$ & $0.246$ & $0.002$ & & $0.47$ & $[0.230 , 0.270]$\\
\multirow{2}{*}{15} & $s$ &$03^\text{\textrm{h}}32^\text{\textrm{m}}19.0^\text{\textrm{s}}$&$-28^\text{\textrm{d}}02^\text{\textrm{m}}29^\text{\textrm{s}}$& $8$ & $0.2778$ & $0.0009$ & \multirow{2}{*}{$498^{+260}_{-96}$} & $0.52$ & $[0.265 , 0.295]$ & \multirow{2}{*}{1}\\
& $p$ & \nodata & \nodata & \nodata & \nodata & \nodata & & \nodata & \nodata\\
\multirow{2}{*}{16} & $s$ &$03^\text{\textrm{h}}32^\text{\textrm{m}}52.7^\text{\textrm{s}}$&$-27^\text{\textrm{d}}45^\text{\textrm{m}}01^\text{\textrm{s}}$& $7$ & $0.2771$ & $0.0003$ & \multirow{2}{*}{$200^{+104}_{-42}$} & $0.52$ & $[0.265 , 0.295]$ & \multirow{2}{*}{3}\\
& $p$ &$03^\text{\textrm{h}}32^\text{\textrm{m}}52.9^\text{\textrm{s}}$&$-27^\text{\textrm{d}}44^\text{\textrm{m}}57^\text{\textrm{s}}$& $11$ & $0.275$ & $0.003$ & & $0.50$ & $[0.260 , 0.300]$\\
\multirow{2}{*}{17} & $s$ &$03^\text{\textrm{h}}32^\text{\textrm{m}}57.1^\text{\textrm{s}}$&$-27^\text{\textrm{d}}37^\text{\textrm{m}}03^\text{\textrm{s}}$& $10$ & $0.2772$ & $0.0003$ & \multirow{2}{*}{$170^{+79}_{-32}$} & $0.52$ & $[0.265 , 0.295]$ & \multirow{2}{*}{3}\\
& $p$ & \nodata & \nodata & \nodata & \nodata & \nodata & & \nodata & \nodata\\
\multirow{2}{*}{18} & $s$ &$03^\text{\textrm{h}}33^\text{\textrm{m}}16.1^\text{\textrm{s}}$&$-28^\text{\textrm{d}}01^\text{\textrm{m}}28^\text{\textrm{s}}$& $10$ & $0.3334$ & $0.0002$ & \multirow{2}{*}{$108^{+43}_{-19}$} & $0.54$ & $[0.320 , 0.350]$ & \multirow{2}{*}{3}\\
& $p$ &$03^\text{\textrm{h}}33^\text{\textrm{m}}15.8^\text{\textrm{s}}$&$-28^\text{\textrm{d}}01^\text{\textrm{m}}29^\text{\textrm{s}}$& $14$ & $0.348$ & $0.001$ & & $0.44$ & $[0.310 , 0.360]$\\
\multirow{2}{*}{19} & $s$ &$03^\text{\textrm{h}}32^\text{\textrm{m}}16.5^\text{\textrm{s}}$&$-27^\text{\textrm{d}}49^\text{\textrm{m}}37^\text{\textrm{s}}$& $11$ & $0.3372$ & $0.0005$ & \multirow{2}{*}{$332^{+161}_{-161}$} & $0.54$ & $[0.320 , 0.350]$ & \multirow{2}{*}{3}\\
& $p$ &$03^\text{\textrm{h}}32^\text{\textrm{m}}15.0^\text{\textrm{s}}$&$-27^\text{\textrm{d}}49^\text{\textrm{m}}11^\text{\textrm{s}}$& $10$ & $0.345$ & $0.002$ & & $0.44$ & $[0.310 , 0.360]$\\
\multirow{2}{*}{20} & $s$ &$03^\text{\textrm{h}}33^\text{\textrm{m}}06.4^\text{\textrm{s}}$&$-28^\text{\textrm{d}}02^\text{\textrm{m}}02^\text{\textrm{s}}$& $11$ & $0.4074$ & $0.0007$ & \multirow{2}{*}{$342^{+156}_{-63}$} & $0.69$ & $[0.390 , 0.420]$ & \multirow{2}{*}{3}\\
& $p$ &$03^\text{\textrm{h}}33^\text{\textrm{m}}05.4^\text{\textrm{s}}$&$-28^\text{\textrm{d}}02^\text{\textrm{m}}10^\text{\textrm{s}}$& $12$ & $0.407$ & $0.003$ & & $0.50$ & $[0.380 , 0.430]$\\
\multirow{2}{*}{21\tablenotemark{$\star$}} & $s$ &$03^\text{\textrm{h}}32^\text{\textrm{m}}43.3^\text{\textrm{s}}$&$-27^\text{\textrm{d}}51^\text{\textrm{m}}53^\text{\textrm{s}}$& $10$ & $0.4154$,$0.4233$ & $0.0001$,$0.0002$ & \multirow{2}{*}{$204^{+194}_{-52}$,$102^{+52}_{-19}$} & $0.63$ & $[0.405 , 0.435]$ & \multirow{2}{*}{3,3}\\
& $p$ &$03^\text{\textrm{h}}32^\text{\textrm{m}}42.4^\text{\textrm{s}}$&$-27^\text{\textrm{d}}52^\text{\textrm{m}}18^\text{\textrm{s}}$& $11$ & $0.423$ & $0.002$ & & $0.55$ & $[0.395 , 0.445]$\\
\multirow{2}{*}{22} & $s$ &$03^\text{\textrm{h}}32^\text{\textrm{m}}14.9^\text{\textrm{s}}$&$-27^\text{\textrm{d}}41^\text{\textrm{m}}37^\text{\textrm{s}}$& $13$ & $0.4214$ & $0.0009$ & \multirow{2}{*}{$654^{+79}_{-79}$} & $0.63$ & $[0.405 , 0.435]$ & \multirow{2}{*}{1}\\
& $p$ &$03^\text{\textrm{h}}32^\text{\textrm{m}}11.3^\text{\textrm{s}}$&$-27^\text{\textrm{d}}41^\text{\textrm{m}}31^\text{\textrm{s}}$& $18$ & $0.422$ & $0.002$ & & $0.55$ & $[0.395 , 0.445]$\\
\multirow{2}{*}{23} & $s$ &$03^\text{\textrm{h}}33^\text{\textrm{m}}04.7^\text{\textrm{s}}$&$-27^\text{\textrm{d}}39^\text{\textrm{m}}24^\text{\textrm{s}}$& $10$ & $0.4224$ & $0.0004$ & \multirow{2}{*}{$406^{+191}_{-78}$} & $0.63$ & $[0.405 , 0.435]$ & \multirow{2}{*}{1}\\
& $p$ &$03^\text{\textrm{h}}33^\text{\textrm{m}}04.4^\text{\textrm{s}}$&$-27^\text{\textrm{d}}39^\text{\textrm{m}}32^\text{\textrm{s}}$& $16$ & $0.419$ & $0.002$ & & $0.55$ & $[0.395 , 0.445]$\\
\multirow{2}{*}{24\tablenotemark{$\star$}} & $s$ &$03^\text{\textrm{h}}33^\text{\textrm{m}}08.3^\text{\textrm{s}}$&$-27^\text{\textrm{d}}46^\text{\textrm{m}}26^\text{\textrm{s}}$& $13$ & $0.5218$,$0.5337$ & $0.0007$,$0.0005$ & \multirow{2}{*}{$370^{+134}_{-62}$,$172^{+149}_{-40}$} & $0.40$ & $[0.505 , 0.535]$ & \multirow{2}{*}{3,3}\\
& $p$ &$03^\text{\textrm{h}}33^\text{\textrm{m}}11.0^\text{\textrm{s}}$&$-27^\text{\textrm{d}}46^\text{\textrm{m}}23^\text{\textrm{s}}$& $19$ & $0.528$ & $0.002$ & & $0.39$ & $[0.495 , 0.545]$\\
\multirow{2}{*}{25} & $s$ &$03^\text{\textrm{h}}33^\text{\textrm{m}}04.7^\text{\textrm{s}}$&$-27^\text{\textrm{d}}52^\text{\textrm{m}}07^\text{\textrm{s}}$& $11$ & $0.5200$ & $0.0007$ & \multirow{2}{*}{$432^{+185}_{-185}$} & $0.40$ & $[0.505 , 0.535]$ & \multirow{2}{*}{4}\\
& $p$ &$03^\text{\textrm{h}}33^\text{\textrm{m}}05.4^\text{\textrm{s}}$&$-27^\text{\textrm{d}}52^\text{\textrm{m}}20^\text{\textrm{s}}$& $23$ & $0.526$ & $0.002$ & & $0.39$ & $[0.495 , 0.545]$\\
\multirow{2}{*}{26} & $s$ &$03^\text{\textrm{h}}32^\text{\textrm{m}}39.6^\text{\textrm{s}}$&$-27^\text{\textrm{d}}37^\text{\textrm{m}}12^\text{\textrm{s}}$& $7$ & $0.5233$ & $0.0002$ & \multirow{2}{*}{$122^{+55}_{-22}$} & $0.42$ & $[0.505 , 0.535]$ & \multirow{2}{*}{3}\\
& $p$ &$03^\text{\textrm{h}}32^\text{\textrm{m}}39.6^\text{\textrm{s}}$&$-27^\text{\textrm{d}}37^\text{\textrm{m}}12^\text{\textrm{s}}$& $7$ & $0.520$ & $0.003$ & & $0.42$ & $[0.495 , 0.545]$\\
\multirow{2}{*}{27\tablenotemark{$\star$}} & $s$ &$03^\text{\textrm{h}}32^\text{\textrm{m}}50.4^\text{\textrm{s}}$&$-27^\text{\textrm{d}}44^\text{\textrm{m}}55^\text{\textrm{s}}$& $12$ & $0.5334,0.5215$ & $0.0006,0.0010$ & \multirow{2}{*}{$359^{+129}_{-60}$,$277^{+327}_{-61}$} & $0.42$ & $[0.520 , 0.550]$ & \multirow{2}{*}{3,3}\\
& $p$ &$03^\text{\textrm{h}}32^\text{\textrm{m}}48.0^\text{\textrm{s}}$&$-27^\text{\textrm{d}}44^\text{\textrm{m}}38^\text{\textrm{s}}$& $16$ & $0.530$ & $0.003$ & & $0.37$ & $[0.510 , 0.560]$\\
\multirow{2}{*}{28\tablenotemark{$\star$}} & $s$ &$03^\text{\textrm{h}}33^\text{\textrm{m}}20.4^\text{\textrm{s}}$&$-27^\text{\textrm{d}}43^\text{\textrm{m}}33^\text{\textrm{s}}$& $17$ & $0.5205,0.5332$ & $0.0006,0.0008$ & \multirow{2}{*}{$411^{+82}_{-82}$,$175^{+219}_{-41}$} & $0.40$ & $[0.505 , 0.535]$ & \multirow{2}{*}{5,3}\\
& $p$ &$03^\text{\textrm{h}}33^\text{\textrm{m}}20.5^\text{\textrm{s}}$&$-27^\text{\textrm{d}}43^\text{\textrm{m}}40^\text{\textrm{s}}$& $32$ & $0.523$ & $0.002$ & & $0.39$ & $[0.495 , 0.545]$\\
\multirow{2}{*}{29} & $s$ &$03^\text{\textrm{h}}32^\text{\textrm{m}}28.1^\text{\textrm{s}}$&$-27^\text{\textrm{d}}36^\text{\textrm{m}}03^\text{\textrm{s}}$& $9$ & $0.5220$ & $0.0007$ & \multirow{2}{*}{$370^{+154}_{-67}$} & $0.40$ & $[0.505 , 0.535]$ & \multirow{2}{*}{3}\\
& $p$ &$03^\text{\textrm{h}}32^\text{\textrm{m}}28.7^\text{\textrm{s}}$&$-27^\text{\textrm{d}}36^\text{\textrm{m}}04^\text{\textrm{s}}$& $10$ & $0.530$ & $0.003$ & & $0.39$ & $[0.495 , 0.545]$\\
\multirow{2}{*}{30} & $s$ &$03^\text{\textrm{h}}33^\text{\textrm{m}}29.8^\text{\textrm{s}}$&$-27^\text{\textrm{d}}58^\text{\textrm{m}}45^\text{\textrm{s}}$& $9$ & $0.5436$ & $0.0007$ & \multirow{2}{*}{$410^{+169}_{-74}$} & $0.42$ & $[0.520 , 0.550]$ & \multirow{2}{*}{4}\\
& $p$ & \nodata & \nodata & \nodata & \nodata & \nodata & & \nodata & \nodata\\
\multirow{2}{*}{31} & $s$ &$03^\text{\textrm{h}}31^\text{\textrm{m}}22.9^\text{\textrm{s}}$&$-27^\text{\textrm{d}}58^\text{\textrm{m}}11^\text{\textrm{s}}$& $17$ & $0.5264$ & $0.0006$ & \multirow{2}{*}{$454^{+105}_{-105}$} & $0.40$ & $[0.505 , 0.535]$ & \multirow{2}{*}{5}\\
& $p$ & \nodata & \nodata & \nodata & \nodata & \nodata & & \nodata & \nodata\\
\multirow{2}{*}{32} & $s$ &$03^\text{\textrm{h}}31^\text{\textrm{m}}45.2^\text{\textrm{s}}$&$-27^\text{\textrm{d}}50^\text{\textrm{m}}33^\text{\textrm{s}}$& $10$ & $0.5615$ & $0.0007$ & \multirow{2}{*}{$379^{+147}_{-68}$} & $0.45$ & $[0.545 , 0.575]$ & \multirow{2}{*}{3}\\
& $p$ &$03^\text{\textrm{h}}31^\text{\textrm{m}}46.2^\text{\textrm{s}}$&$-27^\text{\textrm{d}}50^\text{\textrm{m}}32^\text{\textrm{s}}$& $10$ & $0.553$ & $0.004$ & & $0.47$ & $[0.535 , 0.585]$\\

%% file: Tab2.tex
\multirow{2}{*}{33} & $s$ &$03^\text{\textrm{h}}32^\text{\textrm{m}}02.6^\text{\textrm{s}}$&$-27^\text{\textrm{d}}56^\text{\textrm{m}}18^\text{\textrm{s}}$& $9$ & $0.6195$ & $0.0003$ & \multirow{2}{*}{$171^{+63}_{-29}$} & $0.58$ & $[0.605 , 0.635]$ & \multirow{2}{*}{3}\\
& $p$ &$03^\text{\textrm{h}}32^\text{\textrm{m}}05.7^\text{\textrm{s}}$&$-27^\text{\textrm{d}}56^\text{\textrm{m}}58^\text{\textrm{s}}$& $12$ & $0.629$ & $0.003$ & & $0.57$ & $[0.590 , 0.650]$\\
\multirow{2}{*}{34\tablenotemark{$\star$}} & $s$ &$03^\text{\textrm{h}}31^\text{\textrm{m}}50.7^\text{\textrm{s}}$&$-27^\text{\textrm{d}}43^\text{\textrm{m}}45^\text{\textrm{s}}$& $8$ & $0.6201$,$0.6255$ & $0.0006$,$0.0006$ & \multirow{2}{*}{$209^{+128}_{-42}$,$143^{+169}_{-32}$} & $0.58$ & $[0.605 , 0.635]$ & \multirow{2}{*}{3,3}\\
& $p$ &$03^\text{\textrm{h}}31^\text{\textrm{m}}52.4^\text{\textrm{s}}$&$-27^\text{\textrm{d}}43^\text{\textrm{m}}49^\text{\textrm{s}}$& $10$ & $0.620$ & $0.004$ & & $0.57$ & $[0.590 , 0.650]$\\
\multirow{2}{*}{35} & $s$ &$03^\text{\textrm{h}}32^\text{\textrm{m}}26.2^\text{\textrm{s}}$&$-27^\text{\textrm{d}}57^\text{\textrm{m}}31^\text{\textrm{s}}$& $8$ & $0.6195$ & $0.0008$ & \multirow{2}{*}{$383^{+161}_{-71}$} & $0.58$ & $[0.605 , 0.635]$ & \multirow{2}{*}{2}\\
& $p$ & \nodata & \nodata & \nodata & \nodata & \nodata & & \nodata & \nodata\\
\multirow{2}{*}{36} & $s$ &$03^\text{\textrm{h}}33^\text{\textrm{m}}15.8^\text{\textrm{s}}$&$-27^\text{\textrm{d}}46^\text{\textrm{m}}46^\text{\textrm{s}}$& $11$ & $0.6253$ & $0.0003$ & \multirow{2}{*}{$164^{+49}_{-49}$} & $0.62$ & $[0.610 , 0.640]$ & \multirow{2}{*}{3}\\
& $p$ &$03^\text{\textrm{h}}33^\text{\textrm{m}}15.6^\text{\textrm{s}}$&$-27^\text{\textrm{d}}46^\text{\textrm{m}}33^\text{\textrm{s}}$& $13$ & $0.626$ & $0.004$ & & $0.62$ & $[0.595 , 0.655]$\\
\multirow{2}{*}{37} & $s$ &$03^\text{\textrm{h}}32^\text{\textrm{m}}39.1^\text{\textrm{s}}$&$-27^\text{\textrm{d}}47^\text{\textrm{m}}06^\text{\textrm{s}}$& $7$ & $0.6204$ & $0.0006$ & \multirow{2}{*}{$247^{+129}_{-48}$} & $0.58$ & $[0.605 , 0.635]$ & \multirow{2}{*}{3}\\
& $p$ & \nodata & \nodata & \nodata & \nodata & \nodata & & \nodata & \nodata\\
\multirow{2}{*}{38} & $s$ &$03^\text{\textrm{h}}32^\text{\textrm{m}}43.1^\text{\textrm{s}}$&$-27^\text{\textrm{d}}56^\text{\textrm{m}}55^\text{\textrm{s}}$& $13$ & $0.6197$ & $0.0008$ & \multirow{2}{*}{$471^{+75}_{-75}$} & $0.58$ & $[0.605 , 0.635]$ & \multirow{2}{*}{2}\\
& $p$ & \nodata & \nodata & \nodata & \nodata & \nodata & & \nodata & \nodata\\
\multirow{2}{*}{39} & $s$ &$03^\text{\textrm{h}}32^\text{\textrm{m}}28.4^\text{\textrm{s}}$&$-27^\text{\textrm{d}}55^\text{\textrm{m}}40^\text{\textrm{s}}$& $9$ & $0.6626$ & $0.0029$ & \multirow{2}{*}{$1674^{+654}_{-303}$} & $0.39$ & $[0.655 , 0.685]$ & \multirow{2}{*}{1}\\
& $p$ &$03^\text{\textrm{h}}32^\text{\textrm{m}}29.9^\text{\textrm{s}}$&$-27^\text{\textrm{d}}55^\text{\textrm{m}}31^\text{\textrm{s}}$& $12$ & $0.674$ & $0.002$ & & $0.34$ & $[0.640 , 0.700]$\\
\multirow{2}{*}{40} & $s$ &$03^\text{\textrm{h}}31^\text{\textrm{m}}33.1^\text{\textrm{s}}$&$-27^\text{\textrm{d}}48^\text{\textrm{m}}42^\text{\textrm{s}}$& $13$ & $0.6781$ & $0.0008$ & \multirow{2}{*}{$459^{+67}_{-67}$} & $0.39$ & $[0.665 , 0.695]$ & \multirow{2}{*}{5}\\
& $p$ &$03^\text{\textrm{h}}31^\text{\textrm{m}}34.5^\text{\textrm{s}}$&$-27^\text{\textrm{d}}48^\text{\textrm{m}}36^\text{\textrm{s}}$& $12$ & $0.688$ & $0.004$ & & $0.36$ & $[0.650 , 0.710]$\\
\multirow{2}{*}{41} & $s$ &$03^\text{\textrm{h}}32^\text{\textrm{m}}15.6^\text{\textrm{s}}$&$-27^\text{\textrm{d}}49^\text{\textrm{m}}42^\text{\textrm{s}}$& $10$ & $0.6666$ & $0.0008$ & \multirow{2}{*}{$431^{+73}_{-73}$} & $0.39$ & $[0.655 , 0.685]$ & \multirow{2}{*}{4}\\
& $p$ &$03^\text{\textrm{h}}32^\text{\textrm{m}}15.3^\text{\textrm{s}}$&$-27^\text{\textrm{d}}49^\text{\textrm{m}}39^\text{\textrm{s}}$& $10$ & $0.676$ & $0.003$ & & $0.34$ & $[0.640 , 0.700]$\\
\multirow{2}{*}{42} & $s$ &$03^\text{\textrm{h}}31^\text{\textrm{m}}54.8^\text{\textrm{s}}$&$-27^\text{\textrm{d}}42^\text{\textrm{m}}01^\text{\textrm{s}}$& $13$ & $0.6661$ & $0.0007$ & \multirow{2}{*}{$398^{+154}_{-154}$} & $0.39$ & $[0.655 , 0.685]$ & \multirow{2}{*}{5}\\
& $p$ &$03^\text{\textrm{h}}31^\text{\textrm{m}}56.7^\text{\textrm{s}}$&$-27^\text{\textrm{d}}42^\text{\textrm{m}}12^\text{\textrm{s}}$& $9$ & $0.666$ & $0.002$ & & $0.34$ & $[0.640 , 0.700]$\\
\multirow{2}{*}{43} & $s$ &$03^\text{\textrm{h}}32^\text{\textrm{m}}10.7^\text{\textrm{s}}$&$-27^\text{\textrm{d}}59^\text{\textrm{m}}45^\text{\textrm{s}}$& $9$ & $0.6808$ & $0.0008$ & \multirow{2}{*}{$399^{+160}_{-70}$} & $0.39$ & $[0.665 , 0.695]$ & \multirow{2}{*}{4}\\
& $p$ &$03^\text{\textrm{h}}32^\text{\textrm{m}}10.4^\text{\textrm{s}}$&$-27^\text{\textrm{d}}59^\text{\textrm{m}}49^\text{\textrm{s}}$& $10$ & $0.685$ & $0.005$ & & $0.36$ & $[0.650 , 0.710]$\\
\multirow{2}{*}{44} & $s$ &$03^\text{\textrm{h}}32^\text{\textrm{m}}34.2^\text{\textrm{s}}$&$-28^\text{\textrm{d}}00^\text{\textrm{m}}41^\text{\textrm{s}}$& $7$ & $0.6700$ & $0.0007$ & \multirow{2}{*}{$324^{+145}_{-59}$} & $0.42$ & $[0.655 , 0.685]$ & \multirow{2}{*}{3}\\
& $p$ &$03^\text{\textrm{h}}32^\text{\textrm{m}}34.4^\text{\textrm{s}}$&$-28^\text{\textrm{d}}00^\text{\textrm{m}}39^\text{\textrm{s}}$& $9$ & $0.681$ & $0.002$ & & $0.36$ & $[0.640 , 0.700]$\\
\multirow{2}{*}{45} & $s$ &$03^\text{\textrm{h}}31^\text{\textrm{m}}36.0^\text{\textrm{s}}$&$-27^\text{\textrm{d}}57^\text{\textrm{m}}38^\text{\textrm{s}}$& $11$ & $0.6773$ & $0.0009$ & \multirow{2}{*}{$495^{+165}_{-165}$} & $0.39$ & $[0.665 , 0.695]$ & \multirow{2}{*}{5}\\
& $p$ &$03^\text{\textrm{h}}31^\text{\textrm{m}}35.7^\text{\textrm{s}}$&$-27^\text{\textrm{d}}57^\text{\textrm{m}}31^\text{\textrm{s}}$& $20$ & $0.678$ & $0.002$ & & $0.36$ & $[0.650 , 0.710]$\\
\multirow{2}{*}{46} & $s$ &$03^\text{\textrm{h}}31^\text{\textrm{m}}29.9^\text{\textrm{s}}$&$-27^\text{\textrm{d}}51^\text{\textrm{m}}12^\text{\textrm{s}}$& $8$ & $0.6775$ & $0.0005$ & \multirow{2}{*}{$252^{+99}_{-43}$} & $0.39$ & $[0.665 , 0.695]$ & \multirow{2}{*}{3}\\
& $p$ & \nodata & \nodata & \nodata & \nodata & \nodata & & \nodata & \nodata\\
\multirow{2}{*}{47} & $s$ &$03^\text{\textrm{h}}31^\text{\textrm{m}}51.0^\text{\textrm{s}}$&$-27^\text{\textrm{d}}50^\text{\textrm{m}}19^\text{\textrm{s}}$& $27$ & $0.6787$ & $0.006$ & \multirow{2}{*}{$549^{+68}_{-68}$} & $0.39$ & $[0.665 , 0.695]$ & \multirow{2}{*}{5}\\
& $p$ &$03^\text{\textrm{h}}31^\text{\textrm{m}}51.1^\text{\textrm{s}}$&$-27^\text{\textrm{d}}50^\text{\textrm{m}}11^\text{\textrm{s}}$& $35$ & $0.680$ & $0.002$ & & $0.36$ & $[0.650 , 0.710]$\\
\multirow{2}{*}{48} & $s$ &$03^\text{\textrm{h}}32^\text{\textrm{m}}21.6^\text{\textrm{s}}$&$-27^\text{\textrm{d}}51^\text{\textrm{m}}45^\text{\textrm{s}}$& $10$ & $0.6714$ & $0.0020$ & \multirow{2}{*}{$1050^{+214}_{-214}$} & $0.39$ & $[0.655 , 0.685]$ & \multirow{2}{*}{1}\\
& $p$ & \nodata & \nodata & \nodata & \nodata & \nodata & & \nodata & \nodata\\
\multirow{2}{*}{49} & $s$ &$03^\text{\textrm{h}}33^\text{\textrm{m}}17.8^\text{\textrm{s}}$&$-27^\text{\textrm{d}}58^\text{\textrm{m}}25^\text{\textrm{s}}$& $10$ & $0.6838$ & $0.0003$ & \multirow{2}{*}{$151^{+39}_{-39}$} & $0.49$ & $[0.670 , 0.700]$ & \multirow{2}{*}{3}\\
& $p$ &$03^\text{\textrm{h}}33^\text{\textrm{m}}15.2^\text{\textrm{s}}$&$-27^\text{\textrm{d}}58^\text{\textrm{m}}19^\text{\textrm{s}}$& $15$ & $0.682$ & $0.004$ & & $0.41$ & $[0.655 , 0.715]$\\
\multirow{2}{*}{50} & $s$ &$03^\text{\textrm{h}}31^\text{\textrm{m}}20.8^\text{\textrm{s}}$&$-27^\text{\textrm{d}}58^\text{\textrm{m}}40^\text{\textrm{s}}$& $14$ & $0.6795$ & $0.0005$ & \multirow{2}{*}{$329^{+67}_{-67}$} & $0.39$ & $[0.665 , 0.695]$ & \multirow{2}{*}{3}\\
& $p$ & \nodata & \nodata & \nodata & \nodata & \nodata & & \nodata & \nodata\\
\multirow{2}{*}{51} & $s$ &$03^\text{\textrm{h}}32^\text{\textrm{m}}25.4^\text{\textrm{s}}$&$-28^\text{\textrm{d}}02^\text{\textrm{m}}31^\text{\textrm{s}}$& $13$ & $0.6799$ & $0.0017$ & \multirow{2}{*}{$1045^{+316}_{-316}$} & $0.42$ & $[0.665 , 0.695]$ & \multirow{2}{*}{1}\\
& $p$ &$03^\text{\textrm{h}}32^\text{\textrm{m}}28.6^\text{\textrm{s}}$&$-28^\text{\textrm{d}}02^\text{\textrm{m}}15^\text{\textrm{s}}$& $8$ & $0.686$ & $0.005$ & & $0.36$ & $[0.650 , 0.710]$\\
\multirow{2}{*}{52\tablenotemark{$\star$}} & $s$ &$03^\text{\textrm{h}}32^\text{\textrm{m}}06.1^\text{\textrm{s}}$&$-27^\text{\textrm{d}}55^\text{\textrm{m}}01^\text{\textrm{s}}$& $29$ & $0.6846$,$0.6732$ & $0.0007$,$0.0005$ & \multirow{2}{*}{$500^{+66}_{-66}$,$246^{+49}_{-49}$}  & $0.39$ & $[0.665 , 0.695]$ & \multirow{2}{*}{5,3}\\
& $p$ &$03^\text{\textrm{h}}32^\text{\textrm{m}}03.4^\text{\textrm{s}}$&$-27^\text{\textrm{d}}55^\text{\textrm{m}}11^\text{\textrm{s}}$& $29$ & $0.681$ & $0.002$ & & $0.35$ & $[0.650 , 0.710]$\\
\multirow{2}{*}{53} & $s$ &$03^\text{\textrm{h}}31^\text{\textrm{m}}37.6^\text{\textrm{s}}$&$-27^\text{\textrm{d}}45^\text{\textrm{m}}32^\text{\textrm{s}}$& $17$ & $0.6796$ & $0.0005$ & \multirow{2}{*}{$331^{+203}_{-203}$} & $0.39$ & $[0.665 , 0.695]$ & \multirow{2}{*}{3}\\
& $p$ &$03^\text{\textrm{h}}31^\text{\textrm{m}}38.7^\text{\textrm{s}}$&$-27^\text{\textrm{d}}45^\text{\textrm{m}}34^\text{\textrm{s}}$& $14$ & $0.685$ & $0.002$ & & $0.36$ & $[0.650 , 0.710]$\\
\multirow{2}{*}{54} & $s$ &$03^\text{\textrm{h}}32^\text{\textrm{m}}15.1^\text{\textrm{s}}$&$-27^\text{\textrm{d}}58^\text{\textrm{m}}04^\text{\textrm{s}}$& $7$ & $0.7360$ & $0.0004$ & \multirow{2}{*}{$154^{+71}_{-71}$} & $0.38$ & $[0.720 , 0.750]$ & \multirow{2}{*}{3}\\
& $p$ & \nodata & \nodata & \nodata & \nodata & \nodata & & \nodata & \nodata\\
\multirow{2}{*}{55} & $s$ &$03^\text{\textrm{h}}31^\text{\textrm{m}}59.0^\text{\textrm{s}}$&$-27^\text{\textrm{d}}54^\text{\textrm{m}}40^\text{\textrm{s}}$& $8$ & $0.7357$ & $0.0010$ & \multirow{2}{*}{$438^{+178}_{-78}$} & $0.38$ & $[0.720 , 0.750]$ & \multirow{2}{*}{4}\\
& $p$ &$03^\text{\textrm{h}}31^\text{\textrm{m}}58.4^\text{\textrm{s}}$&$-27^\text{\textrm{d}}54^\text{\textrm{m}}32^\text{\textrm{s}}$& $9$ & $0.737$ & $0.005$ & & $0.39$ & $[0.700 , 0.770]$\\
\multirow{2}{*}{56} & $s$ &$03^\text{\textrm{h}}32^\text{\textrm{m}}18.6^\text{\textrm{s}}$&$-27^\text{\textrm{d}}47^\text{\textrm{m}}04^\text{\textrm{s}}$& $11$ & $0.7341$ & $0.0009$ & \multirow{2}{*}{$474^{+172}_{-172}$} & $0.38$ & $[0.720 , 0.750]$ & \multirow{2}{*}{5}\\
& $p$ &$03^\text{\textrm{h}}32^\text{\textrm{m}}18.8^\text{\textrm{s}}$&$-27^\text{\textrm{d}}46^\text{\textrm{m}}60^\text{\textrm{s}}$& $30$ & $0.740$ & $0.002$ & & $0.39$ & $[0.700 , 0.770]$\\
\multirow{2}{*}{57} & $s$ &$03^\text{\textrm{h}}31^\text{\textrm{m}}48.9^\text{\textrm{s}}$&$-27^\text{\textrm{d}}45^\text{\textrm{m}}39^\text{\textrm{s}}$& $11$ & $0.7334$ & $0.0009$ & \multirow{2}{*}{$451^{+78}_{-78}$} & $0.38$ & $[0.720 , 0.750]$ & \multirow{2}{*}{5}\\
& $p$ &$03^\text{\textrm{h}}31^\text{\textrm{m}}50.1^\text{\textrm{s}}$&$-27^\text{\textrm{d}}45^\text{\textrm{m}}45^\text{\textrm{s}}$& $13$ & $0.743$ & $0.005$ & & $0.39$ & $[0.700 , 0.770]$\\
\multirow{2}{*}{58} & $s$ &$03^\text{\textrm{h}}32^\text{\textrm{m}}46.2^\text{\textrm{s}}$&$-27^\text{\textrm{d}}41^\text{\textrm{m}}47^\text{\textrm{s}}$& $12$ & $0.7311$ & $0.0007$ & \multirow{2}{*}{$392^{+97}_{-97}$} & $0.38$ & $[0.720 , 0.750]$ & \multirow{2}{*}{5}\\
& $p$ &$03^\text{\textrm{h}}32^\text{\textrm{m}}46.0^\text{\textrm{s}}$&$-27^\text{\textrm{d}}41^\text{\textrm{m}}26^\text{\textrm{s}}$& $20$ & $0.733$ & $0.003$ & & $0.39$ & $[0.700 , 0.770]$\\
\multirow{2}{*}{59} & $s$ &$03^\text{\textrm{h}}32^\text{\textrm{m}}09.1^\text{\textrm{s}}$&$-27^\text{\textrm{d}}56^\text{\textrm{m}}05^\text{\textrm{s}}$& $7$ & $0.7362$ & $0.0008$ & \multirow{2}{*}{$399^{+191}_{-78}$} & $0.38$ & $[0.720 , 0.750]$ & \multirow{2}{*}{4}\\
& $p$ &$03^\text{\textrm{h}}32^\text{\textrm{m}}10.4^\text{\textrm{s}}$&$-27^\text{\textrm{d}}56^\text{\textrm{m}}09^\text{\textrm{s}}$& $13$ & $0.744$ & $0.005$ & & $0.39$ & $[0.700 , 0.770]$\\
\multirow{2}{*}{60} & $s$ &$03^\text{\textrm{h}}33^\text{\textrm{m}}21.5^\text{\textrm{s}}$&$-27^\text{\textrm{d}}46^\text{\textrm{m}}11^\text{\textrm{s}}$& $9$ & $0.8350$ & $0.0011$ & \multirow{2}{*}{$480^{+191}_{-83}$} & $0.75$ & $[0.820 , 0.850]$ & \multirow{2}{*}{5}\\
& $p$ &$03^\text{\textrm{h}}33^\text{\textrm{m}}23.8^\text{\textrm{s}}$&$-27^\text{\textrm{d}}46^\text{\textrm{m}}32^\text{\textrm{s}}$& $10$ & $0.841$ & $0.006$& & $0.80$ & $[0.800 , 0.870]$\\
\multirow{2}{*}{61} & $s$ &$03^\text{\textrm{h}}32^\text{\textrm{m}}48.4^\text{\textrm{s}}$&$-27^\text{\textrm{d}}38^\text{\textrm{m}}47^\text{\textrm{s}}$& $7$ & $0.8374$ & $0.0006$ & \multirow{2}{*}{$264^{+116}_{-47}$} & $0.75$ & $[0.820 , 0.850]$ & \multirow{2}{*}{3}\\
& $p$ & \nodata & \nodata & \nodata & \nodata & \nodata & & \nodata & \nodata\\
\multirow{2}{*}{62} & $s$ &$03^\text{\textrm{h}}32^\text{\textrm{m}}49.4^\text{\textrm{s}}$&$-27^\text{\textrm{d}}41^\text{\textrm{m}}11^\text{\textrm{s}}$& $9$ & $0.9666$ & $0.0004$ & \multirow{2}{*}{$142^{+76}_{-28}$} & $0.75$ & $[0.950 , 0.980]$ & \multirow{2}{*}{3}\\
& $p$ &$03^\text{\textrm{h}}32^\text{\textrm{m}}50.6^\text{\textrm{s}}$&$-27^\text{\textrm{d}}40^\text{\textrm{m}}11^\text{\textrm{s}}$& $10$ & $0.969$ & $0.004$ & & $0.84$ & $[0.930 , 1.000]$\\

%% file: TabM.tex
1	&	5	&	0.1253	&	0.65 & 4.9 & 8.3 & $\checkmark$ \\
5	&	5	&	0.1267	&	0.87 & 6.5 & 20.1 & $\checkmark$ \\
25	&	4	&	0.52	&	0.54 & 1.4 & 7.0 & $\star$ \\
28	&	5	&	0.5205	&	0.51 & 1.4 & 6.0 & $\checkmark$ \\
30	&	4	&	0.5436	&	0.50 & 1.3 & 5.8 & $\dagger$ \\
31	&	5	&	0.5264	&	0.56 & 1.5 & 8.1 & $\dagger$ \\
40	&	5	&	0.6781	&	0.52 & 1.2 & 7.7 & $\checkmark$ \\
41	&	4	&	0.6666	&	0.49 & 1.2 & 6.3 & $\times$ \\
42	&	5	&	0.6661	&	0.45 & 1.1 & 5.0 & $\checkmark$ \\
43	&	4	&	0.6808	&	0.45 & 1.1 & 5.0 & $\checkmark$ \\
45	&	5	&	0.6773	&	0.56 & 1.3 & 9.6 & $\dagger$ \\
47	&	5	&	0.6787	&	0.62 & 1.5 & 13.1 & $\checkmark$ \\
52	&	5	&	0.6846	&	0.56 & 1.3 & 9.8 & $\checkmark$ \\
55	&	4	&	0.7357	&	0.48 & 1.1 & 6.4 & $\checkmark$ \\
56	&	5	&	0.7341	&	0.52 & 1.2 & 8.1 & $\checkmark$ \\
57	&	5	&	0.7334	&	0.49 & 1.1 & 7.0 & $\times$ \\
58	&	5	&	0.7311	&	0.43 & 1.0 & 4.6 & $\checkmark$ \\
59	&	4	&	0.7362	&	0.44 & 1.0 & 4.9 & $\times$ \\
60	&	5	&	0.835	&	0.50 & 1.1 & 8.0 & $\star$ \\

%% file: Clustering.bbl
\begin{thebibliography}{}
\bibitem[Adami et al.(2005)]{ami05}
  Adami, C., Mazure, A., Ilbert, O., et al., 2005, \aap, 443, 805
\bibitem[Arnouts et al.(2001)]{avb01}
  Arnouts, S., Vandame, B., Benoist, C., et al., 2001, \aap, 379, 740
\bibitem[Balestra et al.(2010)]{bmp10}
  Balestra, I., Mainieri, V., Popesso, P., et al., 2010, \aap, 512, 12
\bibitem[Beers et al.(1990)]{bfg90}
  Beers, T. C., Flynn, K., \& Gebhardt, K., 1990, \aj, 100, 32
\bibitem[Caballero \& Dinis(2008)]{cd08}
  Caballero, J. A., \& Dinis, L., 2008, Astronomische Nachrichten, 329, 801
\bibitem[Cardamone et al.(2010)]{cdm10}
  Cardamone, C. N., van Dokkum, P. G.; Urry, C. M., et al., 2010, \apjs, 189, 270
\bibitem[Comastri et al.(2011)]{cri11}
  Comastri, A., Ranalli, P., Iwasawa, K., et al., 2011, \aap, 526, L9
\bibitem[Cooper et al.(2012)]{cyd12}
  Cooper, M. C., Yan, R., Dickinson, M., et al., 2012, \mnras, 425, 2116
\bibitem[Dahlen et al.(2010)]{dmd10}
  Dahlen, T., Mobasher, B.; Dickinson, M., et al., 2010, \apj, 724, 425
\bibitem[D\'{\i}az--S\'anchez et al.(2007)]{dia07}
  D\'{\i}az--S\'anchez, A., Villo-P\'erez, I., P\'erez-Garrido, A., \& Rebolo, R., 2007, \mnras, 337, 516
\bibitem[Erben et al.(2005)]{tsd05}
  Erben, T., Schirmer, M., Dietrich, J. P., et al., 2005, Astronomische Nachrichten, 326, 432
\bibitem[Ester et al.(1996)]{eks96}
  Ester, M., Kriegal, H. P., \& Sander, J., 1996, Proc. 2nd Int. Conf. on Knowledge Discovery and Data Mining (KDD'96). AAAI Press, Menlo Park, CA, pp. 226-231, August 1996
\bibitem[Evrard(2004)]{e04}
  Evrard, A. E., 2004, p. 1, Clusters of Galaxies: Probes of Cosmological Structure and Galaxy Evolution, from the Carnegie Observatories Centennial Symposia. Published by Cambridge University Press, as part of the Carnegie Observatories Astrophysics Series. Edited by J.S. Mulchaey, A. Dressler, and A. Oemler 
\bibitem[Finoguenov et al.(2007)]{fgh07}
  Finoguenov, A., Guzzo, L., Hasinger, G., et al., 2007, \apjs, 172, 182
\bibitem[Finoguenov et al.(2010)]{fwt10}
  Finoguenov, A., Watson, M. G., Tanaka, M., et al., 2010, \mnras, 403, 2063
\bibitem[Fixsen et al.(1996)]{fcg96}
  Fixsen, D. J., Cheng, E. S., Gales, J. M., et al., 1996, \apj, 473, 576
\bibitem[Giacconi et al.(2002)]{gzw02}
  Giacconi, R., Zirm, A., Wang, J., et al., 2002, \apjs, 139, 369
\bibitem[Gilli et al.(2003)]{gcd03}
  Gilli, R., Cimatti, A., Daddi, E., et al., 2003, \apj, 592, 721
\bibitem[Gooch(1996)]{g96}
  Gooch, R., 1996, Astronomical Data Analysis Software and Systems V, A.S.P. Conference Series, Vol. 101, 1996, George H. Jacoby and Jeannette Barnes, eds., p. 80
\bibitem[Hildebrandt et al.(2006)]{hed06}
  Hildebrandt, H., Erben, T., Dietrich, J. P., et al., 2006, \aap, 452, 1121
\bibitem[Ilbert et al.(2006)]{iam06}
  Ilbert, O., Arnouts, S., McCracken, H. J., et al., 2006, \aap, 457, 841
\bibitem[Kang \& Im(2009)]{ki09}
  Kang, E. \& Im, M., 2009, \apj, 691, 33
\bibitem[Menci et al.(2008)]{mrg08}
  Menci, N., Rosati, P., Gobat, R., et al., 2008, \apj, 685, 863
\bibitem[Nuza et al.(2013)]{nsp13}
  Nuza, S. E., S\'anchez, A. G.; Prada, F., et al., 2013, \mnras, 432, 743
\bibitem[Old et al.(2013)]{ogp13}
  Old, L., Gray, M. E., \& Pearce, F. R., 2013, \mnras, 434, 2606
\bibitem[Peebles(1980)]{p80}
  Peebles, P. J. E., 1980, The large-scale structure of the universe, Research supported by the National Science Foundation. Princeton, N.J., Princeton University Press
\bibitem[Popesso et al.(2009)]{pdn09}
  Popesso, P., Dickinson, M., Nonino, M., et al., 2009, \aap, 494, 443
\bibitem[Ranalli et al.(2013)]{rcv13}
  Ranalli, P., Comastri, A., Vignali, C., et al., 2013, \aap, 555, 42
\bibitem[Rizzo et al.(2004)]{rab04}
  Rizzo, D., Adami, C., Bardelli, S., et al., 2004, \aap, 413, 453
\bibitem[Salimbeni et al.(2009)]{scp09}
  Salimbeni, S.,  Castellano, M., Pentericci, L., et al., 2009, \aap, 501, 865
\bibitem[Struble \& Rood(1991)]{sr91}
  Struble, M., \& Rood, H., 1991, \apjs, 77, 363
\bibitem[Takizawa et al.(2010)]{tnm10}
  Takizawa, M., Nagino, R., \& Matsushita, K., 2010, \pasj, 62, 951
\bibitem[Tramacere \& Vecchio(2012)]{tv12}
  Tramacere, A. \& Vecchio, C., 2012, AIP Conf. Proc. 1505, pp. 705-708
\bibitem[Trevese et al.(2007)]{tcf07}
  Trevese, D., Castellano, M., Fontana, A., \& Giallongo, E., 2007, \aap, 463, 853
\bibitem[Voit(2005)]{v05}
  Voit, G. M., 2005, RvMP, 77, 207
\bibitem[Wagstaff et al.(2013)]{wtl13}
  Wagstaff, K., Tang, B., Lazio, T. J., \& Spolaor, S., (2013), American Astronomical Society, AAS Meeting \#221, \#154.02
\bibitem[Way et al.(2005)]{wqi05}
  Way, M. J., Quintana, H., Infante, L., Lambas, D. G., \& Muriel, H., 2005, \aj, 130, 2012
\bibitem[Wolf et al.(2004)]{wmk04}
  Wolf, C.,  Meisenheimer, K., Kleinheinrich, M., et al., 2004, \aap, 421, 913
\end{thebibliography}
